\documentclass[]{jfm}
\usepackage{multirow}
\usepackage{graphicx}
\usepackage{newtxtext}
\usepackage{newtxmath}
\usepackage{natbib}
\usepackage{hyperref}
\usepackage{subfigure}
\usepackage{xcolor}
\hypersetup{
    colorlinks = true,
    urlcolor   = blue,
    citecolor  = red,
}

\newcommand{\RomanNumeralCaps}[1]

\title{The interscale behaviour of uncertainty in
	three-dimensional Navier-Stokes turbulence}

\author{Jin GE\aff{1}
  \corresp{\email{jin.ge@cnrs.fr}},
  Joran
  ROLLAND\aff{1} \corresp{\email{joran.rolland@centralelille.fr}}
 \and John
 Christos
 VASSILICOS\aff{1}\corresp{\email{john-christos.vassilicos@cnrs.fr}}}

\affiliation{\aff{1}Univ. Lille, CNRS, ONERA, Arts et Métiers ParisTech,
	Centrale Lille, UMR 9014 - LMFL - Laboratoire de Mécanique des
	Fluides de Lille - Kampé de Feriet, F-59000 Lille, France}

\begin{document}
\maketitle

\begin{abstract}
We derive the scale-by-scale uncertainty energy budget equation and
demonstrate theoretically and computationally the presence of a
self-similar equilibrium cascade of decorrelation in an inertial range
of scales during the time range of power law growth of uncertainty in
statistically stationary homogeneous turbulence. This cascade is
predominantly inverse and driven by compressions of the reference
field's relative deformation tensor and their aligments with the
uncertainty velocity field. Three other subdominant cascade mechanisms
are also present, two of which are forward and also dominated by
compressions and one of which, the weakest and the only non-linear one
of the four, is inverse. The uncertainty production and dissipation
scalings which may follow from the self-similar equilibrium cascade of
decorrelation lead to power law growths of the uncertainty integral
scale and the average uncertainty energy which are also investigated.
Compressions are not only key to chaoticity, as previously shown, but
also to stochasticity.
\end{abstract}


\section{Introduction}
\label{sec:Introduction}

There is a consensus that infinite long-time predictions for turbulent fields are impossible \citep{lesieur1987turbulence}. \citet{lorenz1963deterministic}'s pioneering work demonstrated the extreme sensitivity of non-linear systems to initial conditions leading to the presence of chaos and stange attractors. This uncertainty also exists in fully developed turbulent flow solutions of the Navier-Stokes equations, characterized by non-linearity and a vast range of scales \citep{deissler1986navier}.

Incompressible Navier-Stokes turbulence is characterized by its
time-dependent velocity field. Within the Eulerian framework, the
uncertainty on a velocity field $\boldsymbol{u}^{(1)}(\boldsymbol{x},t)$ is quantified
by the point-to-point velocity difference from another realisation,
$\boldsymbol{u}^{(2)}(\boldsymbol{x},t)$, which is initially infinitesimally close to
$\boldsymbol{u}^{(1)}(\boldsymbol{x},t)$. This difference defines the uncertainty
field $\boldsymbol{w}(\boldsymbol{x},t) \equiv \boldsymbol{u}^{(2)}(\boldsymbol{x},t) -
\boldsymbol{u}^{(1)}(\boldsymbol{x},t)$. The average uncertainty is then evaluated
based on this field's average kinetic energy $\left\langle
E_{\Delta}\right\rangle \equiv \left\langle \left| \boldsymbol{w} \right|^2 /
2 \right\rangle$, where the brackets $\left\langle \right\rangle$
denote a spatial average over $\boldsymbol{x}$. At very early times, when the
two realisations are infinitesimally close, $\boldsymbol{w}$ contains
information from the smallest scales of motion only, given that this
is where the small initial uncertainty is introduced. Using the incompressible Navier-Stokes equations, \citet{ge2023production} derived the time-evolution equation for
$\left\langle E_{\Delta}\right\rangle$ in the case of
homogeneous/periodic incompressible turbulence:
\begin{equation}
	\label{eq:single point uncertainty equation}
	\frac{{\rm d}}{{\rm d}t}\left\langle E_{\Delta}\right\rangle=\left\langle P_{\Delta}\right\rangle-\left\langle \varepsilon_{\Delta}\right\rangle+\left\langle F_{\Delta}\right\rangle. 
\end{equation}

The average uncertainty energy of homogeneous/periodic turbulence is
therefore determined by three mechanisms: internal production of
uncertainty ($P_{\Delta}$), dissipation of uncertainty
($\varepsilon_{\Delta}$) and external input rate of uncertainty
($F_{\Delta}$) via external forcing applied to the Navier-Stokes local
momentum balance (exact expressions for $P_{\Delta}$,
$\varepsilon_{\Delta}$ and $F_{\Delta}$ can be found in the following
section). \citet{ge2023production} proved that the internal
production of uncertainty results from local compressions whereas
local stretchings reduce uncertainty. Hence, uncertainty grows in
Navier-Stokes periodic/homogeneous turbulence because local
compressions overwhelm local stretchings in their effects on
uncertainty. In fact, uncertainty grows in non-decaying
homogeneous/periodic turbulence only after a short initial time
dominated by average dissipation $\left\langle
\varepsilon_{\Delta}\right\rangle$ during which average production
$\left\langle P_{\Delta}\right\rangle$ builds up and soon overwhelms
average dissipation. Then, in statistically stationary homogeneous
turbulence, the average uncertainty energy $\left\langle
E_{\Delta}\right\rangle$ grows exponentially, i.e.
\begin{equation} \label{eq:exponential growth of uncertainty}
	\frac{{\rm d}~}{{\rm d}t}\left\langle
	E_{\Delta}\right\rangle\sim\lambda\left\langle
	E_{\Delta}\right\rangle,
\end{equation}
in agreement with the strange attractor concept of maximal Lyapunov
exponent $\lambda$ in the presence of chaotic dynamics. This chaotic
growth ends when the energy spectrum of
$\boldsymbol{w}(\boldsymbol{x},t)$ grows to eventually equal the sum
of energy spectra of the statistically stationary reference and
perturbed fields at dissipative range wavenumbers.

Following this chaotic regime, the uncertainty continues to grow
because of uncertainty production by local compressions as
demonstrated by \citet{ge2023production}. As the Direct
Numerical Simulations (DNS) of \citet{ge2023production} have
demonstrated, the integral length scale $l_{\Delta}$ of the
uncertainty field $\boldsymbol{w}(\boldsymbol{x},t)$ grows too. A time is therefore
expected when $l_{\Delta}$ grows above the Taylor length
$l_{\lambda}^{(1)}$ of the statistically stationary and homogeneous
reference field $\boldsymbol{u}^{(1)}(\boldsymbol{x},t)$. From then on, the
uncertainty field may be considered to have its own growing range of
inertial length scales of eddy motions between $l_{\lambda}^{(1)}$ and
$l_{\Delta}$. This multi-scale nature of the uncertainty field changes
the way $\boldsymbol{u}^{(1)}(\boldsymbol{x},t)$ and $\boldsymbol{u}^{(2)}(\boldsymbol{x},t)$
gradually decorrelate and $\left\langle E_{\Delta}\right\rangle$ and
$l_{\Delta}$ grow in time.

When decorrelation between $\boldsymbol{u}^{(1)}$ and
$\boldsymbol{u}^{(2)}$ occurs within an inertial range, the
uncertainty is predominantly due to stochasticity
\citep{lorenz1969predictability, thalabard2020butterfly} rather than
chaoticity. The aforementioned early-time exponential growth of
uncertainty, often associated with the `butterfly effect' as described
by \citet{gleick2008chaos}, is driven by chaoticity. According to
equation (\ref{eq:exponential growth of uncertainty}), this chaotic
growth can be reduced by sufficiently reducing the initial
uncertainty, i.e. the initial difference between
$\boldsymbol{u}^{(2)}(\boldsymbol{x},t=0)$ and
$\boldsymbol{u}^{(1)}(\boldsymbol{x},t=0)$. However, as
\citet{lorenz1969predictability} pointed out, the predictability of
multi-scale velocity fields cannot be enhanced merely by reducing the
amplitude of initial uncertainty. The uncertainty field
$\boldsymbol{w}$ contains a wide range of scales and progressively
contaminates larger scales \citep{lorenz1969predictability,
  leith1971atmospheric}. In conditions of infinite Reynolds number
turbulence, \citet{leith1972predictability} and
\citet{metais1986statistical} used the Eddy Damped Quasi-Normal
Markovian (EDQNM) model to demonstrate strong inverse cascade of
uncertainty in both statistically stationary and decaying homogeneous
turbulence. As shown by \citet{chen2022scalings}, inverse cascades of
turbulent energy require predominance of stretching over compressive
motions on inter-scale energy transfers.

Our first aim here is to investigate how the uncertainty production
mechanism by local-in-space compressions articulates with inverse
uncertainty energy transfers in scale-space by local-in-scales
stretching motions. We address this question with the same tools used
by \citet{chen2022scalings} to relate compression
and stretching motions to forward and inverse inter-scale energy
transfers/cascades respectively. We therefore use the incompressible
Navier-Stokes equations to derive the evolution equation for the
two-point structure function of the uncertainty field $\boldsymbol{w}$ in
statistically homogeneous turbulence. This equation parallels the
Kármán-Howarth-Monin-Hill (KHMH) equation
\citep{de1938statistical,monin2013statistical,hill2002exact} which
governs the scale-by-scale energy of the velocity field $\boldsymbol{u}^{(1)}$
or $\boldsymbol{u}^{(2)}$ and which has been widely used to analyse numerical
simulation and experimental data for various turbulent flows
\citep{marati2004energy,danaila2012yaglom,togni2015physical,valente2015energy,cimarelli2016cascades,portela2017turbulence}.
From the KHMH-like equation that we derive for the uncertainty field,
we identify the terms that contribute to the backscatter of
uncertainty and how they depend on stretching/compressing motions of
the reference field $\boldsymbol{u}^{(1)}$ and the uncertainty field
$\boldsymbol{w}$. Our second aim is to attempt to relate this backscatter of
uncertainty to the growth of $l_{\Delta}$ and $\left\langle
E_{\Delta}\right\rangle$ in the time range where the velocity
difference field has a multi-scale structure over a range of scales
between the Taylor length $l_{\lambda}^{(1)}$ and the uncertainty
field's correlation length $l_{\Delta}$.

In the next section we derive a KHMH equation for the uncertainty field and apply it to
the case of periodic/homogeneous turbulence. In this section
\ref{sec:uncertainty transport equation} we are led to analyse
production and interscale transfers of uncertainty in terms of
relative deformation rates of the reference and uncertainty fields at
different scales. In section \ref{sec:Production/dissipation scaling of uncertainty energy} we introduce the equilibrium model of uncertainty growth in
the stochastic time regime and derive from it two alternative power
law growths in time for $\left\langle E_{\Delta}\right\rangle$ and
$l_{\Delta}$ based on two alternative uncertainty production and
dissipation scalings. In section \ref{sec:Numerical steups} we present
the DNS of stationary periodic turbulence that we use to numerically
integrate the reference and uncertainty fields. In section
\ref{sec:Numerical results} we apply the KHMH and relative deformation
tensor framework to these fields to study uncertainty interscale
transfers and the time evolution of $\left\langle
E_{\Delta}\right\rangle$ and $l_{\Delta}$. We conclude in section \ref{sec:Conclusion}. 

\section{\label{sec:uncertainty transport equation}Uncertainty transport in periodic/homogeneous turbulence}

In this section we use a two-point framework to study uncertainty
transfer through scales and then apply this framework to
periodic/homogeneous turbulence. We define the two-point uncertainty
half difference
$\delta\boldsymbol{w}\equiv\left(\boldsymbol{w}^{+}-\boldsymbol{w}^{-}\right)/2$ where
$\boldsymbol{w}^{+}\equiv\boldsymbol{w}(\boldsymbol{x}^{+},t)$ and
$\boldsymbol{w}^{-}\equiv\boldsymbol{w}(\boldsymbol{x}^{-},t)$ are the uncertainty velocities
at spatial positions $\boldsymbol{x}^{+}=\boldsymbol{X}+\boldsymbol{r}/2$ and
$\boldsymbol{x}^{-}=\boldsymbol{X}-\boldsymbol{r}/2$ respectively, and derive a KHMH equation
for the uncertainty field $\boldsymbol{w}$, i.e. the transport equation for
$\left|\delta\boldsymbol{w}\right|^{2}$. Following \citet{germano2007direct}
and \citet{beaumard2024scale} we also derive the transport equation
for $\left|\overline{\boldsymbol{w}}\right|^{2}$, where $\overline{\boldsymbol{w}}$ is
the two-point uncertainty half sum
$\overline{\boldsymbol{w}}\equiv\left(\boldsymbol{w}^{+}+\boldsymbol{w}^{-}\right)/2$. Evidently,
$\left|\delta\boldsymbol{w}\right|^{2}$ and
$\left|\overline{\boldsymbol{w}}\right|^{2}$ are complementary since
$\left|\delta\boldsymbol{w}\right|^{2}+\left|\overline{\boldsymbol{w}}\right|^{2}=(E_{\Delta}^{+}+E_{\Delta}^{-})$
where $E_{\Delta}^{+}\equiv\left|\boldsymbol{w}(\boldsymbol{x}^{+},t)\right|^{2}/2$
and $E_{\Delta}^{-}\equiv\left|\boldsymbol{w}(\boldsymbol{x}^{-},t)\right|^{2}/2$ are
instantaneous single-point uncertainty energies at local positions
$\boldsymbol{x}^{+}$ and $\boldsymbol{x}^{-}$ respectively. Comparing the two
transport equations allows to identify production and transport terms
appearing in both equations.
	\subsection{\label{sec:Transport equation for uncertainty}Transport equation for uncertainty}
Both reference flow $\boldsymbol{u}^{(1)}(\boldsymbol{x},t)$ and perturbed
flow $\boldsymbol{u}^{(2)}(\boldsymbol{x},t)$ are governed by the same
incompressible Navier-Stokes equation, which is
\begin{equation}
	\label{eq:NS equation}
	\begin{aligned}			
		\frac{\partial~}{\partial t}u^{(m)}_{i}+u^{(m)}_{j}\frac{\partial~}{\partial x_{j}}u^{(m)}_{i}=&-\frac{\partial~}{\partial x_{i}}p^{(m)}+\nu\frac{\partial^{2}~}{\partial x_{j}\partial x_{j}}u^{(m)}_{i}+f^{(m)}_{i},\\
		&\frac{\partial~}{\partial x_{i}}u^{(m)}_{i}=0,
	\end{aligned}
\end{equation}
where the number $m=1$ or $2$ in the superscript parentheses indicates
reference or perturbed field respectively, $\nu$ is the fluid's
kinematic viscosity; $p^{(m)}$ is the pressure field divided by the
fluid's mass density and $\boldsymbol{f}$ is the external
forcing. Considering equations (\ref{eq:NS equation}) at points
$\boldsymbol{x}^{+}\equiv\boldsymbol{X}+\boldsymbol{r}/2$ and
$\boldsymbol{x}^{-}\equiv\boldsymbol{X}-\boldsymbol{r}/2$
respectively, taking the difference and sum of these equations at
these two points and changing variables from $\boldsymbol{x}^{+}$ and
$\boldsymbol{x}^{-}$ to spatial centroid $\boldsymbol{X}$ and
separation vector $\boldsymbol{r}$, we obtain
	\begin{small}
	\begin{subequations}
		\label{eq:NS difference/average}
		\begin{align}
			&\frac{\partial~}{\partial t}\delta u_{i}^{(m)}+\overline{u}^{(m)}_{j}\frac{\partial~}{\partial X_{j}}\delta u_{i}^{(m)}+2\delta u^{(m)}_{j}\frac{\partial~}{\partial r_{j}}\delta u_{i}^{(m)}=-\frac{\partial~}{\partial X_{i}}\delta p^{(m)}
			+\frac{\nu}{2}\frac{\partial^{2}~}{\partial X_{j}\partial X_{j}}\delta u_{i}^{(m)}+2\nu\frac{\partial^{2}~}{\partial r_{j}\partial r_{j}}\delta u_{i}^{(m)}+\delta f_{i}^{(m)},&\\
			&\frac{\partial~}{\partial t}\overline{u}_{i}^{(m)}+\overline{u}^{(m)}_{j}\frac{\partial~}{\partial X_{j}}\overline{u}_{i}^{(m)}+2\delta u_{j}^{(m)}\frac{\partial~}{\partial r_{j}}\overline{u}_{i}^{(m)}=-\frac{\partial~}{\partial X_{i}}\overline{p}^{(m)}
			+\frac{\nu}{2}\frac{\partial^{2}~}{\partial X_{j}\partial X_{j}}\overline{u}_{i}^{(m)}+2\nu\frac{\partial^{2}~}{\partial r_{j}\partial r_{j}}\overline{u}_{i}^{(m)}+\overline{f}_{i}^{(m)},&
		\end{align}
	\end{subequations}
\end{small}
with the incompressibility conditions    
\begin{equation*}
	\begin{aligned}
		\frac{\partial~}{\partial X_{i}}\overline{u}_{i}^{(m)}=0,\quad \frac{\partial~}{\partial r_{i}}\overline{u}_{i}^{(m)}=0,\quad
		\frac{\partial~}{\partial X_{i}}\delta u_{i}^{(m)}=0,\quad \frac{\partial~}{\partial r_{i}}\delta u_{i}^{(m)}=0
	\end{aligned}
\end{equation*}
where terms starting with $\delta$ such as $\delta u_{i}^{(m)}$ are
two-point velocity half differences and terms with overbars such as
$\overline{u}_{i}^{(m)}$ are two-point velocity half sums defined
similarly to $\delta\boldsymbol{w}$ and $\overline{\boldsymbol{w}}$ in the first
paragraph of this section. Taking the difference of equations
(\ref{eq:NS difference/average}) for $m=1$ and equations (\ref{eq:NS
	difference/average}) for $m=2$, and using
$\frac{1}{2}\left(q^{(2)+}\pm
q^{(2)-}\right)-\frac{1}{2}\left(q^{(1)+}\pm
q^{(1)-}\right)=\frac{1}{2}\left(q^{(2)+}-q^{(1)+}\right)\pm\frac{1}{2}\left(q^{(2)-}-q^{(1)-}\right)$
which is valid for any quantity $q$, we now obtain
	\begin{small}
	\begin{subequations}
		\label{eq:difference NS difference/average}
		\begin{align}
			&\frac{\partial~}{\partial t}\delta w_{i}+\overline{u}^{(1)}_{j}\frac{\partial~}{\partial X_{j}}\delta w_{i}+2\delta u^{(1)}_{j}\frac{\partial~}{\partial r_{j}}\delta w_{i}+\overline{w}_{j}\frac{\partial~}{\partial X_{j}}\delta w_{i}+2\delta w_{j}\frac{\partial~}{\partial r_{j}}\delta w_{i}
			+\overline{w}_{j}\frac{\partial~}{\partial X_{j}}\delta u^{(1)}_{i}+2\delta w_{j}\frac{\partial~}{\partial r_{j}}\delta u^{(1)}_{i}\notag&\\&\qquad\qquad\qquad\qquad\qquad\qquad\qquad\qquad\qquad\qquad=-\frac{\partial~}{\partial X_{i}}\delta h
			+\frac{\nu}{2}\frac{\partial^{2}~}{\partial X_{j}\partial X_{j}}\delta w_{i}+2\nu\frac{\partial^{2}~}{\partial r_{j}\partial r_{j}}\delta w_{i}+\delta g_{i},\label{eq:difference NS difference}&\\
			\notag\\
			&\frac{\partial~}{\partial t}\overline{w}_{i}+\overline{u}^{(1)}_{j}\frac{\partial~}{\partial X_{j}}\overline{w}_{i}+2\delta u^{(1)}_{j}\frac{\partial~}{\partial r_{j}}\overline{w}_{i}+\overline{w}_{j}\frac{\partial~}{\partial X_{j}}\overline{w}_{i}+2\delta w_{j}\frac{\partial~}{\partial r_{j}}\overline{w}_{i}
			+\overline{w}_{j}\frac{\partial~}{\partial X_{j}}\overline{u}^{(1)}_{i}+2\delta w_{j}\frac{\partial~}{\partial r_{j}}\overline{u}^{(1)}_{i}\notag&\\&\qquad\qquad\qquad\qquad\qquad\qquad\qquad\qquad\qquad\qquad=-\frac{\partial~}{\partial X_{i}}\overline{h}
			+\frac{\nu}{2}\frac{\partial^{2}~}{\partial X_{j}\partial X_{j}}\overline{w}_{i}+2\nu\frac{\partial^{2}~}{\partial r_{j}\partial r_{j}}\overline{w}_{i}+\overline{g}_{i},\label{eq:difference NS average}&
		\end{align}
	\end{subequations}
\end{small}where $h\equiv p^{(2)}-p^{(1)}$ is the pressure uncertainty field
(divided by the fluid's mass density) and
$\boldsymbol{g}\equiv\boldsymbol{f}^{(2)}-\boldsymbol{f}^{(1)}$ is the difference
between the forcings applied to the reference and perturbed
fields. Similarly, the incompressibility conditions are
\begin{equation*}
	\begin{aligned}
		\frac{\partial~}{\partial X_{i}}\overline{w}_{i}=0,\quad \frac{\partial~}{\partial r_{i}}\overline{w}_{i}=0,\quad
		\frac{\partial~}{\partial X_{i}}\delta w_{i}=0,\quad \frac{\partial~}{\partial r_{i}}\delta w_{i}=0.
	\end{aligned}
\end{equation*} 
By taking the scalar product of equation (\ref{eq:difference NS
	difference}) with $2\delta w_{i}$ and of equation
(\ref{eq:difference NS average}) with $2\overline{w}_{i}$ we arrive at
\begin{subequations}
	\label{eq:difference KHMH difference/average}
	\begin{align}
		\frac{\partial~}{\partial t}\underbrace{\left|\delta\boldsymbol{w}\right|^{2}}_{\mathcal{A}_{\delta}}&+\underbrace{\frac{\partial~}{\partial X_{j}}\overline{u}^{(1)}_{j}\left|\delta\boldsymbol{w}\right|^{2}+\frac{\partial~}{\partial X_{j}}\overline{w}_{j}\left|\delta\boldsymbol{w}\right|^{2}}_{-\mathcal{T}_{\delta}}+\underbrace{2\frac{\partial~}{\partial r_{j}}\delta u^{(1)}_{j}\left|\delta\boldsymbol{w}\right|^{2}+2\frac{\partial~}{\partial r_{j}}\delta w_{j}\left|\delta\boldsymbol{w}\right|^{2}}_{-\Pi_{\delta}}\notag&\\&\qquad\qquad\qquad\qquad+\underbrace{2\delta w_{i}\overline{w}_{j}\frac{\partial~}{\partial X_{j}}\delta u^{(1)}_{i}+4\delta w_{i}\delta w_{j}\frac{\partial~}{\partial r_{j}}\delta u^{(1)}_{i}}_{-\mathcal{P}_{\delta}}=\notag&\\&\underbrace{-2\frac{\partial~}{\partial X_{i}}\delta w_{i}\delta h}_{\mathcal{T}_{\delta, p}}
		+\underbrace{\frac{\nu}{2}\frac{\partial^{2}~}{\partial X_{j}\partial X_{j}}\left|\delta\boldsymbol{w}\right|^{2}+2\nu\frac{\partial^{2}~}{\partial r_{j}\partial r_{j}}\left|\delta\boldsymbol{w}\right|^{2}}_{\mathcal{D}_{\delta}}-\frac{1}{2}\left(\varepsilon_{\Delta}^{+}+\varepsilon_{\Delta}^{-}\right)+\underbrace{2\delta g_{i}\delta w_{i}}_{\mathcal{F}_{\delta}},\label{eq:difference KHMH difference}&\\
		\notag\\
		\frac{\partial~}{\partial t}\underbrace{\left|\overline{\boldsymbol{w}}\right|^{2}}_{\mathcal{A}_{X}}&+\underbrace{\frac{\partial~}{\partial X_{j}}\overline{u}^{(1)}_{j}\left|\overline{\boldsymbol{w}}\right|^{2}+\frac{\partial~}{\partial X_{j}}\overline{w}_{j}\left|\overline{\boldsymbol{w}}\right|^{2}}_{-\mathcal{T}_{X}}+\underbrace{2\frac{\partial~}{\partial r_{j}}\delta u^{(1)}_{j}\left|\overline{\boldsymbol{w}}\right|^{2}+2\frac{\partial~}{\partial r_{j}}\delta w_{j}\left|\overline{\boldsymbol{w}}\right|^{2}}_{-\Pi_{X}}\notag&\\&\qquad\qquad\qquad\qquad
		+\underbrace{2\overline{w}_{i}\overline{w}_{j}\frac{\partial~}{\partial X_{j}}\overline{u}^{(1)}_{i}+4\overline{w}_{i}\delta w_{j}\frac{\partial~}{\partial r_{j}}\overline{u}^{(1)}_{i}}_{-\mathcal{P}_{X}}=\notag&\\&\underbrace{-2\frac{\partial~}{\partial X_{i}}\overline{w}_{i}\overline{h}}_{\mathcal{T}_{X, p}}
		+\underbrace{\frac{\nu}{2}\frac{\partial^{2}~}{\partial X_{j}\partial X_{j}}\left|\overline{\boldsymbol{w}}\right|^{2}+2\nu\frac{\partial^{2}~}{\partial r_{j}\partial r_{j}}\left|\overline{\boldsymbol{w}}\right|^{2}}_{\mathcal{D}_{X}}-\frac{1}{2}\left(\varepsilon_{\Delta}^{+}+\varepsilon_{\Delta}^{-}\right)+\underbrace{2\overline{g}_{i}\overline{w}_{i}}_{\mathcal{F}_{X}},\label{eq:difference KHMH average}&
	\end{align}  
\end{subequations}
where $\varepsilon_{\Delta}=\nu\frac{\partial w_{i}}{\partial
	x_{j}}\frac{\partial w_{i}}{\partial x_{j}}$ is the single-point
uncertainty dissipation rate and the superscripts $\pm$ indicate where
it is evaluated ($\boldsymbol{x}^{+}$ or $\boldsymbol{x}^{-}$). Equations
(\ref{eq:difference KHMH difference/average}) describe the inter-space
and inter-scale behaviours of uncertainty in $\boldsymbol{X}$ and $\boldsymbol{r}$
spaces; they consist of the following eight sets of terms:
\begin{enumerate}[i)]
	\item $\frac{\partial~}{\partial t}\mathcal{A}_{\delta}$ and
	$\frac{\partial~}{\partial t}\mathcal{A}_{X}$ represent the rate of
	change in time of $\left|\delta\boldsymbol{w}\right|^{2}$ and
	$\left|\overline{\boldsymbol{w}}\right|^{2}$ respectively.
	\item $\mathcal{T}_{\delta}$ and $\mathcal{T}_{X}$ are inter-space
	transport rates of $\left|\delta\boldsymbol{w}\right|^{2}$ and
	$\left|\overline{\boldsymbol{w}}\right|^{2}$ respectively as they are
	conservative terms in $\boldsymbol{X}$-space.
	\item $\Pi_{\delta}$ and $\Pi_{X}$ are inter-scale transfer rates of
	$\left|\delta\boldsymbol{w}\right|^{2}$ and
	$\left|\overline{\boldsymbol{w}}\right|^{2}$ respectively as they are
	conservative tems in $\boldsymbol{r}$-space. In particular,
	$\Pi_{\delta}=\Pi_{\delta,\text{ref}}+\Pi_{\delta,\text{err}}$ where
	$\Pi_{\delta,\text{ref}}\equiv-2\frac{\partial~}{\partial r_{j}}\delta
	u^{(1)}_{j}\left|\delta\boldsymbol{w}\right|^{2}$ represents inter-scale
	transfer rate of $\left|\delta\boldsymbol{w}\right|^{2}$ by the reference
	field and $\Pi_{\delta,\text{err}}\equiv-2\frac{\partial~}{\partial
		r_{j}}\delta w_{j}\left|\delta\boldsymbol{w}\right|^{2}$ represents
	inter-scale transfer rate of $\left|\delta\boldsymbol{w}\right|^{2}$ by
	itself.
	\item $\mathcal{P}_{\delta}$ and $\mathcal{P}_{X}$ appear as two-point
	production rates but consist of one-point production of uncertainty
	and transfer rates as shown in the following dedicated sub-section.
	\item $\mathcal{T}_{\delta, p}$ and $\mathcal{T}_{X, p}$ are the
	velocity-pressure terms having the form of pressure transport rates
	through $\boldsymbol{X}$-space.
	\item $\mathcal{D}_{\delta}$ and $\mathcal{D}_{X}$ are viscous
	diffusion rates of $\left|\delta\boldsymbol{w}\right|^{2}$ and
	$\left|\overline{\boldsymbol{w}}\right|^{2}$ respectively.
	\item
	$\frac{1}{2}\left(\varepsilon_{\Delta}^{+}+\varepsilon_{\Delta}^{-}\right)$
	is the average of the single-point dissipation rates at points
	$\boldsymbol{x}^{+}$ and $\boldsymbol{x}^{-}$ and is also the dissipation rate of
	both $\left|\delta\boldsymbol{w}\right|^{2}$ and
	$\left|\overline{\boldsymbol{w}}\right|^{2}$.
	\item $\mathcal{F}_{\delta}$ and $\mathcal{F}_{X}$ are the
	input/output rates induced by the external forcings.
\end{enumerate}
\subsection{Uncertainty production and transfers}
In this subsection, we start the analysis of the important terms
$\mathcal{P}_{\delta}$ in equation (\ref{eq:difference KHMH
	difference}) and $\mathcal{P}_{X}$ in equation (\ref{eq:difference
	KHMH average}). Straightforward manipulations involving changes of
variables from $\boldsymbol{X}$ and $\boldsymbol{r}$ to $\boldsymbol{x}^{+}$ and $\boldsymbol{x}^{-}$
and incompressibility lead to
\begin{equation}
	\label{eq:lasttwoterms of I in terms of delta}
	\mathcal{P}_{\delta}=\frac{1}{2}\left(P_{\Delta}^{+}+P_{\Delta}^{-}\right)
	+\underbrace{\frac{1}{4}\frac{\partial~}{\partial X_{j}}\left(w_{i}^{-}w_{j}^{+}u_{i}^{(1)+}+w_{i}^{+}w_{j}^{-}u_{i}^{(1)-}\right)}_{\mathcal{T}_{\mathcal{P}_{\delta}}}
	+\underbrace{\frac{1}{2}\frac{\partial~}{\partial r_{j}}\left(w_{i}^{-}w_{j}^{+}u_{i}^{(1)+}-w_{i}^{+}w_{j}^{-}u_{i}^{(1)-}\right)}_{\Pi_{\mathcal{P}_{\delta}}}
\end{equation}
and
\begin{equation}
	\label{eq:singlepoint of Iave}
	\mathcal{P}_{X}=\frac{1}{2}\left(P_{\Delta}^{+}+P_{\Delta}^{-}\right)
	+\underbrace{\frac{1}{4}\frac{\partial~}{\partial X_{j}}\left(-w_{i}^{-}w_{j}^{+}u_{i}^{(1)+}-w_{i}^{+}w_{j}^{-}u_{i}^{(1)-}\right)}_{\mathcal{T}_{\mathcal{P}_{X}}}
	+\underbrace{\frac{1}{2}\frac{\partial~}{\partial r_{j}}\left(-w_{i}^{-}w_{j}^{+}u_{i}^{(1)+}+w_{i}^{+}w_{j}^{-}u_{i}^{(1)-}\right)}_{\Pi_{\mathcal{P}_{X}}}.
\end{equation}
Note the presence of one-point uncertainty production rates
$P_{\Delta}^{+}=-w_{i}^{+}w_{j}^{+}S_{ij}^{(1)+}$ and
$P_{\Delta}^{-}=-w_{i}^{-}w_{j}^{-}S_{ij}^{(1)-}$ where
$S_{ij}=\frac{1}{2}\left(\frac{\partial u_{i}}{\partial
	x_{j}}+\frac{\partial u_{j}}{\partial x_{i}}\right)$ is the local
strain rate. Equations (\ref{eq:lasttwoterms of I in terms of delta})
and (\ref{eq:singlepoint of Iave}) show that $\mathcal{P}_{\delta}$
and $\mathcal{P}_{X}$ are both composed of one-point uncertainty
production rates and conservative divergence terms with respect to
space ($\boldsymbol{X}$) and scale ($\boldsymbol{r}$). $P_{\Delta}$ is the one-point
production rate of uncertainty identified by \citet{ge2023production} who used it to prove that uncertainty is
always produced in the local strain rate's compressive eigenvector
direction(s) and always attenuated in the local strain rate's
stretching eigenvector direction(s). It is, of course, significant
that this uncertainty production/attenuation mechanism is also present
in the two-point scale-by-scale equations (\ref{eq:difference KHMH
	difference/average}).

The conservative divergence terms contributing to
$\mathcal{P}_{\delta}$ and $\mathcal{P}_{X}$ are such that
$\mathcal{T}_{\mathcal{P}_{\delta}}=-\mathcal{T}_{\mathcal{P}_{X}}$
and $\Pi_{\mathcal{P}_{\delta}}=-\Pi_{\mathcal{P}_{X}}$, which
indicates that $\mathcal{T}_{\mathcal{P}_{\delta}}$ and
$\Pi_{\mathcal{P}_{\delta}}$ appear with opposite signs in equations
(\ref{eq:difference KHMH difference}) and (\ref{eq:difference KHMH
	average}) and are therefore terms which transfer energy between
$\left|\delta\boldsymbol{w}\right|^{2}$ and
$\left|\overline{\boldsymbol{w}}\right|^{2}$.  This is consistent with
$\left|\delta\boldsymbol{w}\right|^{2} + \left|\overline{\boldsymbol{w}}\right|^{2} =
\left|\boldsymbol{w}^{+}\right|^{2} + \left|\boldsymbol{w}^{-}\right|^{2}$ and the
resulting fact that the sum of equations (\ref{eq:difference KHMH
	difference}) and (\ref{eq:difference KHMH average}) collapses to the
sum of the one-point equation for $\left|\boldsymbol{w}^{+}\right|^{2}$ and
the one-point equation for $\left|\boldsymbol{w}^{-}\right|^{2}$ where the
production of uncertainty appears as $\mathcal{P}_{\delta} +
\mathcal{P}_{X} = P_{\Delta}^{+} + P_{\Delta}^{-}$.

The transport terms $\mathcal{T}_{\mathcal{P}_{\delta}}$ and
$\Pi_{\mathcal{P}_{\delta}}$ can be rewritten in terms of two-point
velocity half differences and half sums as follows:
\begin{equation}
	\mathcal{T}_{\mathcal{P}_{\delta}}=\frac{1}{2}\frac{\partial~}{\partial X_{j}}\left(\overline{w}_{j}\overline{w}_{i}\overline{u}_{i}^{(1)}+\delta w_{j}\overline{w}_{i}\delta u_{i}^{(1)}-\overline{w}_{j}\delta w_{i}\delta u_{i}^{(1)}-\delta w_{j}\delta w_{i}\overline{u}_{i}^{(1)}\right),
\end{equation}
and
\begin{equation}
	\label{eq:single point of Ir}
	\begin{aligned}
		\Pi_{\mathcal{P}_{\delta}}&=-\frac{\partial~}{\partial
			r_{j}}\left(\overline{w}_{j}\delta
		w_{i}\overline{u}_{i}^{(1)}-\overline{w}_{j}\overline{w}_{i}\delta
		u_{i}^{(1)}+\delta w_{j}\delta w_{i}\delta
		u_{i}^{(1)}-\delta
		w_{j}\overline{w}_{i}\overline{u}_{i}^{(1)}\right)\\ &=\underbrace{2\frac{\partial~}{\partial
				r_{j}}\left(\overline{w}_{j}\overline{w}_{i}\delta{u}_{i}^{(1)}\right)}_{\mathcal{I}_{\delta\uparrow}}\underbrace{-2\frac{\partial~}{\partial
				r_{j}}\left(\delta w_{j}\delta w_{i}\delta
			u_{i}^{(1)}\right)}_{\mathcal{I}_{\delta\downarrow}}-\frac{1}{2}\frac{\partial}{\partial
			X_{j}}\left(w_{j}^{-}\overline{w_{i}u_{i}^{(1)}}\right),
	\end{aligned}
\end{equation}
The important roles of $\mathcal{I}_{\delta\uparrow}$ and
$\mathcal{I}_{\delta\downarrow}$ are discussed in detail in the
following subsection.
\subsection{Scale-by-scale uncertainty energy balance in homogeneous turbulence\label{sec:Scale-by-scale uncertainty energy balance in forced homogeneous turbulence}}

In the present work, the turbulence studied is periodic/homogenous and
the uncertaintly velocity field $\boldsymbol{w}$ is therefore also
periodic/homogenous. As a result, spatial averages over $\boldsymbol{X}$ of
divergence terms in $\boldsymbol{X}$-space vanish. By applying the spatial
average $\left\langle \right\rangle$ to equation (\ref{eq:difference
	KHMH difference}) and using
$\Pi_{\delta}=\Pi_{\delta,\text{ref}}+\Pi_{\delta,\text{err}}$, we
obtain
\begin{equation}
	\label{eq:average difference KHMH difference/average}
	\frac{\partial~}{\partial
		t}\left\langle\mathcal{A}_{\delta}\right\rangle=\left\langle\Pi_{\delta,\text{ref}}\right\rangle+
	\left\langle\Pi_{\delta,\text{err}}\right\rangle +
	\left\langle\mathcal{I}_{\delta\uparrow}\right\rangle+\left\langle\mathcal{I}_{\delta\downarrow}\right\rangle+\left\langle\mathcal{D}_{\delta}\right\rangle+\left\langle
	P_{\Delta}\right\rangle-\left\langle\varepsilon_{\Delta}\right\rangle+\left\langle\mathcal{F}_{\delta}\right\rangle.
\end{equation}
Here, we note that the spatial average of the last term in equation
(\ref{eq:single point of Ir}) is zero
by homogeneity/periodicity.

It is instructive to start by addressing the following two limits:
$\left|\mathbf{r}\right|\to\infty$ and $\boldsymbol{r}\to\boldsymbol{0}$. In the limit
$\left|\mathbf{r}\right|\to\infty$ we have
$\left\langle\mathcal{A}_{\delta}\right\rangle\to\left\langle
E_{\Delta}\right\rangle$ and all transport terms in equation
(\ref{eq:average difference KHMH difference/average}) tend to
zero. Therefore, equation (\ref{eq:average difference KHMH
	difference/average}) reduces to the single-point uncertainty
evolution equation (\ref{eq:single point uncertainty equation}), where
$P_{\Delta}=-w_{i}w_{j}S^{(1)}_{ij}$ is the internal production rate
of uncertainty, $\varepsilon_{\Delta}=\nu\frac{\partial
	w_{i}}{\partial x_{j}}\frac{\partial w_{i}}{\partial x_{j}}$ is the
dissipation rate of uncertainty, and $F_{\Delta}=w_{i}g_{i}$ is the
external input/output rate of uncertainty. In the limit
$\boldsymbol{r}\to\boldsymbol{0}$, all two-point terms in equation (\ref{eq:average
	difference KHMH difference/average}) tend to $0$ except for the
viscous diffusion term $\left\langle\mathcal{D}_{\delta}\right\rangle$
which tends to $\left\langle \varepsilon_{\Delta}\right\rangle$, and
$\mathcal{I}_{\delta\uparrow}$ which tends to $- P_{\Delta}$ as can
readily be worked out from the definition of
$\mathcal{I}_{\delta\uparrow}$ in equation (\ref{eq:single point of
	Ir}). This is an important observation. We return to it in this
subsection's penultimate paragraph.

\citet{chen2022scalings} and \citet{apostolidis2023turbulent} identified the roles of compressions and stretchings in
interscale turbulence energy transfers by considering volume averages
in $\boldsymbol{r}$-space. We apply their approach to the uncertainty field by
applying to each term of equation (\ref{eq:average difference KHMH
	difference/average}) an additional average over spheres of radius
$r$ in $\mathbf{r}$-space. Hence, by defining
$Q^{a}(r)\equiv\frac{3}{4\pi
	r^{3}}\iiint_{\left|\boldsymbol{\rho}\right|<r}Q(\boldsymbol{\rho}){\rm d}\boldsymbol{\rho}$
for any $Q(\boldsymbol{\rho})$, we obtain
\begin{equation}
	\label{eq:scale-by-scale uncertainty budget}
	\frac{\partial~}{\partial
		t}\left\langle\mathcal{A}_{\delta}^{a}\right\rangle=\left\langle\Pi^{a}_{\delta,\text{ref}}\right\rangle+
	\left\langle\Pi^{a}_{\delta,\text{err}}\right\rangle +
	\left\langle\mathcal{I}^{a}_{\delta\uparrow}\right\rangle+\left\langle\mathcal{I}^{a}_{\delta\downarrow}\right\rangle+\left\langle\mathcal{D}^{a}_{\delta}\right\rangle+\left\langle
	P_{\Delta}\right\rangle-\left\langle\varepsilon_{\Delta}\right\rangle+\left\langle\mathcal{F}_{\delta}^{a}\right\rangle.
\end{equation}
The time-derivative term on the left-hand side of equation
(\ref{eq:scale-by-scale uncertainty budget}) represents the rate of
change in time of the uncertainty energy at scales $r$ and smaller. It
takes the form
\begin{equation}
	\label{eq:time-derivative term of budget}
	\frac{\partial~}{\partial t}\left\langle\mathcal{A}_{\delta}^{a}\right\rangle=\frac{3}{4\pi r^{3}}\frac{\partial~}{\partial t}\int_{0}^{r}\rho^{2}\int_{\left|\boldsymbol{\rho}\right|=r}\left\langle \left|\delta \boldsymbol{w}\right|^{2}\right\rangle\mathrm{d}\Omega\mathrm{d}\rho
\end{equation}
where the integral over $\Omega$ is an integral over the solid angle
in $\boldsymbol{r}$-space. Note the presence of the average one-point
uncertainty production rate on the right hand side of equation
(\ref{eq:scale-by-scale uncertainty budget}) which governs the
evolution of the uncertainty energy at scales $r$ and
smaller. (Note also that we are not using an equation
  for an uncertainty energy density per unit scale. Whereas such an
  equation can be derived from the present one, it is not convenient
  for making the points that follow.)

Before addressing the four interscale transfer rates on the right hand
side of this equation (\ref{eq:scale-by-scale uncertainty budget}), we
note that by application of the Gauss divergence theorem, the viscous
diffusion term becomes
\begin{equation}
	\label{eq:viscous diffusion term of budget}
	\left\langle\mathcal{D}^{a}_{\delta}\right\rangle=\frac{3\nu}{2\pi
		r}\frac{\partial}{\partial
		r}\left(\int_{\left|\boldsymbol{\rho}\right|=r}\left\langle
	\left|\delta
	\boldsymbol{w}\right|^{2}\right\rangle\mathrm{d}\Omega\right).
\end{equation}

We start with the interscale transfer term
$\left\langle\Pi^{a}_{\delta,\text{ref}}\right\rangle$. Application of
the Gauss divergence theorem yields
\begin{equation}
	\label{eq:Pi ref term of budget}
	\left\langle\Pi^{a}_{\delta,\text{ref}}\right\rangle=-\frac{3}{2\pi}\int_{\left|\boldsymbol{\rho}\right|=r}\left\langle \frac{\delta \boldsymbol{u}^{(1)}\cdot\widetilde{\boldsymbol{r}}}{r}\left|\delta\boldsymbol{w}\right|^{2}\right\rangle\mathrm{d}\Omega,
\end{equation}
where $\widetilde{\boldsymbol{r}}=\boldsymbol{r}/\left|\boldsymbol{r}\right|$ is the unit
vector in $\boldsymbol{r}$ space. It is clear from this expression (\ref{eq:Pi
	ref term of budget}) that compression events
$\widetilde{\boldsymbol{r}}\cdot\delta \boldsymbol{u}^{(1)}<0$ in the reference field
contribute positively to
$\left\langle\Pi^{a}_{\delta,\text{ref}}\right\rangle$ whereas
stretching events $\widetilde{\boldsymbol{r}}\cdot\delta \boldsymbol{u}^{(1)}>0$ in
the reference field contribute negatively to
$\left\langle\Pi^{a}_{\delta,\text{ref}}\right\rangle$. These
compression and stretching motions are local in $\boldsymbol{r}$-space rather
than physical space and are distinct from the compressions and
stretchings pertaining to the local strain rate which are local in
physical space. Whereas strain rate compressions and stretchings
determine the one-point production of uncertainty as shown by \citet{ge2023production}, the compressive and streching motions defined
by the sign of $\widetilde{\boldsymbol{r}}\cdot\delta \boldsymbol{u}^{(1)}$ determine
the two-point interscale transfer rate
$\left\langle\Pi^{a}_{\delta,\text{ref}}\right\rangle$ and whether it
is a forward transfer from large to small scales or a backwards
transfer from small to large ones (backscatter).

We stress that $\left\langle\Pi^{a}_{\delta,\text{ref}}\right\rangle$
is a linear interscale transfer rate of uncertainty energy at scales
$r$ by two-point differences at such scales in the reference velocity
field. This is in contrast to
$\left\langle\Pi^{a}_{\delta,\text{err}}\right\rangle$ which is a
non-linear interscale transfer rate of uncertainty energy at scales
$r$ by two-point velocity differences at such scales in the
uncertainty velocity field itself as one can see from
\begin{equation}
	\label{eq:Pi err}
	\left\langle\Pi^{a}_{\delta,\text{err}}\right\rangle=-\frac{3}{2\pi}\int_{\left|\boldsymbol{\rho}\right|=r}\left\langle \frac{\delta \boldsymbol{w}\cdot\widetilde{\boldsymbol{r}}}{r}\left|\delta\boldsymbol{w}\right|^{2}\right\rangle\mathrm{d}\Omega
\end{equation}
which follows from application of the Gauss divergence theorem to
$\left\langle\Pi^{a}_{\delta,\text{err}}\right\rangle$. Compressive/stretching
motions in the uncertainty field itself make forward/inverse
contributions to the overall average interscale transfer rate
$\left\langle\Pi^{a}_{\delta,\text{err}}\right\rangle$. 

Application of the Gauss divergence theorem to the other two
interscale transfer rates 
$\left\langle\mathcal{I}^{a}_{\delta\uparrow}\right\rangle$ and 
$\left\langle\mathcal{I}^{a}_{\delta\downarrow}\right\rangle$
yields
\begin{subequations}
	\begin{align}
		&\left\langle\mathcal{I}^{a}_{\delta\uparrow}\right\rangle=\frac{3}{2\pi}\int_{\left|\boldsymbol{\rho}\right|=r}\left\langle\frac{\delta u^{(1)}_{i}\widetilde{r}_{j}}{r}\overline{w}_{i}\overline{w}_{j}\right\rangle\mathrm{d}\Omega=\frac{3}{2\pi}\int_{\left|\boldsymbol{\rho}\right|=r}\left\langle \Xi^{(1)}_{ij}\overline{w}_{i}\overline{w}_{j}\right\rangle\mathrm{d}\Omega,\label{eq:I up term of budget}&\\
		&\left\langle\mathcal{I}^{a}_{\delta\downarrow}\right\rangle=-\frac{3}{2\pi}\int_{\left|\boldsymbol{\rho}\right|=r}\left\langle\frac{\delta u^{(1)}_{i}\widetilde{r}_{j}}{r}\delta{w}_{i}\delta{w}_{j}\right\rangle\mathrm{d}\Omega=-\frac{3}{2\pi}\int_{\left|\boldsymbol{\rho}\right|=r}\left\langle\Xi^{(1)}_{ij}\delta w_{i}\delta w_{j}\right\rangle\mathrm{d}\Omega,\label{eq:I down term of budget}&
	\end{align}
\end{subequations}
where $\Xi_{ij}^{(1)}=\frac{1}{2}\left(\frac{\delta
	u_{i}^{(1)}\widetilde{r}_{j}}{r}+\frac{\delta
	u_{j}^{(1)}\widetilde{r}_{i}}{r}\right)$ represents the relative
deformation rate for two points $\boldsymbol{X}-\boldsymbol{r}/2$ and
$\boldsymbol{X}+\boldsymbol{r}/2$ in the reference flow.
It is observed that equations (\ref{eq:I up term of budget}) and
(\ref{eq:I down term of budget}) have a structural similarity with the
expression for the one-point uncertainty production
$P_{\Delta}=-w_{i}w_{j}S^{(1)}_{ij}$, where the uncertainty is
produced through local strain rate of the reference field. Here, the
uncertainty energy is transfered through scale $r$ by the average
deformation rate of the reference field at that scale.

The tensor $\boldsymbol{\Xi}^{(1)}$ is symmetric and real and therefore has
three real eigenvalues: $\Lambda^{(1)}_1$, $\Lambda^{(1)}_2$, and
$\Lambda^{(1)}_3$ which are functions of $\boldsymbol{X}$, $r$ and
$\widetilde{\boldsymbol{r}}$. The spatial average of its trace,
$\left\langle\Xi_{ii}^{(1)}\right\rangle =
\left\langle\Lambda^{(1)}_{1} + \Lambda^{(1)}_{2} +
\Lambda^{(1)}_{3}\right\rangle$, vanishes because of statistical
homogeneity, i.e.  $\left\langle\Xi_{ii}^{(1)}\right\rangle =
\left\langle\delta\boldsymbol{u}^{(1)} \cdot \widetilde{\boldsymbol{r}}/r\right\rangle
= \frac{1}{2} \left\langle\boldsymbol{u}^{(1)+} \cdot
\widetilde{\boldsymbol{r}}/r\right\rangle - \frac{1}{2}
\left\langle\boldsymbol{u}^{(1)-} \cdot
\widetilde{\boldsymbol{r}}/r\right\rangle=0$. Hence,
$\left\langle\Lambda^{(1)}_{1} + \Lambda^{(1)}_{2} +
\Lambda^{(1)}_{3}\right\rangle =0$ even though $\Lambda^{(1)}_{1} +
\Lambda^{(1)}_{2} + \Lambda^{(1)}_{3}$ does not necessarily vanish.

It can also be shown by simple algebraic manipulations that the
determinant of $\boldsymbol{\Xi}^{(1)}$ is zero, i.e. $\det(\boldsymbol{\Xi}^{(1)}) =
\Lambda_{1}^{(1)}\Lambda_{2}^{(1)}\Lambda_{3}^{(1)} =0$. This implies
that at least one of the three eigenvalues is zero, say
$\Lambda^{(1)}_{2}=0$. We now show in the following paragraph that
$\Lambda^{(1)}_{1}$ and $\Lambda^{(1)}_{3}$ cannot have the same sign.

Using the eigenvectors of $\boldsymbol{\Xi}^{(1)}$ as local (in
$\boldsymbol{X}$-space) orthonormal reference frame, we can write
\begin{equation}
	\Xi^{(1)}_{ij}\delta u^{(1)}_{i}\delta u^{(1)}_{j}=\Lambda^{(1)}_{1}\left|\boldsymbol{e}^{(1)}_{1}\cdot\delta\boldsymbol{u}^{(1)}\right|^{2}+\Lambda_{3}^{(1)}\left|\boldsymbol{e}^{(1)}_{3}\cdot\delta\boldsymbol{u}^{(1)}\right|^{2},
\end{equation}
where $\boldsymbol{e}^{(1)}_{1}$ and $\boldsymbol{e}^{(1)}_{3}$ are the eigenvectors
corresponding to $\Lambda^{(1)}_{1}$ and $\Lambda^{(1)}_{3}$
respectively. Given that $\Xi_{ij}^{(1)}\delta u^{(1)}_{i}\delta
u^{(1)}_{j}=\left(\delta\boldsymbol{u}^{(1)} \cdot
\widetilde{\boldsymbol{r}}/r\right)\left|\delta\boldsymbol{u}^{(1)}\right|^{2}$, we
obtain
\begin{equation}
	\label{eq:Lambda distribution}
	\frac{\Lambda_{1}^{(1)}}{\delta\boldsymbol{u}^{(1)} \cdot \widetilde{\boldsymbol{r}}/r}\frac{(\boldsymbol{e}^{(1)}_{1}\cdot\delta\boldsymbol{u}^{(1)})^{2}}{\left|\delta\boldsymbol{u}^{(1)}\right|^{2}}+\frac{\Lambda^{(1)}_{3}}{\delta\boldsymbol{u}^{(1)} \cdot \widetilde{\boldsymbol{r}}/r}\frac{(\boldsymbol{e}^{(1)}_{3}\cdot\delta\boldsymbol{u}^{(1)})^{2}}{\left|\delta\boldsymbol{u}^{(1)}\right|^{2}}=1.
\end{equation}
Taking into account $\frac{\Lambda^{(1)}_{1}}{\delta\boldsymbol{u}^{(1)} \cdot
	\widetilde{\boldsymbol{r}}/r}+\frac{\Lambda_{3}^{(1)}}{\delta\boldsymbol{u}^{(1)}
	\cdot \widetilde{\boldsymbol{r}}/r}=1$ and
$0\leq\frac{\left|\boldsymbol{e}^{(1)}_{1}\cdot\delta\boldsymbol{u}^{(1)}\right|^{2}}{\left|\delta\boldsymbol{u}^{(1)}\right|^{2}},\frac{\left|\boldsymbol{e}^{(1)}_{3}\cdot\delta\boldsymbol{u}^{(1)}\right|^{2}}{\left|\delta\boldsymbol{u}^{(1)}\right|^{2}}\leq1$,
the only way that (\ref{eq:Lambda distribution}) can hold is for $\Lambda_{1}^{(1)}$ and
$\Lambda_{3}^{(1)}$ to have opposite signs. The order of eigenvalues
is therefore as follows: $\Lambda_{1}^{(1)} \leq \Lambda_{2}^{(1)} = 0
\leq \Lambda_{3}^{(1)}$. Furthermore,
$\left\langle\Lambda^{(1)}_{1}\right\rangle =
-\left\langle\Lambda^{(1)}_{3}\right\rangle$ now follows from
$\left\langle\Lambda^{(1)}_{1} + \Lambda^{(1)}_{2} +
\Lambda^{(1)}_{3}\right\rangle =0$.

We can now rewrite equation (\ref{eq:Pi ref term of budget}) as follows
\begin{equation}
	\label{eq:Pi reference in pricipale axe}
	\left\langle\Pi^{a}_{\delta,\text{ref}}\right\rangle=-\frac{3}{2\pi}\int_{\left|\boldsymbol{\rho}\right|=r}\left\langle
	(\Lambda^{(1)}_{1} + \Lambda^{(1)}_{3})
	\left|\delta\boldsymbol{w}\right|^{2}\right\rangle\mathrm{d}\Omega,
\end{equation}
and we can use the orthonormal reference frame composed of the
eigenvectors of $\boldsymbol{\Xi}^{(1)}$ to rewrite equations (\ref{eq:I up
	term of budget}) and (\ref{eq:I down term of budget}) too:
\begin{subequations}
	\begin{align}
		\left\langle\mathcal{I}^{a}_{\delta\uparrow}\right\rangle=\frac{3}{2\pi}\int_{\left|\boldsymbol{\rho}\right|=r}\left\langle \Lambda_{1}^{(1)}\left|\boldsymbol{e}_{1}^{(1)}\cdot\overline{\boldsymbol{w}}\right|^{2}+\Lambda_{3}^{(1)}\left|\boldsymbol{e}_{3}^{(1)}\cdot\overline{\boldsymbol{w}}\right|^{2} \right\rangle\mathrm{d}\Omega,\label{eq:I up term of budget in eigen base}&\\
		\left\langle\mathcal{I}^{a}_{\delta\downarrow}\right\rangle=-\frac{3}{2\pi}\int_{\left|\boldsymbol{\rho}\right|=r}\left\langle \Lambda_{1}^{(1)}\left|\boldsymbol{e}_{1}^{(1)}\cdot\delta{\boldsymbol{w}}\right|^{2}+\Lambda_{3}^{(1)}\left|\boldsymbol{e}_{3}^{(1)}\cdot\delta{\boldsymbol{w}}\right|^{2} \right\rangle\mathrm{d}\Omega.\label{eq:I down term of budget in eigen base}&
	\end{align}
\end{subequations}
All these three average interscale transfer rates of uncertainty
across scale $r$ depend on the two non-zero eigenvalues of the
reference field's relative deformation tensor
$\Xi_{ij}^{(1)}=\frac{1}{2}\left(\frac{\delta
  u_{i}^{(1)}\widetilde{r}_{j}}{r}+\frac{\delta
  u_{j}^{(1)}\widetilde{r}_{i}}{r}\right)$ and the latter two,
i.e. $\left\langle\mathcal{I}^{a}_{\delta\uparrow}\right\rangle$ and
$\left\langle\mathcal{I}^{a}_{\delta\downarrow}\right\rangle$, also
depend on the alignments of the half difference and the half sum of
the uncertainty velocity with the two corresponding eigenvectors.  The
signs of these three average interscale transfer rates of uncertainty
depend on whether compressions (alignments with
$\boldsymbol{e}_{1}^{(1)}$ and values of $\Lambda^{(1)}_{1}$) or
stretchings (alignments with $\boldsymbol{e}_{3}^{(1)}$ and values of
$\Lambda^{(1)}_{3}$) overwhelm the integrals defining them:
$\left\langle\Pi^{a}_{\delta,\text{ref}}\right\rangle$ and
$\left\langle\mathcal{I}^{a}_{\delta\downarrow}\right\rangle$ are
positive (forward interscale uncertainty transfer) if they are
dominated by compressions and negative (inverse interscale uncertainty
transfer) if they are dominated by stretchings whereas
$\left\langle\mathcal{I}^{a}_{\delta\uparrow}\right\rangle$ is
negative (inverse uncertainty transfer) if it is dominated by
compressions and positive (forward interscale uncertainty transfer) if
it is dominated by stretchings.

As we have already seen from the definitions of
$\mathcal{I}_{\delta\uparrow}$ and $\mathcal{I}_{\delta\downarrow}$ in
equation (\ref{eq:single point of Ir}),
$\mathcal{I}_{\delta\uparrow}\to- P_{\Delta}$ and
$\mathcal{I}_{\delta\downarrow}\to0$ in the limit $\boldsymbol{r}\to\boldsymbol{0}$.
In particular,
$\left\langle\mathcal{I}^{a}_{\delta\uparrow}\right\rangle\to-\left\langle
P_{\Delta}\right\rangle$ in that $\boldsymbol{r}\to\boldsymbol{0}$ limit, and we know
from \citet{ge2023production} that $\left\langle
P_{\Delta}\right\rangle$ is positive at all times in statistically
homogeneous turbulence. We can therefore expect
$\left\langle\mathcal{I}^{a}_{\delta\uparrow}\right\rangle$ to be
negative at small $r$ scales which means that it has to be dominated
by compressive motions of reference velocity differences and therefore
be an inverse interscale transfer of uncertainty at such
scales. Uncertainty is therefore produced by the local reference
field's strain rate compressions and is then transfered from small to
larger scales by compressive motions of reference velocity
differences. If these compressive motions are more generally dominant
and therefore also dominate the integrals on the right hand sides of
(\ref{eq:Pi reference in pricipale axe})
and (\ref{eq:I down term of budget in eigen base}), then
$\left\langle\Pi^{a}_{\delta,\text{ref}}\right\rangle$ and
$\left\langle\mathcal{I}^{a}_{\delta\downarrow}\right\rangle$ will be
positive and will therefore represent forward interscale transfer of
uncertainty. This scenario is confirmed in section \ref{sec:Numerical
	results} with DNS of statistically stationary periodic Navier-Stokes
turbulence.

We end this section with a note on the only one non-linear interscale
transfer rate involved here: $\Pi^{a}_{\delta,\text{err}}$. Equation
(\ref{eq:Pi err}) can be recast in the form
\begin{equation}
	\left\langle\Pi^{a}_{\delta,\text{err}}\right\rangle=-\frac{3}{2\pi}\int_{\left|\boldsymbol{\rho}\right|=r}\left\langle\frac{\delta w_{i}\widetilde{r}_{j}}{r} \delta w_{i}\delta w_{j}\right\rangle\mathrm{d}\Omega=-\frac{3}{2\pi}\int_{\left|\boldsymbol{\rho}\right|=r}\left\langle\Xi_{\Delta ij}\delta w_{i}\delta w_{j}\right\rangle\mathrm{d}\Omega\label{eq:Pi err term of budget}
\end{equation}
where $\Xi_{\Delta ij}=\frac{1}{2}\left(\frac{\delta
	w_{i}\widetilde{r}_{j}}{r}+\frac{\delta
	w_{j}\widetilde{r}_{i}}{r}\right)$ represents the relative
deformation rate for two points $\boldsymbol{X}-\boldsymbol{r}/2$ and
$\boldsymbol{X}+\boldsymbol{r}/2$ in the uncertainty velocity field. The tensor
$\boldsymbol{\Xi}_{\Delta}$ has the same mathematical properties as
$\boldsymbol{\Xi}^{(1)}$, i.e. $\det({\boldsymbol{\Xi}_{\Delta}})=0$,
$\left\langle\Xi_{\Delta ii}\right\rangle=0$ and, using equality
$\Xi_{\Delta ij}\delta w_{i}\delta
w_{j}=\left(\delta\boldsymbol{w}\cdot\widetilde{\boldsymbol{r}}\right)\left|\delta\boldsymbol{w}\right|^{2}$,
the eigenvalues of $\boldsymbol{\Xi}_{\Delta}$ are such that $\Lambda_{\Delta
	1} \leq \Lambda_{\Delta 2} = 0 \leq \Lambda_{\Delta 3}$ and
$\left\langle \Lambda_{\Delta 1} \right\rangle = -\left\langle
\Lambda_{\Delta 3} \right\rangle$. Using the orthonormal reference
frame composed of the eigenvectors of $\boldsymbol{\Xi}_{\Delta}$, equation
(\ref{eq:Pi err term of budget}) can be rewritten as
\begin{equation}
	\label{eq:pi err principale axes}
	\left\langle\Pi^{a}_{\delta,\text{err}}\right\rangle=\frac{3}{2\pi}\int_{\left|\boldsymbol{\rho}\right|=r}\left\langle
	\Lambda_{\Delta 1}\left|\boldsymbol{e}_{\Delta
		1}\cdot\delta{\boldsymbol{w}}\right|^{2}+\Lambda_{\Delta
		3}\left|\boldsymbol{e}_{\Delta 3}\cdot\delta{\boldsymbol{w}}\right|^{2}
	\right\rangle\mathrm{d}\Omega,
\end{equation}
where $\boldsymbol{e}_{\Delta 1}$ and $\boldsymbol{e}_{\Delta 3}$ are
the eigenvectors corresponding to $\Lambda_{\Delta 1}$ and
$\Lambda_{\Delta 3}$, respectively. As mentioned under equation
(\ref{eq:Pi err}), compressive ($\Lambda_{\Delta 1} <0$ and alignments
of $\delta \boldsymbol{w}$ with $\boldsymbol{e}_{\Delta 1}$) and
stretching ($\Lambda_{\Delta 3} >0$ and alignments of $\delta
\boldsymbol{w}$ with $\boldsymbol{e}_{\Delta 3}$) motions in the
uncertainty field make, respectively, forward and inverse
contributions to the overall average interscale uncertainty transfer
rate $\left\langle\Pi^{a}_{\delta,\text{err}}\right\rangle$.

Note that both fields $\boldsymbol{u}^{(1)}$ and $\boldsymbol{u}^{(2)}$ can be taken as the reference field (where the velocity-difference field is $\boldsymbol{w}$ for $\boldsymbol{u}^{(1)}$ and $-\boldsymbol{w}$ for $\boldsymbol{u}^{(2)}$ ) and we therefore must have
\begin{equation}
	\begin{aligned}
		&\left\langle\Pi^{a}_{\delta,\text{ref}}\right\rangle+
		\left\langle\Pi^{a}_{\delta,\text{err}}\right\rangle +
		\left\langle\mathcal{I}^{a}_{\delta\uparrow}\right\rangle+\left\langle\mathcal{I}^{a}_{\delta\downarrow}\right\rangle\\
		&=-\frac{3}{2\pi}\int_{\left|\boldsymbol{\rho}\right|=r}\left\langle \frac{\delta \boldsymbol{u}^{(1)}\cdot\widetilde{\boldsymbol{r}}}{r}\left|\delta\boldsymbol{w}\right|^{2}\right\rangle+\left\langle \frac{\delta \boldsymbol{w}\cdot\widetilde{\boldsymbol{r}}}{r}\left|\delta\boldsymbol{w}\right|^{2}\right\rangle-\left\langle \Xi^{(1)}_{ij}\overline{w}_{i}\overline{w}_{j}\right\rangle+\left\langle\Xi^{(1)}_{ij}\delta w_{i}\delta w_{j}\right\rangle\mathrm{d}\Omega\\
		&=-\frac{3}{2\pi}\int_{\left|\boldsymbol{\rho}\right|=r}\left\langle \frac{\delta \boldsymbol{u}^{(2)}\cdot\widetilde{\boldsymbol{r}}}{r}\left|\delta\boldsymbol{w}\right|^{2}\right\rangle+\left\langle \frac{(-\delta \boldsymbol{w})\cdot\widetilde{\boldsymbol{r}}}{r}\left|\delta\boldsymbol{w}\right|^{2}\right\rangle-\left\langle \Xi^{(2)}_{ij}\overline{w}_{i}\overline{w}_{j}\right\rangle+\left\langle\Xi^{(2)}_{ij}\delta w_{i}\delta w_{j}\right\rangle\mathrm{d}\Omega
	\end{aligned}
\end{equation}
in periodic/homogeneous turbulence. Indeed, $\delta
  \boldsymbol{u}^{(1)}\cdot\widetilde{\boldsymbol{r}}+\delta
  \boldsymbol{w}\cdot\widetilde{\boldsymbol{r}}=\delta
  \boldsymbol{u}^{(2)}\cdot\widetilde{\boldsymbol{r}}$ and
  $\Xi^{(1)}_{ij}\delta w_{i}\delta w_{j}=\Xi^{(2)}_{ij}\delta
  w_{i}\delta w_{j}-\Xi_{\Delta, ij}\delta w_{i}\delta w_{j}$ where
  $-\Xi_{\Delta, ij}\delta w_{i}\delta w_{j}=\frac{(-\delta
    \boldsymbol{w})\cdot\widetilde{\boldsymbol{r}}}{r}\left|\delta\boldsymbol{w}\right|^{2}$. For
  $\left\langle\mathcal{I}^{a}_{\delta\uparrow}\right\rangle$, we have
  $-\left\langle
  \Xi^{(1)}_{ij}\overline{w}_{i}\overline{w}_{j}\right\rangle=-\left\langle
  \Xi^{(2)}_{ij}\overline{w}_{i}\overline{w}_{j}\right\rangle+\left\langle
  \Xi_{\Delta, ij}\overline{w}_{i}\overline{w}_{j}\right\rangle$,
  where $\left\langle \Xi_{\Delta,
    ij}\overline{w}_{i}\overline{w}_{j}\right\rangle=\left\langle\left|\boldsymbol{w}^{+}\right|^{2}(\overline{\boldsymbol{w}}\cdot\widetilde{\boldsymbol{r}}/r)\right\rangle-\left\langle\left|\boldsymbol{w}^{-}\right|^{2}(\overline{\boldsymbol{w}}\cdot\widetilde{\boldsymbol{r}}/r)\right\rangle=0$
  because of the homogeneity of the velocity-difference field. Hence,
  switching the reference field from $\boldsymbol{u}^{(1)}$ to
  $\boldsymbol{u}^{(2)}$ cannot change the physical picture that we
  draw from this section's equations in the remainder of this paper.

%
%

\section{\label{sec:Production/dissipation scaling of uncertainty energy}Equilibrium model of uncertainty growth in the stochastic time regime}


Here, we consider uncertainty growth in a reference flow that is a
high Reynolds number homogeneous/periodic turbulence with a
significant range of excited length scales between its Taylor length
$l_{\lambda}^{(1)}$ and its integral length scale $L^{(1)}$. More
specifically, we consider in that reference flow the stochastic time
range where the uncertainty field's integral length scale $l_{\Delta}$
has grown above $l_{\lambda}^{(1)}$ but still remains below $L^{(1)}$.
Looking at equation (\ref{eq:scale-by-scale uncertainty budget}), we
may neglect the diffusion term at scales $r$ larger than
$l_{\lambda}^{(1)}$ and we may also neglect large-scale forcing at
scales $r$ smaller than $L^{(1)}$, so that this equation reduces to
\begin{equation}
	\label{eq:strcuture scale-by-scale uncertainty budget no diffusion}
	\frac{\partial~}{\partial
		t}\left\langle\mathcal{A}_{\delta}^{a}\right\rangle \approx \left\langle\Pi^{a}_{\delta,\text{ref}}\right\rangle+
	\left\langle\Pi^{a}_{\delta,\text{err}}\right\rangle +
	\left\langle\mathcal{I}^{a}_{\delta\uparrow}\right\rangle+\left\langle\mathcal{I}^{a}_{\delta\downarrow}\right\rangle+\left\langle
	P_{\Delta}\right\rangle-\left\langle\varepsilon_{\Delta}\right\rangle
\end{equation}
in the range $l_{\lambda}^{(1)}\ll r \ll L^{(1)}$.

\begin{figure}
	\centering
	\includegraphics[width=0.6\textwidth]{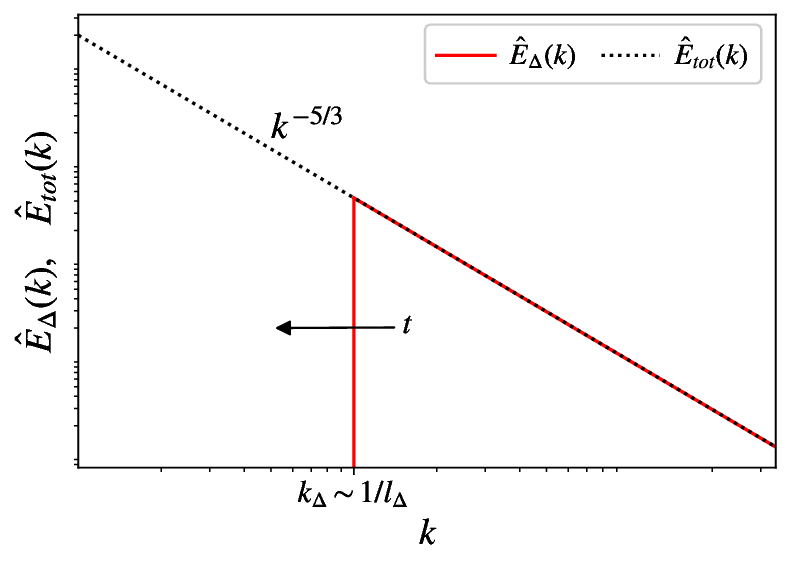}
	\caption{Cartoon model of uncertainty energy in the inertial
		range \cite{leith1972predictability,boffetta2017chaos} in
		the form of a log-log
		plot. $\hat{E}_{tot}(k)\equiv\hat{E}^{(1)}(k)+\hat{E}^{(2)}(k)$
		is the total energy spectrum defined as the sum of the
		energy spectra $\hat{E}^{(1)}(k)$ of the reference field
		$\boldsymbol{u}^{(1)}$ and $\hat{E}^{(2)}(k)$ of the perturbed field
		$\boldsymbol{u}^{(2)}$. $\hat{E}_{tot}(k)$ is represented by the
		black dashed line which is reasonably modeled as a
		Kolmogorov $k^{-5/3}$ power law for high Reynolds number
		homogeneous turbulence. $\hat{E}_{\Delta}(k)$ is the
		uncertainty energy spectrum and is represented by the red
		solid line. The two lines collapse at scales smaller than
		$l_{\Delta}$ and this range of collapsed scales increases
		with time.}
	\label{fig:equilibriumspectra} 
\end{figure}

We revisit the simplified cartoon model of uncertainty energy
evolution considered in previous studies
\citep{lesieur1987turbulence,leith1972predictability,boffetta2017chaos}
and approximately supported by the DNS results of 
\citet{ge2023production}. This model is illustrated in figure
\ref{fig:equilibriumspectra}. According to these studies, most
uncertainty energy ({\it all} uncertainty energy in the cartoon model)
is confined to scales smaller than the the uncertainty field's
integral scale $l_{\Delta}$. This scale $l_{\Delta}$ can be
interpreted as the length over which the velocity-difference field is
correlated, representing a characteristic length scale of the eddies
that contain uncertainty \citep{ge2023production}.
There is therefore a high degree of decorrelation between the
reference and the perturbed flows at scales smaller that $l_{\Delta}$;
the cartoon model assumes this decorrelation to be complete at scales
smaller that $l_{\Delta}$ and these two flows to be perfectly
correlated at scales larger than $l_{\Delta}$. Mathematically, we
write $\left\langle\left|\delta\boldsymbol{w}\right|^{2}\right\rangle
= 0$ for $r > l_{\Delta}$ and
\begin{equation}
	\label{eq:decorrelated 2nd order structure function of DeltaU}
	\int_{\left|\boldsymbol{\rho}\right|=r}\left\langle \left|\delta \boldsymbol{w}\right|^{2}\right\rangle\mathrm{d}\Omega=\int_{\left|\boldsymbol{\rho}\right|=r}\left\langle \left|\delta \boldsymbol{u}^{(1)}\right|^{2}\right\rangle\mathrm{d}\Omega+\int_{\left|\boldsymbol{\rho}\right|=r}\left\langle \left|\delta \boldsymbol{u}^{(2)}\right|^{2}\right\rangle\mathrm{d}\Omega,
\end{equation}
for $r < l_{\Delta}$.

The reference and perturbed velocity fields are both forced and
statistically stationary in time. It therefore follows from
(\ref{eq:time-derivative term of budget}) and (\ref{eq:decorrelated
	2nd order structure function of DeltaU}) that the uncertainty field
is in equilibrium at scales $r < l_{\Delta}$, i.e.
\begin{equation}
	\label{eq:equilibrium}
	\frac{\partial~}{\partial t}\left\langle\mathcal{A}_{\delta}^{a}\right\rangle\approx0
\end{equation}
for $r < l_{\Delta}$. Therefore, equation (\ref{eq:strcuture
	scale-by-scale uncertainty budget no diffusion}) takes the form
\begin{equation}
	\label{eq:simplified strcuture scale-by-scale uncertainty budget}
	\left\langle \varepsilon_{\Delta}\right\rangle-\left\langle
	P_{\Delta}\right\rangle \approx
	\left\langle\Pi^{a}_{\delta,\text{ref}}\right\rangle+
	\left\langle\Pi^{a}_{\delta,\text{err}}\right\rangle +
	\left\langle\mathcal{I}^{a}_{\delta\uparrow}\right\rangle+\left\langle\mathcal{I}^{a}_{\delta\downarrow}\right\rangle
\end{equation}
at scales $r$ in the range $l_{\lambda}^{(1)}\ll r\ll
l_{\Delta}$. Whilst the right hand side of this equation depends in
principle on $r$ the left hand side does not. We therefore have a
self-similar equilibrium cascade where the sum of all interscale
transfer rates on the right hand side is independent of $r$
(i.e. self-similar) and equal to the difference between uncertainty
dissipation and production rates in the range $l_{\lambda}^{(1)}\ll
r\ll l_{\Delta}$.

\subsection{\label{cartoon} Equilibrium scalings and power law time dependencies}

The self-similar equilibrium (\ref{eq:simplified strcuture
	scale-by-scale uncertainty budget}) can be used to predict the time
dependencies of $l_{\Delta}$ and $\left\langle
E_{\Delta}\right\rangle$. The cartoon model we consider here (figure
\ref{fig:equilibriumspectra}) has already been shown to predict the
time dependencies of $l_{\Delta}$ and $\left\langle
E_{\Delta}\right\rangle$ in the stochastic time regime
\citep{lorenz1969predictability,
	lesieur1987turbulence,leith1972predictability,aurell1997predictability,boffetta2017chaos}
but, as we attempt to make clear in this subsection, its predictions
depend on how $\left\langle P_{\Delta}\right\rangle -\left\langle
\varepsilon_{\Delta}\right\rangle$ is evaluated from equation
(\ref{eq:simplified strcuture scale-by-scale uncertainty budget}) at
$r=l_{\Delta}$.

The first step is to write $\left\langle E_{\Delta}\right\rangle
\approx \int_{1/l_{\Delta}}^{\infty} \hat{E}_{tot}(k)$ with
$\hat{E}_{tot}(k)\sim \varepsilon^{2/3}k^{-5/3}$ where $\varepsilon$
is the space-time average of the turbulence dissipation rate of the
reference field (or the perturbed field equivalently). This leads to
the relation $\varepsilon l_{\Delta} \sim \left\langle
E_{\Delta}\right\rangle^{3/2}$ between $l_{\Delta} (t)$ and
$\left\langle E_{\Delta}\right\rangle (t)$. The second step, within
the present framework, is to consider equation (\ref{eq:single point
	uncertainty equation}) without external forcing as the times
considered are too short for large-scale forcing to be significant,
i.e. $\frac{{\rm d}}{{\rm d}t}\left\langle
E_{\Delta}\right\rangle=\left\langle
P_{\Delta}\right\rangle-\left\langle
\varepsilon_{\Delta}\right\rangle$. The DNS results of
\citet{ge2023production} suggest that $\left\langle
P_{\Delta}\right\rangle/\left\langle
\varepsilon_{\Delta}\right\rangle$ is about constant and independent
of time in the stochastic time regime as also confirmed by the DNS in
the present paper's following sections. Finally, the third step is to
evaluate $\left\langle P_{\Delta}\right\rangle-\left\langle
\varepsilon_{\Delta}\right\rangle$ from equation (\ref{eq:simplified
	strcuture scale-by-scale uncertainty budget}) at $r=l_{\Delta}$. If
the right hand side of this equation is dominated by the interscale
transfer rate $\left\langle\Pi^{a}_{\delta,\text{err}}\right\rangle$
which is fully determined by the uncertainty field, then (\ref{eq:simplified strcuture scale-by-scale uncertainty budget}) may
suggest that $\left\langle\Pi^{a}_{\delta,\text{err}}\right\rangle
\sim \left\langle E_{\Delta}\right\rangle^{3/2}/l_{\Delta}$ at $r\sim
l_{\Delta}$, and therefore $\left\langle
P_{\Delta}\right\rangle-\left\langle \varepsilon_{\Delta}\right\rangle
\sim \left\langle E_{\Delta}\right\rangle^{3/2}/l_{\Delta}$. Given
that $\left\langle P_{\Delta}\right\rangle$ and
$\left\langle\varepsilon_{\Delta}\right\rangle$ remain approximately
proportional to each other during the stochastic time range, as we
confirm by DNS in subsection \ref{sec:Time evolution of single-point
	uncertainty}, we may write
\begin{subequations}
	\label{eq:production/disspation scaling law}
	\begin{align}
		\left\langle P_{\Delta}\right\rangle \sim \left\langle
		E_{\Delta}\right\rangle^{3/2}/l_{\Delta},\label{eq:production/disspation scaling law a}\\
		\left\langle\varepsilon_{\Delta}\right\rangle \sim \left\langle
		E_{\Delta}\right\rangle^{3/2}/l_{\Delta}  \label{eq:production/disspation scaling law b}.
	\end{align}
\end{subequations}
It now follows that $\frac{{\rm d}}{{\rm d}t}\left\langle
E_{\Delta}\right\rangle \sim \left\langle
E_{\Delta}\right\rangle^{3/2}/l_{\Delta}$. With $\varepsilon
l_{\Delta} \sim \left\langle E_{\Delta}\right\rangle^{3/2}$ we obtain
the well-known predictions $\left\langle E_{\Delta}\right\rangle \sim
\varepsilon t$, $l_{\Delta}\sim \varepsilon^{1/2} t^{3/2}$
\citep{lorenz1969predictability,
	lesieur1987turbulence,kraichnan1970instability,leith1972predictability,metais1986statistical,
	aurell1997predictability,boffetta2017chaos}.

The linear growth of $\left\langle E_{\Delta}\right\rangle$ has been
reported in DNS of periodic turbulence
\citep{boffetta2017chaos,berera2018chaotic}. However, recent DNS
suggest that this linear growth may depend on the type of forcing
\citep{ge2023production}, particularly if the Reynolds number is not
large enough. No linear growth has been reported in simulations where
$\boldsymbol{u}^{(1)}$ and $\boldsymbol{u}^{(2)}$ are forced by an identical forcing
(i.e. $\boldsymbol{f}^{(2)}(\boldsymbol{x},t)-\boldsymbol{f}^{(1)}(\boldsymbol{x},t)\equiv 0$) so that
the forcing terms in equations (\ref{eq:single point uncertainty
	equation}) and (\ref{eq:simplified strcuture scale-by-scale
	uncertainty budget}) may drop out as required in the previous
paragraph's argument without having to invoke high Reynolds number.


An obvious weakness in the argument leading to the linear growth of
$\left\langle E_{\Delta}\right\rangle$ is the assumption that the
right hand side of equation (\ref{eq:simplified strcuture
	scale-by-scale uncertainty budget}) is dominated by the interscale
transfer rate
$\left\langle\Pi^{a}_{\delta,\text{err}}\right\rangle$. The
alternative assumption is that it should be dominated by the other
interscale transfer rates,
$\left\langle\Pi^{a}_{\delta,\text{ref}}\right\rangle+
\left\langle\mathcal{I}^{a}_{\delta\uparrow}\right\rangle+\left\langle\mathcal{I}^{a}_{\delta\downarrow}\right\rangle$
which are not fully determined by the uncertainty field but also by
the reference velocity field. The respective formulae for these
interscale transfer rates are (\ref{eq:Pi ref term of budget}),
(\ref{eq:I up term of budget}) and (\ref{eq:I down term of budget}),
and we make the bold assumption that the linear contribution of the
reference velocity field to these formulae contributes a scaling
proportional to the r.m.s. velocity of the reference flow $U^{(1)}$
rather than $(\varepsilon l_{\Delta})^{1/3}$ which is smaller than
$U^{(1)}\sim (\varepsilon L^{(1)})^{1/3}$. This enhanced contribution
is a simple naive attempt to qualitatively account for significant
correlations between the uncertainty and the reference fields. The
implication is that all three interscale transfer rates may scale as
$U^{(1)}\left\langle E_{\Delta}\right\rangle/l_{\Delta}$ at $r\sim
l_{\Delta}$. Hence, $\left\langle P_{\Delta}\right\rangle-\left\langle
\varepsilon_{\Delta}\right\rangle \sim U^{(1)}\left\langle
E_{\Delta}\right\rangle/l_{\Delta}$, and by virtue of the
proportionality of $\left\langle P_{\Delta}\right\rangle$ and
$\left\langle\varepsilon_{\Delta}\right\rangle$ in the stochastic time
range, we may write
\begin{subequations}
	\label{eq:production/disspation scaling law2}
	\begin{align}
		\left\langle P_{\Delta}\right\rangle \sim U^{(1)}\left\langle
		E_{\Delta}\right\rangle/l_{\Delta},\label{eq:production/disspation scaling law2 a}\\
		\left\langle\varepsilon_{\Delta}\right\rangle \sim U^{(1)}\left\langle
		E_{\Delta}\right\rangle/l_{\Delta}  \label{eq:production/disspation scaling law2 b}.
	\end{align}
\end{subequations}
The resulting alternative predictions are: $\left\langle
E_{\Delta}\right\rangle \sim (\varepsilon U^{(1)} t)^{2/3}$,
$l_{\Delta}\sim U^{(1)} t$. This alternative prediction implies that
contaminated length scales $l_{\Delta}$ much smaller than $L$ can be
reached much faster than in the case where $l_{\Delta}\sim
\varepsilon^{1/2} t^{3/2}$. 

If we had assumed that the linear contribution of the reference
velocity field to formulae (\ref{eq:Pi ref term of budget}),
(\ref{eq:I up term of budget}) and (\ref{eq:I down term of budget})
contributes a scaling proportional to $(\varepsilon l_{\Delta})^{1/3}$
we would have obtained scalings (\ref{eq:production/disspation scaling
  law}) once again, but without having to assume that the right hand
side of equation (\ref{eq:simplified strcuture scale-by-scale
  uncertainty budget}) is dominated by the interscale transfer rate
$\left\langle\Pi^{a}_{\delta,\text{err}}\right\rangle$. No alternative
time dependencies for $\left\langle E_{\Delta}\right\rangle$ and
$l_{\Delta}$ would have been obtained. The advantage of considering an
alternative lies in the certainty that correlations exist between the
uncertainty and the reference fields which affect the right hand sides
of (\ref{eq:Pi ref term of budget}), (\ref{eq:I up term of budget})
and (\ref{eq:I down term of budget}) in a way that influences their
scalings. Whilst we must leave the task of exploring how these
correlations shape scalings for a future study, it is fair to suspect
that a naive first guess based on $(\varepsilon l_{\Delta})^{1/3}$ and
leading to (\ref{eq:production/disspation scaling law}) may not be the
right answer in general.

	In the following section we describe the DNS used in this paper and in
	section \ref{sec:Numerical results} we present a series of DNS results
	in relation to the theoretical analyses and predictions in this and
	the previous sections.
\section{\label{sec:Numerical steups}Numerical setup}
To study the evolution of average uncertainty energy in
periodic/homogeneous turbulence, we use a fully de-aliased
pseudo-spectral code to perform DNS of forced incompressible
Navier-Stokes turbulence in a periodic box of size
$\mathcal{L}^{3}=(2\pi)^{3}$. Time advancement is achieved with a
second-order Runge-Kutta scheme, and the time step is calculated by
the Courant–Friedrichs–Lewy (CFL) condition with CFL number $0.4$. The
code strategy is detailed by \citet{vincent1991spatial}. We first
generate a statistically stationary reference flow. For the reference
flow's initial condition we use a von K\'arm\'an initial energy
spectrum with the same coefficients as
\citet{https://doi.org/10.48550/arxiv.1306.3408} and random initial
Fourier phases. We integrate the reference flow in time until it
reaches a statistically steady state and then seed it with a small
perturbation at the highest wavenumbers $0.9k_{max}$ to $k_{max}$
(where $k_{max}$ is the highest resolved wavenumber) to create the
perturbed flow at a time which we refer to as $t_{0}=0$. The
generation of the perturbed flow is the same as
\citet{ge2023production} where it is presented in detail.

The reference flow turbulence is sustained by a negative damping
forcing. The forcing function is divergence-free as it depends on the
low wavenumber modes of the velocity in Fourier space as follows
\begin{equation}
	\label{eq:negative damping forcing 1}
	\hat{\boldsymbol{f}}^{(1)}\left(\boldsymbol{k},t\right)=\left\{
	\begin{array}{ccc}
		\frac{\varepsilon_{0}}{2E_{f}^{(1)}}\hat{\boldsymbol{u}}^{(1)}\left(\boldsymbol{k},t\right)&       &\text{if } 0<\left|\boldsymbol{k}\right|\leq k_{f},\\
		0&       &\text{otherwise,}
	\end{array}\right.
\end{equation}
where $\hat{\boldsymbol{f}}$ and $\hat{\boldsymbol{u}}$ are the
Fourier transforms of $\boldsymbol{f}$ and $\boldsymbol{u}$
respectively, $\varepsilon_{0}$ is the preset average turbulence
dissipation rate and $E_{f}$ is the kinetic energy contained in the
forcing bandwidth $0<\left|\boldsymbol{k}\right|\leq k_{f}$. This
forcing has been widely used to simulate statistically steady
homogeneous isotropic turbulence (HIT) on the computer. It offers the
advantage of setting the average turbulence dissipation a priori for
statistically steady turbulence. In the present work, we chose
$k_{f}=2.5$ and $\varepsilon_{0}=0.1$, $0.15$, $0.2$ to obtain
turbulence with different energy levels. The main parameters
characterising the reference flows are given in table \ref{tab:main
	parameters}. All the DNS results presented in this paper are
obtained with $N=1024$.

In the present work, we are mainly interested in the
  evolution of uncertainty without external input. To avoid external
forcing effects on the uncertainty field's evolution (see the previous
section's penultimate paragaph) we ensure that the reference and the
perturbed fields are forced by the exact same forcing, i.e. that
$\boldsymbol{g}$ vanishes. The forcing in the perturbed field is
therefore determined by the velocity in the reference field, i.e.
\begin{equation}
	\label{eq:negative damping forcing 2}
	\hat{\boldsymbol{f}}^{(2)}\left(\boldsymbol{k},t\right)=\hat{\boldsymbol{f}}^{(1)}\left(\boldsymbol{k},t\right)=\left\{
	\begin{array}{ccc}
		\frac{\varepsilon_{0}}{2E_{f}^{(1)}}\hat{\boldsymbol{u}}^{(1)}\left(\boldsymbol{k},t\right)&       &\text{if } 0<\left|\boldsymbol{k}\right|\leq k_{f},\\
		0&       &\text{otherwise,}
	\end{array}\right.
\end{equation}		 
and the uncertainty therefore grows
exclusively by internal production. For consistency
  with our previous work \citep{ge2023production}, we refer to this
  setup as F2. To reduce statistical fluctuations, we repeat the case
  $\varepsilon_{0}=0.1$ six times, based on exactly the same reference
  flow (the top case in table \ref{tab:main parameters}), but with
  initial randomly generated perturbations.

To highlight the influence of external forcing on the
  later-time evolution of uncertainty, we simulate an additional case
  based on the following forcing set up
\begin{equation}
	\hat{\boldsymbol{f}}^{(m)}\left(\boldsymbol{k},t\right)=\left\{
	\begin{array}{ccc}
		\frac{\varepsilon_{0}}{2E_{f}^{(m)}}\hat{\boldsymbol{u}}^{(m)}\left(\boldsymbol{k},t\right)&       &\text{if } 0<\left|\boldsymbol{k}\right|\leq k_{f},\\
		0&       &\text{otherwise,}
	\end{array}\right.
\end{equation}	
with $\varepsilon_{0}=0.1$ and $m=1,2$. In this
  setup, which we refer to as F1, the external forcing applied to the
  reference and perturbed fields depends on the respective fields and
  therefore eventually grows to be different in the two fields. This
  case represents the configuration that has widely been applied in
  previous works
  \citep{berera2018chaotic,boffetta2017chaos,mohan2017scaling,ge2023production,ho2020fluctuations}.

\begin{table}
	\begin{center}
		\def~{\hphantom{0}}
		\begin{tabular}{cccccccccc}
			Case&$N^{3}$&$\nu$&$\left\langle\left\langle\varepsilon\right\rangle\right\rangle_{t}$&$\left\langle U\right\rangle_{t}$&$\left\langle L\right\rangle_{t}$&$ \left\langle T\right\rangle_{t}$&$\left\langle Re_{L}\right\rangle_{t}$&$\left\langle Re_{\lambda}\right\rangle_{t}$&$ \left\langle k_{\max}\eta\right\rangle_{t}$\\ [3pt]
		\multirow{3}{*}{F2}&$1024^{3}$&$4.00\times10^{-4}$&0.1023&0.615&1.010&1.640&1554.7&229.8&1.71\\
		&$1024^{3}$&$4.53\times10^{-4}$&0.1489&0.714&1.073&1.503&1690.9&240.6&1.71\\
		&$1024^{3}$&$4.98\times10^{-4}$&0.2038&0.782&1.035&1.323&1629.0&235.9&1.69\\
		F1&$1024^{3}$&$4.00\times10^{-4}$&0.1023&0.615&1.010&1.640&1554.7&229.8&1.71\\
		\end{tabular}
		\caption{Parameters of the reference flows, where
			$\left\langle \right\rangle$ represents spatial
			averaging, $\left\langle \right\rangle_{t}$
			represents temporal averaging over the entire
			simulation time of the reference flow and
			$\left\langle\left\langle\cdot\right\rangle\right\rangle_{t}$
			represents averaging in both space and time. $N^{3}$
			is the number of grid points of the simulations and
			$k_{\max}=N/3$ is the maximum resolvable wavenumber,
			$\nu$ is the kinematic viscosity, $\varepsilon$ is
			the dissipation, and
			$\eta\equiv\left(\nu^{3}/\left\langle\varepsilon\right\rangle\right)^{1/4}$
			is the Kolmogorov scale. $U\equiv\sqrt{2\left\langle
				E\right\rangle/3}$ (where $E$ is the space-time
			local kinetic energy in the reference flow) is the
			rms velocity and $L\equiv\left(3\pi/4\left\langle
			E\right\rangle\right)\int k^{-1}\hat{E}(k)\rm{d}$$k$
			is the integral length scale. $T\equiv L/U$ is the
			large eddy turnover time. $Re_{L}\equiv UL/\nu$ is
			the Reynolds number and $Re_{\lambda}\equiv
			Ul_{\lambda}/\nu$ is the turbulent Reynolds number
			defined with the Taylor length
			$l_{\lambda}\equiv\sqrt{10\left\langle
				E\right\rangle\nu/\left\langle\varepsilon\right\rangle}$.}
		\label{tab:main parameters}
	\end{center}
\end{table}

	\section{\label{sec:Numerical results}DNS results}

In this section we present our DNS results starting with the
time evolution of single point uncertainty in
subsection \ref{sec:Time evolution of single-point
	uncertainty} followed by the study of the scale-by-scale
uncertainty budget in subsections \ref{sec:Scale-by-scale
	uncertainty energy budget}. In subsection
\ref{sec:Validation of equilibrium dual cascades of
	uncertainty energy} we summarise the previous subsection's
results and present a dual cascades of uncertainty
hypothesis. Finally in subsection \ref{sec:Non-equilibrium
	correction} we use our DNS to investigate the uncertainty
production and dissipation scalings.


\subsection{Time evolution of single point uncertainty\label{sec:Time evolution of single-point uncertainty}}                           

\subsubsection{Average uncertainty energy\label{sec:Average uncertainty energy}}
\begin{figure}
	\centering
	\includegraphics[width=1\textwidth]{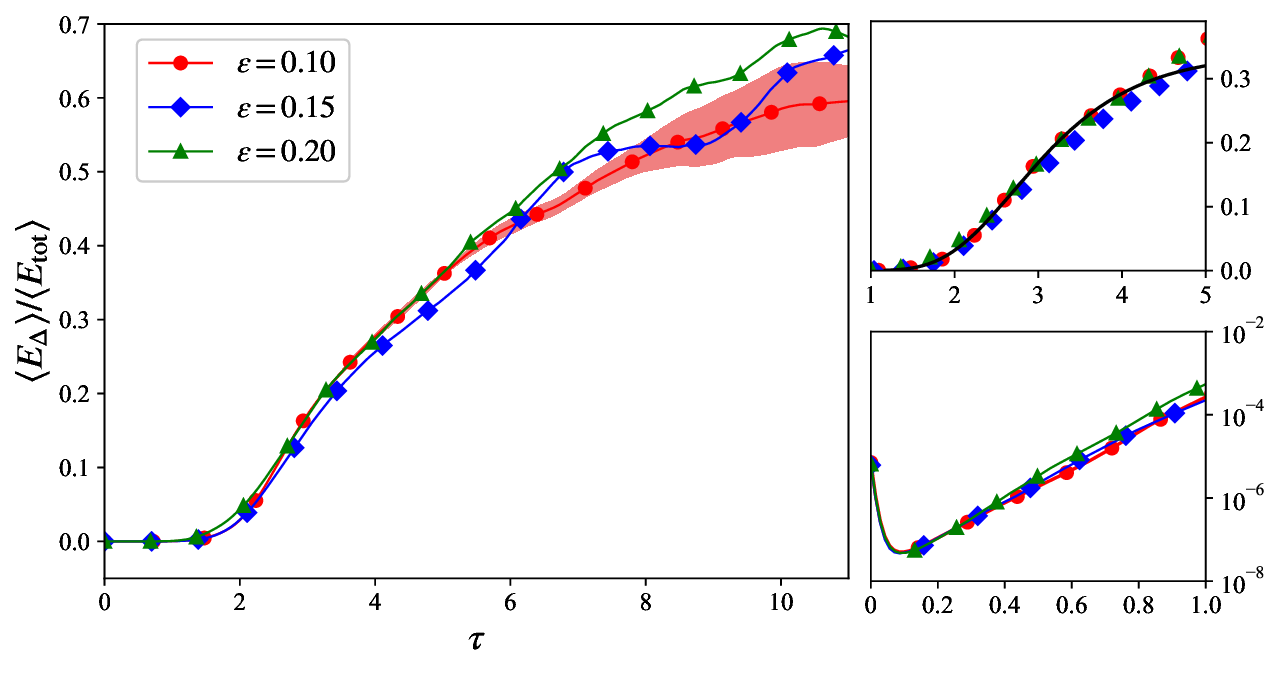}
	\caption{ Time evolutions of average uncertainty energy for F2
          cases. For $\varepsilon_{0}=0.1$,
            $\left\langle E_{\Delta}\right\rangle$ is also averaged
            over an ensemble of six independent realisations, while
            for $\varepsilon_{0}=0.15$ and $0.2$, $\left\langle
            E_{\Delta}\right\rangle$ is only taken from a single
            realisation. The fluctuations of the uncertainty energy
            for $\varepsilon_{0}=0.1$ within one standard deviation
            from the mean are represented by the shaded area.
          Subfigure in bottom right: the initial time
          ($\tau\in[0,1.0]$) growth of average uncertainty energy in
          semilogarithmic plot. Subfigure in top right: the
          uncertainty evolution during $\tau\in[1.0,5.0]$, where the
          exponential of exponential function fit is indicated by a
          solid black line.}
	\label{fig:time evolution of uncertinty} 
\end{figure}

Figure \ref{fig:time evolution of uncertinty} shows the time
evolutions of $\left\langle E_{\Delta}\right\rangle$ for each
F2 case. For $\varepsilon_{0}=0.1$, $\left\langle
  E_{\Delta}\right\rangle$ is averaged over an ensemble of six
  independent realisations, as mentioned in section \ref{sec:Numerical
    steups}, while for $\varepsilon_{0}=0.15$ and $0.2$, $\left\langle
  E_{\Delta}\right\rangle$ is only taken from a single realisation.
$\left\langle E_{\Delta}\right\rangle$ decreases immediately after the
perturbations are seeded because average dissipation
$\left\langle\varepsilon_{\Delta}\right\rangle$ initially overwhelms
average production $\left\langle P_{\Delta}\right\rangle$ in equation
(\ref{eq:single point uncertainty equation}) as already observed and
explained in \citet{ge2023production}. The initial decrease regime is
followed by exponental growth of $\left\langle
E_{\Delta}\right\rangle$ as shown in the bottom right plot of figure
\ref{fig:time evolution of uncertinty}. This is the self-similar
chaotic regime of uncertainty growth studied in detail by
\citet{ge2023production}. As time proceeds, the exponential growth of
uncertainty energy transits into the exponential of exponential growth
regime identified in \citet{ge2023production}, as shown in the top
right plot of figure \ref{fig:time evolution of uncertinty}. The time
ranges for each one of these regimes are given in table \ref{tab:time
  characteristics} in terms of normalised time $\tau\equiv
t/\left\langle T^{(1)}\right\rangle_{t}$ ($T^{(1)}$ is $T$ in table
\ref{tab:main parameters}) with origin $\tau =0$ when the high
wavenumber small perturbation is introduced into the flow. Note that
the time ranges are the same for all three F2 cases
even though they have different levels of turbulence dissipation and
turbulence intensity. As shown by \citet{ge2023production} these three
time ranges are longer/shorter in terms of normalised time $\tau$ for
lower/higher Reynolds numbers.


In the present work, we are particularly interested in the growth of
uncertainty after the exponential of exponential growth regime which
ends at about $\tau =4$ and before the saturation time when the
uncertainty energy stops growing any further because the reference and
perturbed fields have decorrelated as much as they can. It is within
this time range that we may find the stochastic time range where
$\left\langle E_{\Delta}\right\rangle$ and $l_{\Delta}$ both grow as
power laws of time. We have run three F2 simulations
  with three different values of $\varepsilon_0$ till $\tau$ close to
  16 (not shown here) and have established that the saturation time is
  at about $\tau =10$ (see table \ref{tab:time characteristics}). We
  therefore run the remaining five F2 simulations for $\varepsilon =
  0.1$ till $\tau =11$ (see large left plot in figure \ref{fig:time
  evolution of uncertinty} ) and look for power law behaviours in the
time range between $\tau = 4.0$ and $\tau=10$. We
have two alternative predictions for $\left\langle
E_{\Delta}\right\rangle$: $\left\langle E_{\Delta}\right\rangle \sim
\varepsilon t$ on the basis of scalings (\ref{eq:production/disspation
  scaling law a})-(\ref{eq:production/disspation scaling law b}) and
$\left\langle E_{\Delta}\right\rangle \sim (\varepsilon U^{(1)}
t)^{2/3}$ on the basis of scalings (\ref{eq:production/disspation
  scaling law2 a})-(\ref{eq:production/disspation scaling law2
  b}). Note that these power law predictions are equivalent to
$\left\langle E_{\Delta}\right\rangle/\left\langle
E_{tot}\right\rangle \sim \tau$ and $\left\langle
E_{\Delta}\right\rangle/\left\langle E_{tot}\right\rangle \sim
\tau^{2/3}$, respectively, on account of $\varepsilon \sim
U^{(1)3}/L^{(1)}$. Therefore, the collapse of the three different
curves in figure \ref{fig:time evolution of uncertinty} does not
favour any of the two power laws. To evaluate which
  power-law scaling better describes the uncertainty growth, we apply
  power law fits to the time evolution of $\left\langle
  E_{\Delta}\right\rangle$ and also use finite-size Lyapunov exponent
  (FSLE) analysis following \cite{boffetta2002predictability}.

Taking into account the translation and offset caused
  by the transition between different growth regimes, the applied
  curve fitting models are $a(\tau-b)+c$ (linear) and
  $a(\tau-b)^{2/3}+c$ ($2/3$ power law).  As shown by the
coefficients of determination in figure \ref{fig:time evolution of
  uncertintyfit}, a $2/3$ power law fit over the time range
$\tau_s \le \tau \le 10$ is rather better than a
linear fit over this time range for any $\tau_s$ between 4 and 5 and
for all three F2 cases with different turbulence and
intensity levels. However, this conclusion may be
  contingent on the choice of time range and/or forcing
  configuration. We therefore now try the F1 forcing setup and also
  the FSLE approach.

\begin{figure}
	\centering 
	\subfigure[]{
		\label{fig:growthcertaintyenergyComparsion}
		\includegraphics[width=0.48\textwidth]{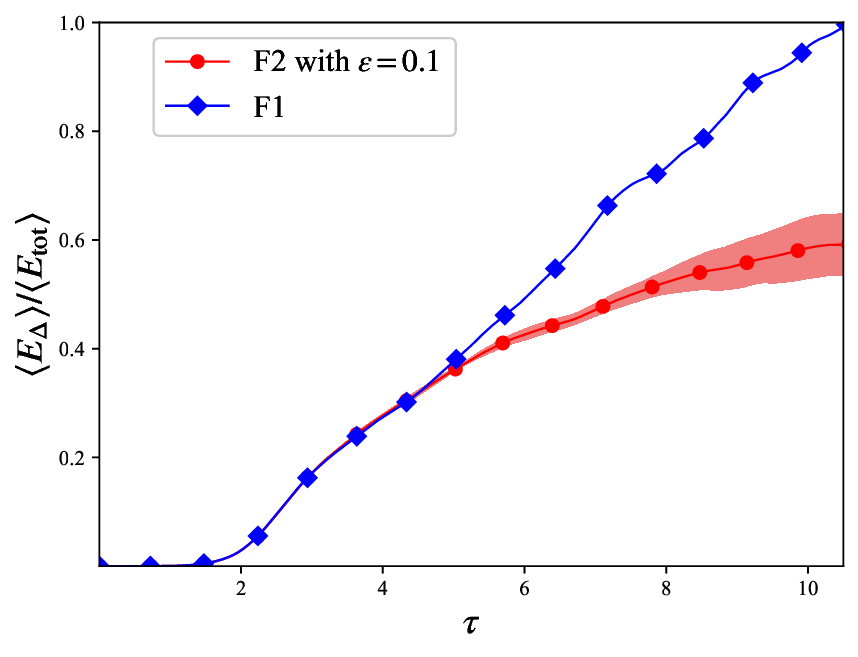}}
	\subfigure[]{
		\label{fig:FSLEenergyComparsion}
		\includegraphics[width=0.48\textwidth]{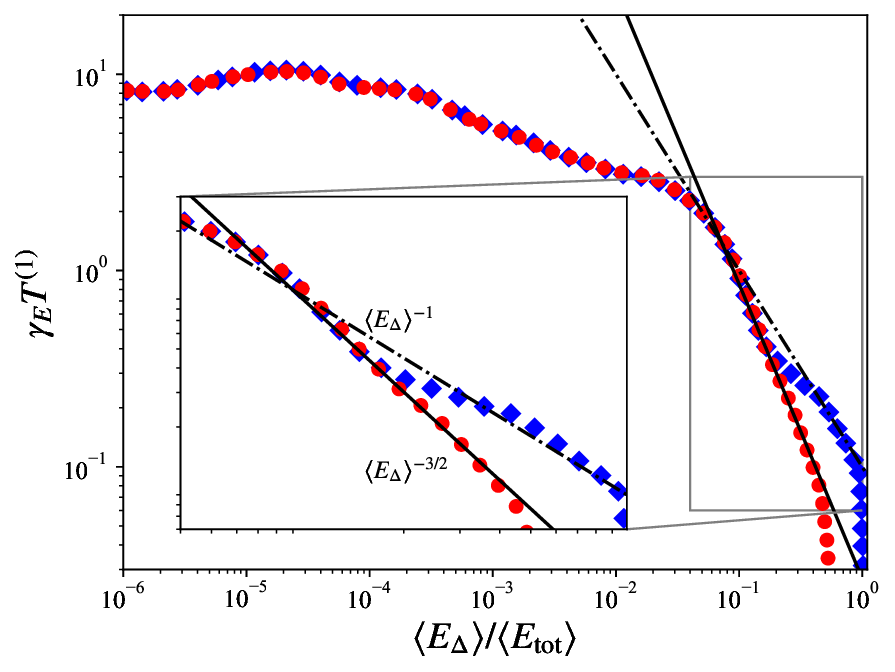}}
	\caption{(a) Evolution of uncertainty energy for cases F1 and F2 with $\varepsilon_{0}=0.1$. (b) Evolution of FSLE for cases F1 and F2 - $\varepsilon_{0}=0.1$.}
\end{figure}

Figure \ref{fig:growthcertaintyenergyComparsion}
  compares the uncertainty energy evolutions for the F1 and F2 setups,
  both with $\varepsilon_{0}=0.1$. Although there is only one single
  realisation for F1 and six realisations for F2, it is clearly
  observed that F1 significantly and systematically deviates from F2
  starting from $\tau$ between $4$ and $5$, which corresponds to the
  start of the power-law growth fitted for F2 in the previous
  paragraph. Linear growth of uncertainty energy has been reported in
  previous works for F1 in the stochastic time regime
  \citep{leith1972predictability,boffetta2017chaos,ge2023production}. Therefore,
  this systematic deviation between F1 and F2 may indicate the
  critical role of external input and the possibility of a new
  uncertainty scaling in the stochastic time regime. Indeed, we
  already made the point in section \ref{sec:Production/dissipation
    scaling of uncertainty energy} that different correlations between
  reference and uncertainty fields can, in principle, lead to
  different scalings.

For an additional and perhaps less ambiguous
  quantification of the stochastic time regime's power law, we now use
  the finite-size Lyapunov exponent (FSLE)
  \citep{boffetta2002predictability} defined as
\begin{equation}
	\label{eq:FSLE uncertainty}
		\gamma_{E}(\left\langle
                E_{\Delta}\right\rangle)=\frac{1}{\left\langle
                  T_{a}\right\rangle_{e}}\left\langle\ln\left(\frac{\left\langle
                  E_{\Delta}\right\rangle(t+T_{a})}{\left\langle
                  E_{\Delta}\right\rangle(t)}\right)\right\rangle_{e},
\end{equation}
where $\left\langle\cdot\right\rangle_{e}$ denotes
  the ensemble average. For case F1, only a single realization is available, so no ensemble averaging is applied. For case F2, the ensemble average is performed over six independent realizations. $T_{a}$ is the time required for
  $\left\langle E_{\Delta}\right\rangle$ to grow by a factor
  $a$. Here, we set $a=2$ for $\left\langle
  E_{\Delta}\right\rangle/\left\langle E_{tot}\right\rangle\leq0.2$,
  and for larger uncertainty energies we take $T_a$ constant and equal
  to the $T_a$ which corresponds to $\left\langle
  E_{\Delta}\right\rangle/\left\langle
  E_{tot}\right\rangle=0.2$. (Other choices of $a$ and $T_a$ were also
  tried without change in our conclusions.) The two candidate scalings
  $\left\langle E_{\Delta}\right\rangle \sim \varepsilon t$ and
  $\left\langle E_{\Delta}\right\rangle \sim (\varepsilon U^{(1)}
  t)^{2/3}$ lead to $\gamma_{E}(\left\langle
  E_{\Delta}\right\rangle)\sim\varepsilon\left\langle
  E_{\Delta}\right\rangle^{-1}$ and $\gamma_{E}(\left\langle
  E_{\Delta}\right\rangle)\sim \varepsilon U^{(1)}\left\langle
  E_{\Delta}\right\rangle^{-3/2}$ respectively. Figure
  \ref{fig:FSLEenergyComparsion} shows the evolution of $\gamma_{E}$
  as a function of $\left\langle E_{\Delta}\right\rangle$ in cases F1
  with $\varepsilon_{0}=0.1$ (average over one realisation) and F2
  with $\varepsilon_{0}=0.1$ (average over six realisations). The
  stochastic time range deviation between F1 and F2 appears clearly in
  the figure, with the former following $\gamma_{E}(\left\langle
  E_{\Delta}\right\rangle)\sim\left\langle
  E_{\Delta}\right\rangle^{-1}$ and the latter following
  $\gamma_{E}(\left\langle E_{\Delta}\right\rangle)\sim \left\langle
  E_{\Delta}\right\rangle^{-3/2}$.

Both the direct power law fits of $\left\langle
  E_{\Delta}\right\rangle (\tau)$ and the FSLE analysis show that 2/3
  power law better describes the stochastic time growth of uncertainty
  without forcing-generated uncertainty. This may also suggest an
indirect preference of our F2 data for the uncertainty dissipation
scalings (\ref{eq:production/disspation scaling law2
  a})-(\ref{eq:production/disspation scaling law2 b}) rather than
(\ref{eq:production/disspation scaling law
  a})-(\ref{eq:production/disspation scaling law b}), a point to which
we return in subsection \ref{sec:Non-equilibrium correction} at the
end of this section.

\begin{figure}
	\centering
	\includegraphics[width=0.7\textwidth]{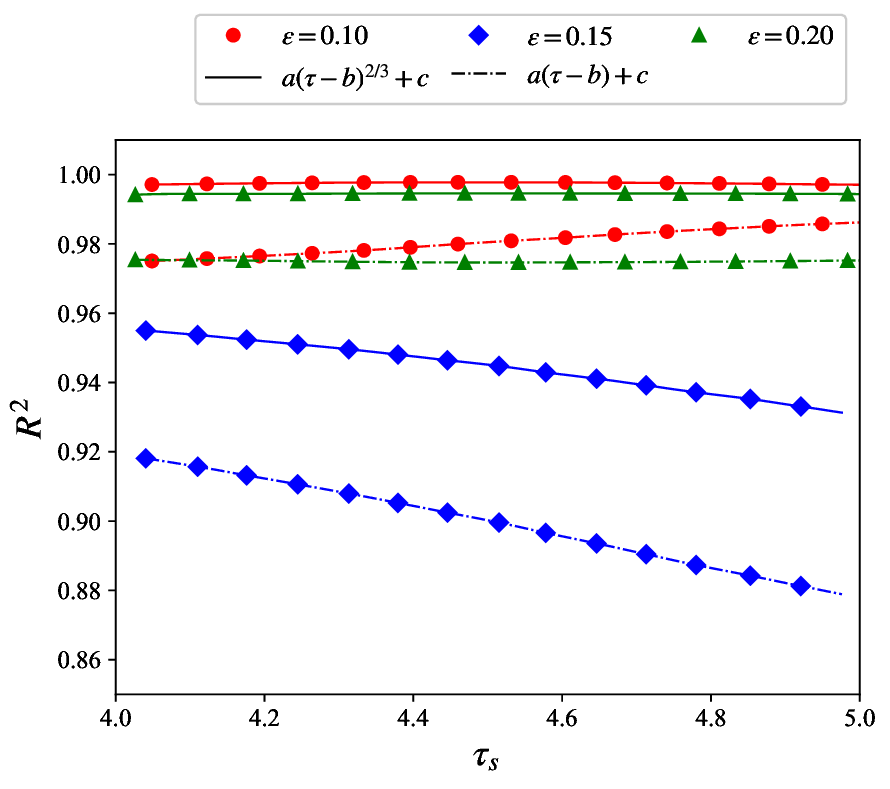}
	\caption{$R^{2}$ coefficients from curve fitting models
          $a(\tau-b)+c$ and $a(\tau-b)^{2/3}+c$ for the evolution of
          $\left\langle E_{\Delta}\right\rangle$ in figure
          \ref{fig:time evolution of uncertinty} over the time
          interval $\tau \in [\tau_{s}, 10]$.}
	\label{fig:time evolution of uncertintyfit} 
\end{figure}

	\begin{table}
	\begin{center}
		\def~{\hphantom{0}}
		\begin{tabular}{lccccccccc}
			$N^{3}$&Uncertainty regime&Time interval $\tau$\\
			\multirow{5}*{$1024^{3}$}&Initial decrease&$[0,0.1]$\\
			~&Exponential growth&$[0.1,1.0]$\\
			~&Exponential of exponential growth&$[1.0,4.0]$\\
			~&Transient growth&$[4.0,5.0]$\\
			~&Power-law growth&$[5.0,10]$\\
			~&Saturation&$[10,+\infty]$\\
		\end{tabular}
		\caption{Time ranges of different uncertainty
			evolution regimes.}
		\label{tab:time characteristics}
	\end{center}
\end{table}

\subsubsection{Characteristic length of uncertainty}



Deciding between $\left\langle E_{\Delta}\right\rangle/\left\langle
E_{tot}\right\rangle \sim \tau$ and $\left\langle
E_{\Delta}\right\rangle/\left\langle E_{tot}\right\rangle \sim
\tau^{2/3}$ is more convincing if the data also favour the consistent
time-dependence of $l_{\Delta}$ in the same time range,
i.e. $l_{\Delta}\sim \varepsilon^{1/2} t^{3/2}$ in the case of linear
time-dependence of $\left\langle E_{\Delta}\right\rangle$ and
$l_{\Delta}\sim U^{(1)} t$ in the case of $2/3$ power law
time-dependence of $\left\langle E_{\Delta}\right\rangle$.

\begin{figure}
	\centering
	\includegraphics[width=0.7\textwidth]{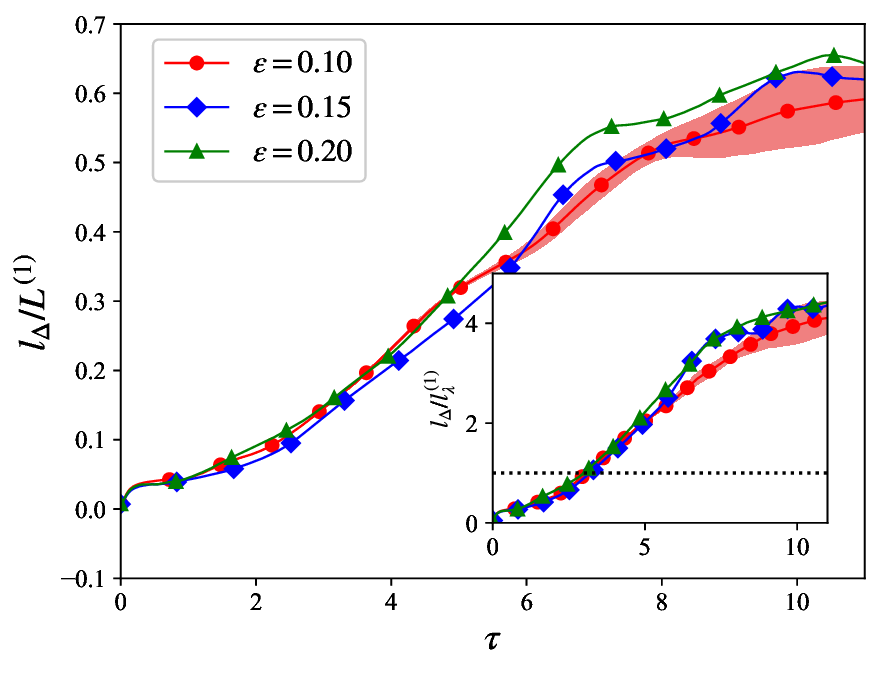}
	\caption{ Time evolution of
          $l_{\Delta}/L^{(1)}$ for F2
          cases. Inset: Time evolution of
            $l_{\Delta}/l_{\lambda}^{(1)}$. For $\varepsilon_{0}=0.1$,
            $l_{\Delta}$ is averaged over an ensemble of six
            independent realisations, while for $\varepsilon_{0}=0.15$
            and $0.2$, $l_{\Delta}$ is only taken from a single
            realisation. The fluctuations of $l_{\Delta}$ for
            $\varepsilon_{0}=0.1$ within one standard deviation from
            the mean are represented by the shaded area. Note that
          the two power law predictions for $l_{\Delta}$ are
          equivalent to $l_{\Delta}/l_{\lambda}^{(1)}\sim
          \tau^{3/2}\sqrt{U^{(1)}L^{(1)}/\nu}$ and
          $l_{\Delta}/l_{\lambda}^{(1)}\sim \tau
          \sqrt{U^{(1)}L^{(1)}/\nu}$ on account of $\varepsilon \sim
          U^{(1)3}/L^{(1)}$. The three different F2 cases have more or
          less the same Reynolds number.}
	\label{fig:LDelta} 
\end{figure}

Figure \ref{fig:LDelta} shows the time evolution of $l_{\Delta}$ for
each one of F2 cases. In this figure, $l_{\Delta}$ is normalized by
the reference field's integral length scale
  $L^{(1)}$, as well as the Taylor length $l_{\lambda}^{(1)}$ of the
reference flow in order to easily check whether $l_{\Delta}$ is
sufficently larger than $l_{\lambda}^{(1)}$ during the power law time
range as this is a condition for the validity of the equilibrium model
of uncertainty growth on which the power law predictions are based.

\begin{figure}
	\centering
	\includegraphics[width=0.7\textwidth]{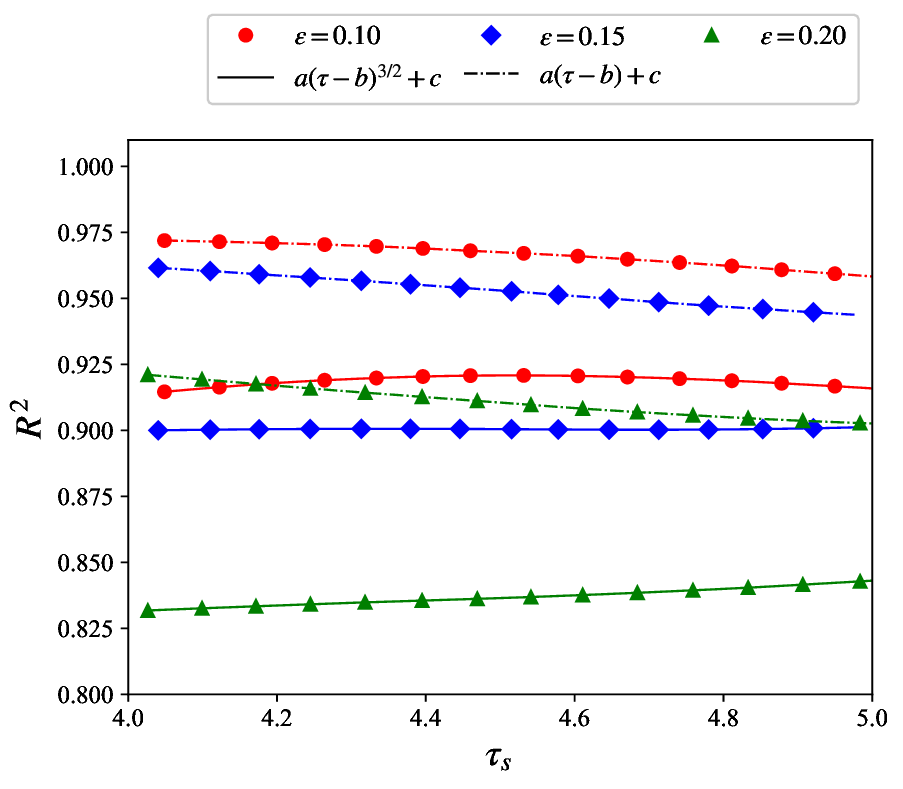}
	\caption{$R^{2}$ coefficients from curve fitting models
          $a(\tau-b)+c$ and $a(\tau-b)^{3/2}+c$ for the evolution of
          $l_{\Delta}$ in figure \ref{fig:LDelta} over the time
          interval $\tau \in [\tau_{s}, 10]$.}
	\label{fig:LDeltafit} 
\end{figure}

The evolution of $l_{\Delta}$, in particular during the early time
when the chaoticity dominates the growth of uncertainty, has been
already discussed in our previous work \citep{ge2023production}: at
the very initial times when uncertainty dissipates, $l_{\Delta}$ is
correspondingly increasing until the chaotic regime (from
$\tau\approx0.1$ to $\tau\approx1.0$ in the present simulations) where
$\left\langle E_{\Delta}\right\rangle$ grows exponentially
and $l_{\Delta}$ reaches a constant plateau as shown by
\citet{ge2023production}. After the chaotic exponential growth of
$\left\langle E_{\Delta}\right\rangle$, $l_{\Delta}$ increases again.
At $\tau\approx4.0$, close to the start of the power law regime
according to our previous subsection's results on $\left\langle
E_{\Delta}\right\rangle$, $l_{\Delta}/l^{(1)}_{\lambda}$ has reached a
value of approximately $1.5$ (see figure \ref{fig:LDelta}). We may
therefore consider that the decorrelation process has entered an
inertial range stochastic regime during which
$l_{\Delta}/l^{(1)}_{\lambda}$ grows even further as a power law to
reach about 4.5 at approximate saturation time $\tau=10$ after which
$l_{\Delta}/l^{(1)}_{\lambda}$ oscillates around the same value
for all three F2 cases.
The coefficients of determination in figure \ref{fig:LDeltafit}
suggest that $l_{\Delta}/L^{(1)}(\tau)$ is marginally better fitted by
a linear function of $\tau$ than by a 2/3 power law over the time
range $\tau_s \le \tau \le 10$, for any $\tau_s$
between 4 and 5 and for all three F2 cases inspite of their different
turbulence dissipation and intensity levels. Similar
  to section \ref{sec:Average uncertainty energy}, the applied curve
  fitting models are $a(\tau-b)^{2/3}+c$ ($2/3$ power law) and
  $a(\tau-b)+c$ (linear) to adapt the translation and offset caused by
  the transition between different growth regimes.

\begin{figure}
	\centering 
	\subfigure[]{
		\label{fig:LDeltaComparsion}
		\includegraphics[width=0.48\textwidth]{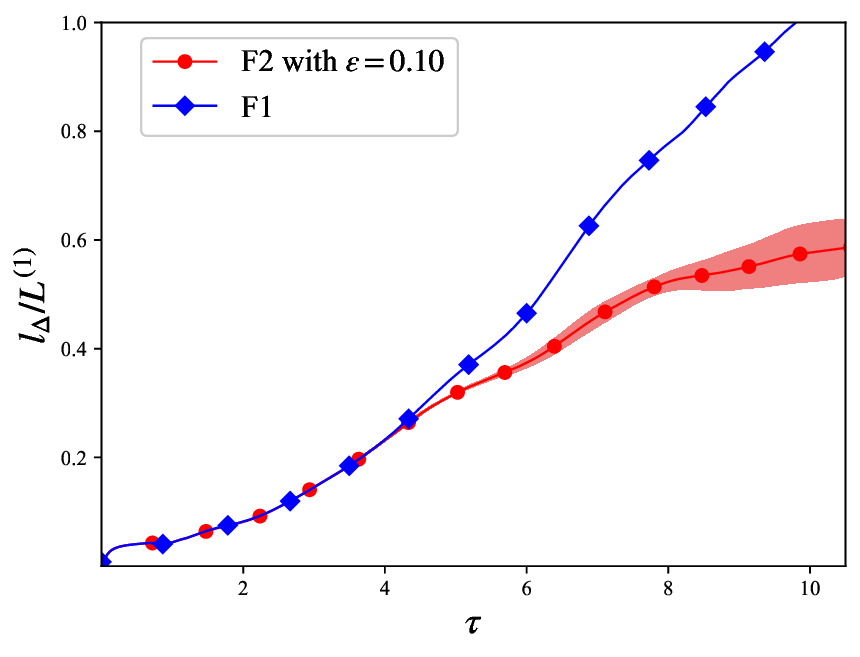}}
	\subfigure[]{
		\label{fig:FSLElDeltaComparsion}
		\includegraphics[width=0.48\textwidth]{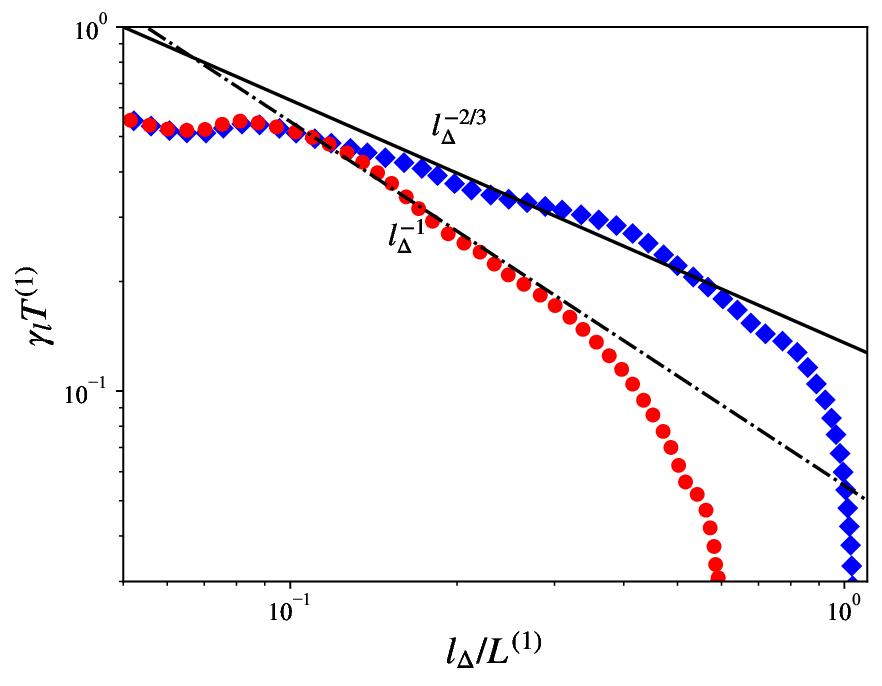}}
	\caption{(a) Evolution of $l_{\Delta}$ for cases F1 and F2 with $\varepsilon_{0}=0.1$. (b) Evolution of FSLE for cases F1 and F2 - $\varepsilon_{0}=0.1$.}
\end{figure}

 Figure \ref{fig:LDeltaComparsion} compares
   $l_{\Delta}$ growths and FSLEs for $l_{\Delta}$ for F1 and F2 with
   $\varepsilon_{0}=0.1$ in both cases. The deviation between F1 and
   F2 is once again observed to happen at $\tau$ between $4$ and
   $5$. We define the FSLE for $l_{\Delta}$ as
\begin{equation}
	\label{eq:FSLE lDelta}
	\gamma_{l}(l_{\Delta})=\frac{1}{\left\langle T_{a}\right\rangle_{e}}\left\langle\ln\left(\frac{ l_{\Delta}(t+T_{a})}{ l_{\Delta}(t)}\right)\right\rangle_{e}.
\end{equation} 
Here $a$ is set to be $2$ when $ l_{\Delta}/
  l_{\Delta}^{(1)}\leq1$. When $l_{\Delta}/ l_{\Delta}^{(1)}>1$,
  $T_{a}$ is taken as a constant and equal to the value of $T_{a}$ for
  $l_{\Delta}/ l_{\Delta}^{(1)}=1$.  The two candidate scalings
  $l_{\Delta} \sim t^{3/2}$ and $l_{\Delta} \sim t$ yield
  $\gamma_{l}(l_{\Delta})\sim l_{\Delta}^{-2/3}$ and
  $\gamma_{l}(l_{\Delta})\sim l_{\Delta}^{-1}$ respectively. Figure
  \ref{fig:FSLElDeltaComparsion} shows the evolution of $\gamma_{l}$
  as a function of $l_{\Delta}$ in case F1 ($\varepsilon_{0}=0.1$ and
  average over one realisation) and F2 with $\varepsilon_{0}=0.1$
  (average over six realisations). The results suggest that
  $l_{\Delta}$ follows $\gamma_{l}(l_{\Delta})\sim l_{\Delta}^{-2/3}$
  for F1 but follows $\gamma_{l}(l_{\Delta})\sim l_{\Delta}^{-1}$ for
  F2. A coherent picture of a power law time range is therefore
emerging where $l_{\Delta} \sim t$ and $\left\langle
E_{\Delta}\right\rangle \sim t^{2/3}$ are both better supported by our
data than $l_{\Delta} \sim t^{3/2}$ and $\left\langle
E_{\Delta}\right\rangle \sim t$ for the stochastic
  time regime of uncertainty growth without forcing-generated
  uncertainty.



\subsubsection{Ratio between uncertainty production and dissipation}
\begin{figure}
	\centering
	\includegraphics[width=0.7\textwidth]{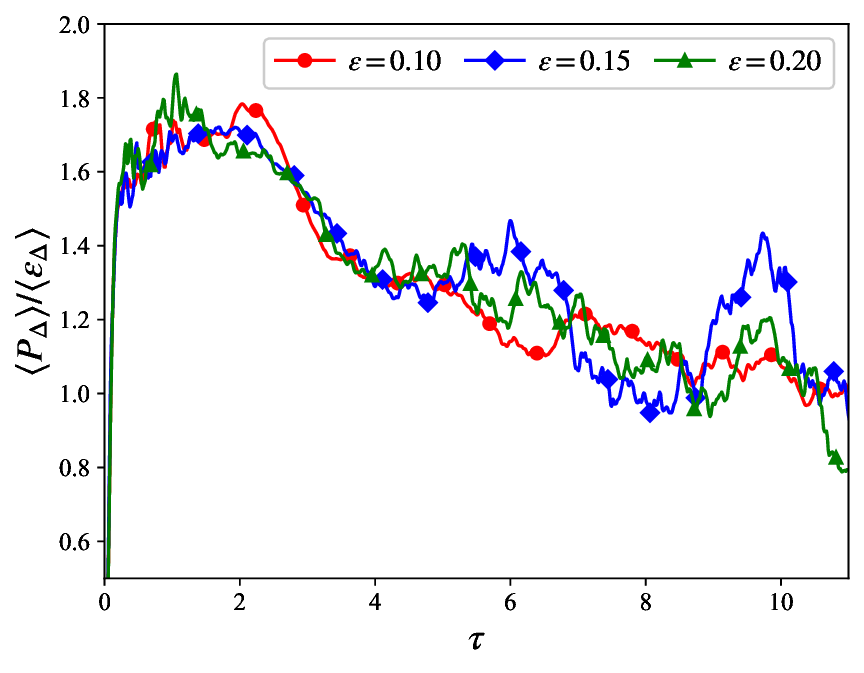}
	\caption{Time evolutions of ratio $\left\langle
          P_{\Delta}\right\rangle/\left\langle\varepsilon_{\Delta}\right\rangle$
          for F2 cases. For $\varepsilon_{0}=0.1$,
            ensemble averaging over six realisations is applied,
            whereas for the other two values of $\varepsilon_0$, the
            data are taken from one single realisation.  }
	\label{fig:P-Epsilon.eps} 
\end{figure}

We close subsection \ref{sec:Time evolution of single-point
  uncertainty} by addressing the proportionality between $\left\langle
P_{\Delta}\right\rangle$ and
$\left\langle\varepsilon_{\Delta}\right\rangle$ which is assumed in
the argument of section \ref{sec:Production/dissipation scaling of
  uncertainty energy}. \citet{ge2023production} have shown a quasi
time-constant ratio $\left\langle
P_{\Delta}\right\rangle/\left\langle\varepsilon_{\Delta}\right\rangle$
during time ranges which may be interpreted as stochastic time ranges
in $N=512$ cases. Figure \ref{fig:P-Epsilon.eps} shows the time
evolutions of ratio $\left\langle
P_{\Delta}\right\rangle/\left\langle\varepsilon_{\Delta}\right\rangle$
for our three F2 cases. It is observed that the ratio oscillates
around a certain value, which may be considered statistically steady,
in the stochastic time range $5\le \tau \le 10$. For
the case where $\varepsilon_{0}=0.10$, this ensemble
  averaged value is measured to be $1.17\pm0.08$, which is more
stable than the other cases because of the ensemble
  average. This may be why the case $\varepsilon_{0}=0.10$ returns a
better fit to the theoretical predictions than the other two
cases. For the cases where $\varepsilon_{0}=0.15$ and
$\varepsilon_{0}=0.20$, the time average of the ratio during the
stochastic time range, as well as its corresponding standard
deviation, is $1.19\pm0.16$ and $1.15\pm0.11$, respectively. Our DNS
therefore supports the proportionality between $\left\langle
P_{\Delta}\right\rangle$ and
$\left\langle\varepsilon_{\Delta}\right\rangle$ in the stochastic time
range.

\begin{figure}
	\centering 
	\includegraphics[width=0.9\textwidth]{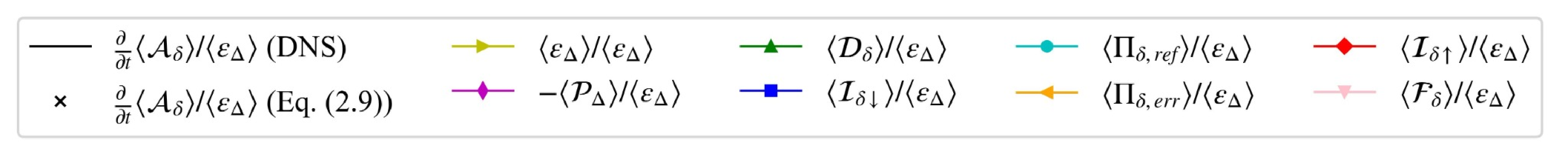}
	\subfigure[Exponential growth regime: $\tau=0.58$]{
		\label{fig:KHMHequation f2 1}
		\includegraphics[width=0.48\textwidth]{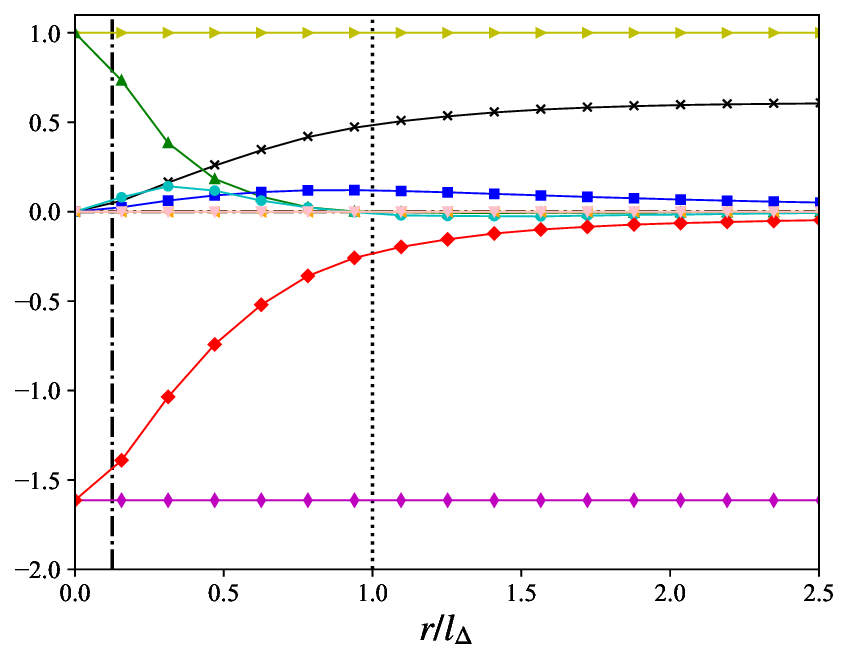}}
	\subfigure[Exponential of exponential growth regime: $\tau=2.66$]{
		\label{fig:KHMHequation f2 2}
		\includegraphics[width=0.48\textwidth]{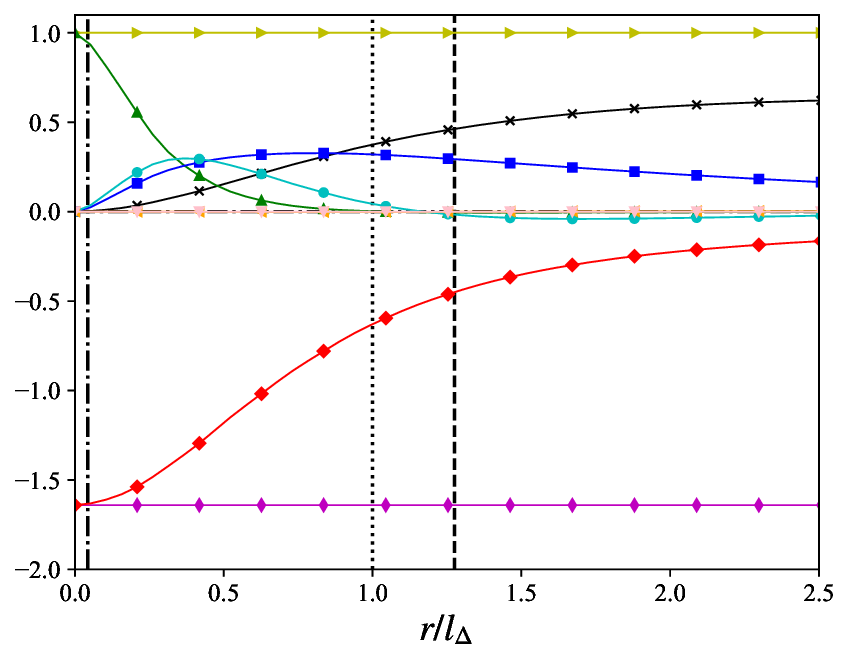}}
	\subfigure[Early power-law growth regime: $\tau=5.43$]{
		\label{fig:KHMHequation f2 3}
		\includegraphics[width=0.48\textwidth]{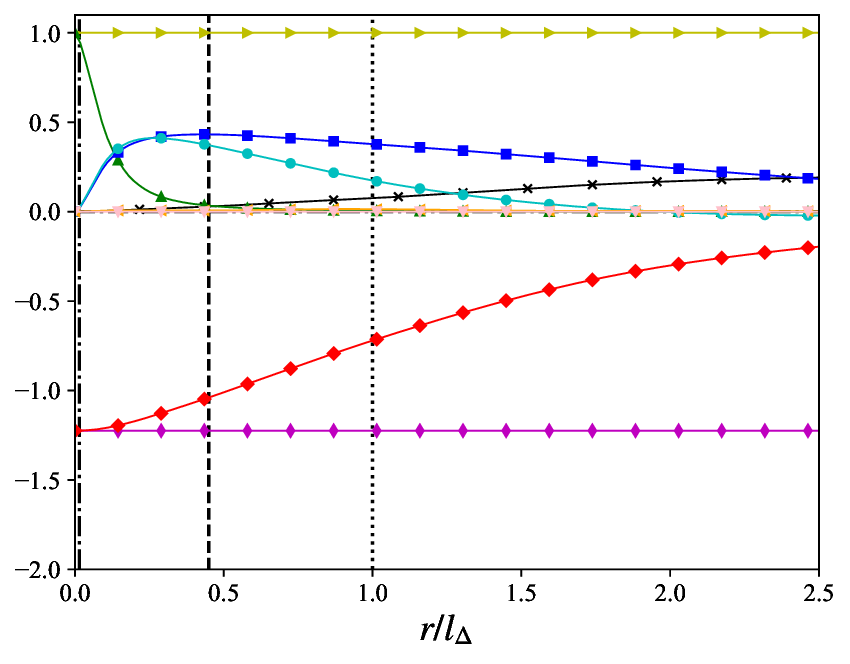}}
	\subfigure[Late power-law growth regime: $\tau=9.56$]{
		\label{fig:KHMHequation f2 4}
		\includegraphics[width=0.48\textwidth]{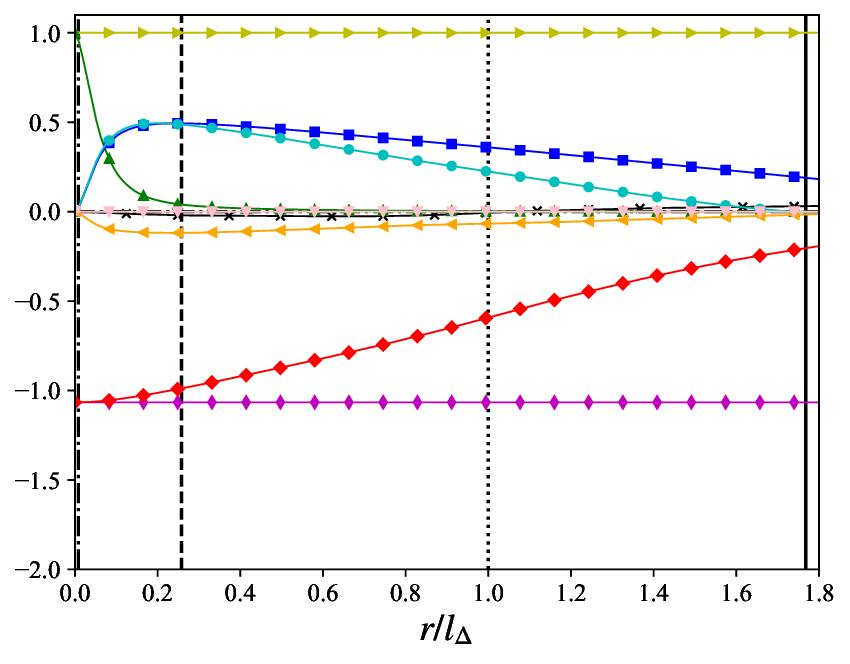}}
	\caption{For case F2 and $\varepsilon_{0}=0.1$, the scale
          evolutions of each sphere-area-averaged term in equation
          (\ref{eq:average difference KHMH difference/average}) at
          different times corresponding to different uncertainty
          growth regimes. The Kolmogorov scale $\eta^{(1)}$, Taylor
          scale $l_{\lambda}^{(1)}$ and integral length scale
          $L^{(1)}$ are plotted as dash-dotted, dashed and solid lines
          respectively. Averaging includes ensemble
            averaging over six realisations in all statistics plotted
            here.}
	\label{fig:KHMHequation} 
\end{figure}

\subsection{Scale-by-scale uncertainty energy budget \label{sec:Scale-by-scale uncertainty energy budget}}	

The dependence on scale $r$ of each term in equation (\ref{eq:average
	difference KHMH difference/average}) at different times, including
the time-derivative of the uncertainty's second order structure
function
$\partial\left\langle\mathcal{A}_{\delta}^{a}\right\rangle/\partial t$
obtained directly from the DNS, are presented in figure
\ref{fig:KHMHequation}. All these terms are averaged over the surface
of a sphere of radius $r= \vert \boldsymbol{r}\vert$. We have adopted this
strategy as it is numerically much cheaper than evaluating every term
in equation (\ref{eq:scale-by-scale uncertainty budget}) by
integrating over the volume of a sphere of radius $r$ for various $r$
and times $t$. We only present the ensemble averaged terms obtained for case F2
with $\varepsilon_{0}=0.1$, but results for the other two F2
cases are qualitatively identical albeit with some quantitative
differences which have no baring on the conclusions and statements
made in this subsection and paper in general. We checked that the
sphere-area-averaged
$\partial\left\langle\mathcal{A}_{\delta}\right\rangle/\partial t$
agrees well with its value obtained from the sphere-area-averaged
equation (\ref{eq:average difference KHMH difference/average}) for all
F2 cases, all $r$ and all times tried, see examples in figure
\ref{fig:KHMHequation}. Furthermore, at $r=0$ the viscous diffusion
term $\left\langle\mathcal{D}_{\delta}\right\rangle$ equals
$\left\langle \varepsilon_{\Delta}\right\rangle$ and the interscale
transfer rate $\left\langle\mathcal{I}_{\delta\uparrow}\right\rangle$
equals $-\left\langle P_{\Delta}\right\rangle$ at all times (see
figure \ref{fig:KHMHequation}) and in all cases as required by theory
(see subsection \ref{sec:Scale-by-scale uncertainty energy balance in
	forced homogeneous turbulence}).

\begin{figure}
	\centering \subfigure[]{
		\label{fig:eigenvalue of xi}
		\includegraphics[width=0.48\textwidth]{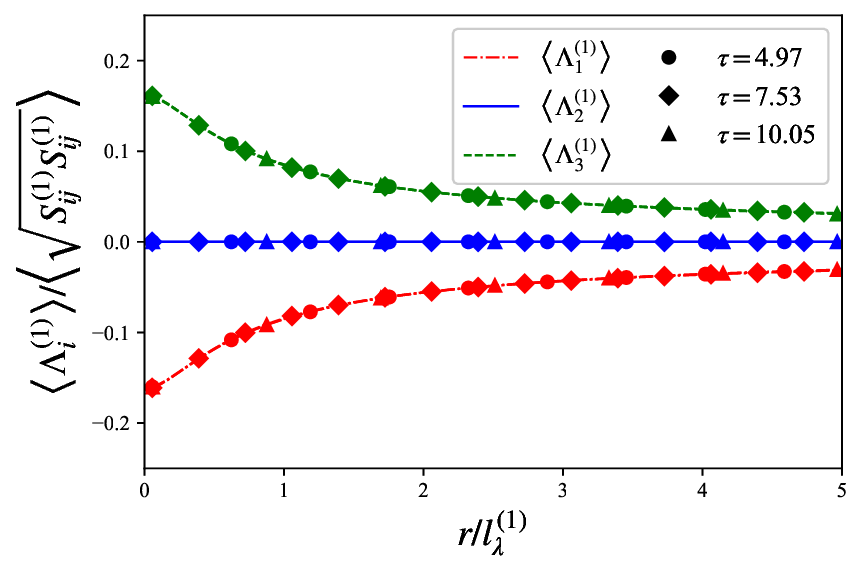}}
	\subfigure[]{
		\label{fig::eigenvalue of deltaxi}
		\includegraphics[width=0.48\textwidth]{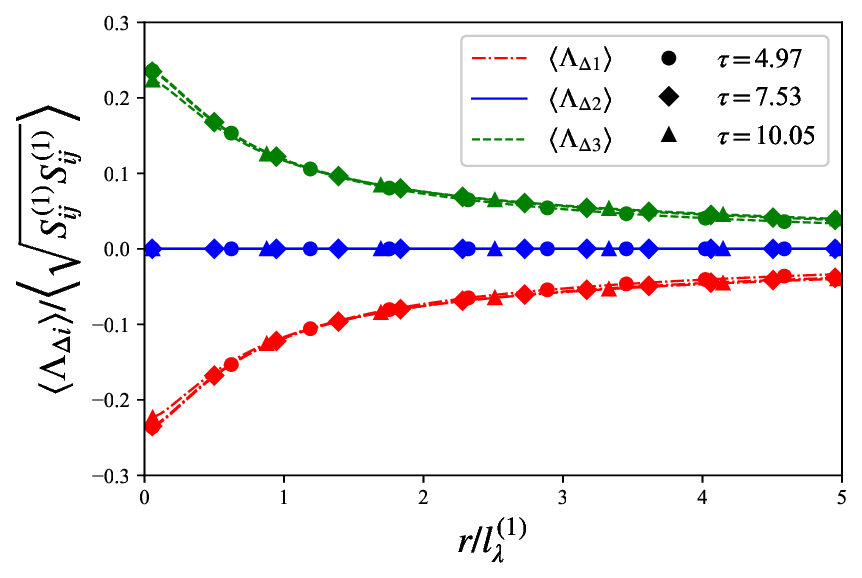}}
	\caption{Dependence of the spatially averaged eigenvalues of
		$\boldsymbol{\Xi}^{(1)}$ and $\boldsymbol{\Xi}_{\Delta}$ on scale $r$ for
		three different times at the start, middle and end of the
		power-law growth regime. For reasons to do with
		computational expense, here and in the following two
		figures, $\boldsymbol{r}=r \boldsymbol{e}_{x}$ where $\boldsymbol{e}_{x}$ is the
		unit vector of one of the three orthonormal directions of
		the simulation field. }
	\label{fig:eigenvalueXi} 
\end{figure}

\begin{figure}
	\centering \subfigure[]{
		\label{fig:averagealignment of deltaw in xi}
		\includegraphics[width=0.48\textwidth]{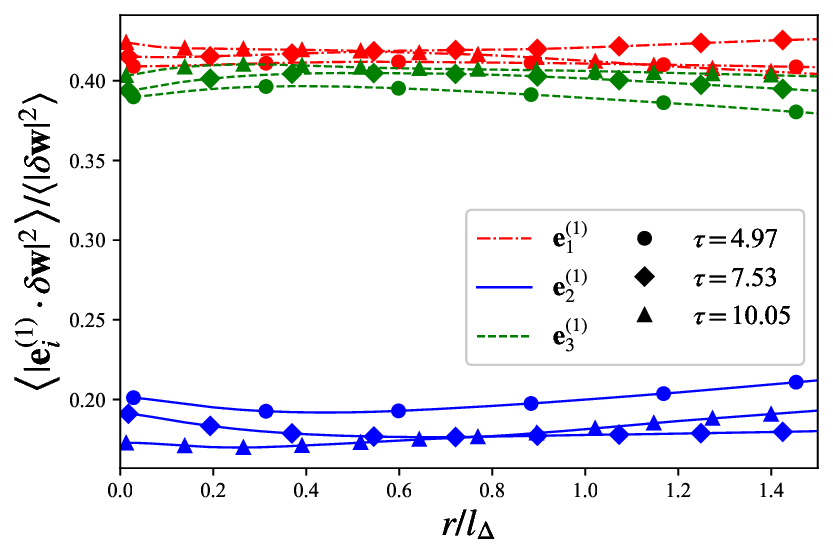}}
	\subfigure[]{
		\label{fig:averagealignment of overlinew in xi}
		\includegraphics[width=0.48\textwidth]{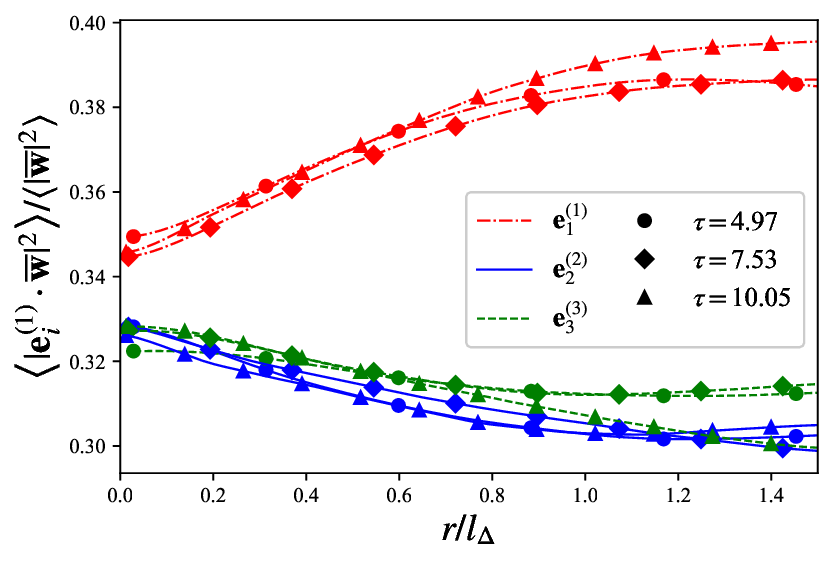}}
	\subfigure[]{
		\label{fig:averagealignment of deltaew in deltaxi}
		\includegraphics[width=0.48\textwidth]{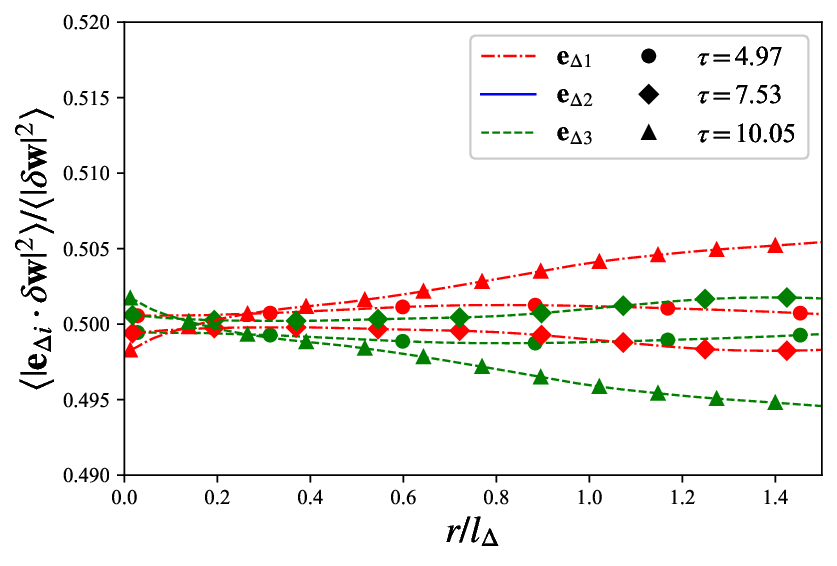}}
	\caption{Evolution of the spatially averaged alignment of (a)
		$\delta\boldsymbol{w}$ and (b) $\overline{\boldsymbol{w}}$ with
		$\boldsymbol{\Xi}^{(1)}$, and (c) the spatially averaged alignment
		of $\delta\boldsymbol{w}$ with $\boldsymbol{\Xi}_{\Delta}$ in scale $r$ and
		time. For the definition of $r$ in this plot see the caption
		of figure \ref{fig:eigenvalueXi}.}
	\label{fig:averagealignment} 
\end{figure}

Figure \ref{fig:KHMHequation} presents four plots, each one
corresponding to a different time: (a) a time within the chaotic
exponential growth regime, (b) a time within the exponential of
exponential growth regime, (c) an early time and (d) a late time
within the stochastic power-law growth regime. The first major
observation is that the interscale transfer rate
$\left\langle\mathcal{I}_{\delta\uparrow}\right\rangle$ makes the
largest absolute magnitude contribution to the right hand side of the
sphere-area-averaged equation (\ref{eq:average difference KHMH
	difference/average}) at all times for $r< l_{\Delta}$. It is
negative at all scales $r$ and therefore, as explained in subsection
\ref{sec:Scale-by-scale uncertainty energy balance in forced
	homogeneous turbulence}, represents inverse interscale transfer
caused by compression overwhelming stretching contributions to
$\left\langle\mathcal{I}_{\delta\uparrow}\right\rangle$. Whilst the
average eigenvalues of the reference field's relative deformation rate
tensor $\boldsymbol{\Xi}^{(1)}$ take positive, zero and negative values and
sum up to zero at all $r$ as predicted in subsection
\ref{sec:Scale-by-scale uncertainty energy balance in forced
	homogeneous turbulence} (see figure \ref{fig:eigenvalue of xi}), the
two-point uncertainty half sum velocity $\overline{\boldsymbol{w}}$ is
predominantly aligned with the compressive eigenvector of the
reference field's relative deformation rate tensor (see figure
\ref{fig:averagealignment of overlinew in xi}). This preferential
alignment significantly contributes to the negative sign of the
sphere-area-averaged
$\left\langle\mathcal{I}_{\delta\uparrow}\right\rangle$, see equation
(\ref{eq:I up term of budget in eigen base}). As argued in section
\ref{sec:Scale-by-scale uncertainty energy balance in forced
	homogeneous turbulence},
$\left\langle\mathcal{I}_{\delta\uparrow}\right\rangle$ represents the
principal mechanism whereby uncertainty energy produced locally in
$\boldsymbol{X}$-space by compressions of the reference field's local strain
rate tensor is transfered to larger scales, and it does so by
compressive motions of the reference field's relative deformation rate
tensor $\boldsymbol{\Xi}^{(1)}$. Note the source of uncertainty energy
transfer at $r=0$,
$\left\langle\mathcal{I}_{\delta\uparrow}\right\rangle=-\left\langle
P_{\Delta}\right\rangle$.

The second major observation is that the interscale transfer rates
$\left\langle\mathcal{I}_{\delta\downarrow}\right\rangle$ and
$\left\langle\Pi_{\delta,\text{ref}}\right\rangle$ are both positive
at all times and for all accessible scales $r$ (figure
\ref{fig:KHMHequation}), except at $r=0$ where they vanish. These
interscale transfer rates therefore represent forward interscale
transfers also caused by compression overwhelming stretching
contributions as explained in subsection \ref{sec:Scale-by-scale
  uncertainty energy balance in forced homogeneous
  turbulence}. Compressive motions are therefore responsible both for
the inverse transfer by
$\left\langle\mathcal{I}_{\delta\uparrow}\right\rangle$ and the
forward transfers by
$\left\langle\mathcal{I}_{\delta\downarrow}\right\rangle$ and
$\left\langle\Pi_{\delta,\text{ref}}\right\rangle$. A contributor to
the compressive effect on
$\left\langle\mathcal{I}_{\delta\downarrow}\right\rangle$, as shown by
equation (\ref{eq:I down term of budget in eigen base}), is the
preferential alignment of the two-point uncertainty half difference
velocity $\delta {\boldsymbol{w}}$ with the compressive eigenvector of
the reference field's relative deformation tensor
$\boldsymbol{\Xi}^{(1)}$, which is marginally the case compared to the
alignment with this tensor's stretching eigenvector as shown in figure
\ref{fig:averagealignment of deltaw in xi}. The forward interscale
transfer rates
$\left\langle\mathcal{I}_{\delta\downarrow}\right\rangle$ and
$\left\langle\Pi_{\delta,\text{ref}}\right\rangle$ are
compression-driven primarily (in fact completely in the case of
$\left\langle\Pi_{\delta,\text{ref}}\right\rangle$, see equation
(\ref{eq:Pi reference in pricipale axe}) because the moduli of $\delta
{\boldsymbol w}$ and of its projections on the compressive and
stretching eigenvectors are preferentially correlated with compressive
rather than stretching eigenvalues of the reference field's relative
deformation tensor $\boldsymbol{\Xi}^{(1)}$ (see equations
((\ref{eq:Pi reference in pricipale axe}) and (\ref{eq:I down term of
  budget in eigen base})).

\begin{figure}
	\centering \subfigure[]{
		\label{fig:alignment of deltaw in xi}
		\includegraphics[width=0.48\textwidth]{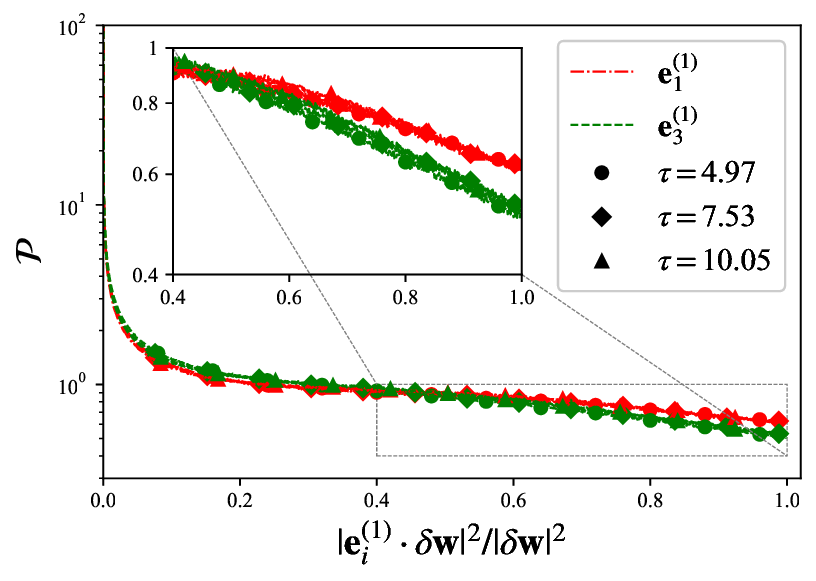}}
	\subfigure[]{
		\label{fig:alignment of overlinew in xi}
		\includegraphics[width=0.48\textwidth]{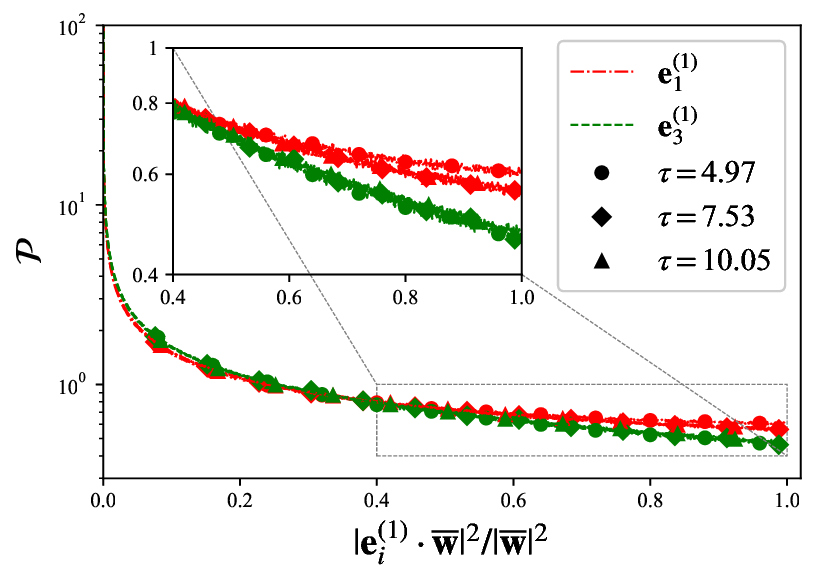}}
	\subfigure[]{
		\label{fig:alignment of deltaew in deltaxi}
		\includegraphics[width=0.48\textwidth]{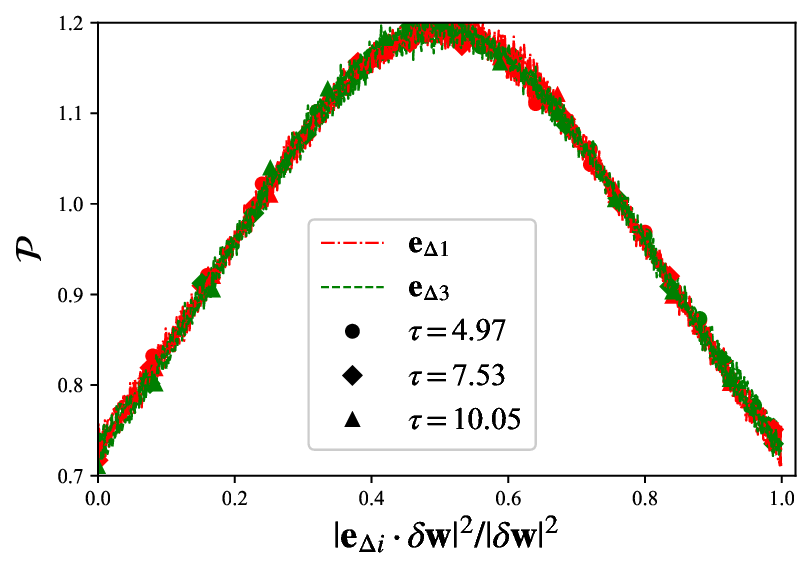}}
	\caption{At the scale $r=l_{\Delta}$, the probability
		distribution function (PDF) of the alignment coefficient for
		(a)
		$\left|\boldsymbol{e}^{(1)}_{i}\cdot\delta\boldsymbol{w}\right|^{2}/\left|\delta
		\boldsymbol{w}\right|^{2}$, (b)
		$\left|\boldsymbol{e}^{(1)}_{i}\cdot\overline{\boldsymbol{w}}\right|^{2}/\left|\overline{\boldsymbol{w}}\right|^{2}$
		and (c) $\left|\boldsymbol{e}_{\Delta
			i}\cdot\delta\boldsymbol{w}\right|^{2}/\left|\delta
		\boldsymbol{w}\right|^{2}$ at different time. (For the definition of
		$r$ in this plot see the caption of figure
		\ref{fig:eigenvalueXi}.)}
	\label{fig:alignment} 
\end{figure}

The one remaining interscale transfer rate is the one for the
nonlinear interscale tranfer fully within the uncertainty field, i.e.
$\left\langle\Pi_{\delta,\text{err}}\right\rangle$. As can be seen in
(figure \ref{fig:KHMHequation}),
$\left\langle\Pi_{\delta,\text{err}}\right\rangle$ is negative and is
hardly detectable except towards the end of the power law regime. It
therefore corresponds to an inverse nonlinear interscale transfer
driven by stretching motions as explained in subsection
\ref{sec:Scale-by-scale uncertainty energy balance in forced
	homogeneous turbulence}. Equation (\ref{eq:pi err principale axes}) shows that the effects of
stretching/compressive motions on
$\left\langle\Pi_{\delta,\text{err}}\right\rangle$ occur via the
eigenvalues of $\boldsymbol{\Xi}_{\Delta}$ (the uncertainty field's relative
deformation rate) and the alignments of $\delta {\boldsymbol w}$ with the
eigenvectors of $\boldsymbol{\Xi}_{\Delta}$. These alignments are presented in
figure \ref{fig:averagealignment of deltaew in deltaxi} which shows
that all the uncertainty energy aligns with both $\boldsymbol{e}_{\Delta1}$
and $\boldsymbol{e}_{\Delta3}$. There is only a very slight preference for
alignment with the stretching direction, which is hardly visible in
the probability distributions of the alignments of uncertainty energy
to $\boldsymbol{e}_{\Delta1}$ and $\boldsymbol{e}_{\Delta3}$ in figure
\ref{fig:alignment of deltaew in deltaxi}, which actually appear to
collapse throughout the stochastic time range. This slight preference
results in $\left\langle\Pi_{\delta,\text{ref}}\right\rangle$ being
less than zero, though small, if not negligible, compared to the three
other interscale transfer rates.



Looking now at the evolution through the different time regimes in
figure \ref{fig:KHMHequation}, we note that
$\left\langle\mathcal{I}_{\delta\downarrow}\right\rangle$,
$\left\langle\Pi_{\delta,\text{ref}}\right\rangle$ and
$-\left\langle\Pi_{\delta,\text{err}}\right\rangle$ grow gradually
over time. Their contributions are small and may be considered
negligible during the chaotic exponential growth regime, so that
$\left\langle\mathcal{I}_{\delta\uparrow}\right\rangle$ dominates
interscale transfers at scales $r< l_{\Delta}$. In this regime,
$\partial \left\langle\mathcal{A}_{\delta}\right\rangle/\partial t$ is
significantly non-zero at $r$ values above the Kolmogorov length. At
scales below $l_{\Delta}$ this time derivative is determined mainly by
the competing influences of molecular diffusion minus dissipation
$\left\langle\mathcal{D}_{\delta}\right\rangle - \left\langle
\varepsilon_{\Delta}\right\rangle$ which causes
$\left\langle\mathcal{A}_{\delta}\right\rangle$ to decrease and
interscale transfer rate plus production
$\left\langle\mathcal{I}_{\delta\uparrow}\right\rangle + \left\langle
P_{\Delta}\right\rangle$ which causes it to increase.  At scales
$l_{\Delta}$ and larger, molecular diffusion is effectively zero and
all interscale transfers are between small and negligible which is
consistent with the fact, shown by \citet{ge2023production},
that $l_{\Delta}$ remains approximately constant during the chaotic
exponential growth regime.

After the exponential growth of $\left\langle
E_{\Delta}\right\rangle$, the interscale transfer rates
$\left\langle\mathcal{I}_{\delta\uparrow}\right\rangle$ and
$\left\langle\mathcal{I}_{\delta\downarrow}\right\rangle$ are no
longer negligible at any $r$ between $l_{\Delta}$ and $2l_{\Delta}$,
as shown in figure \ref{fig:KHMHequation}(b),(c),(d), with
$\left|\left\langle\mathcal{I}_{\delta\uparrow}\right\rangle(l_{\Delta})\right|>\left|\left\langle\mathcal{I}_{\delta\downarrow}\right\rangle(l_{\Delta})\right|$,
suggesting that uncertainty energy is transported across $l_{\Delta}$
to larger scales, thereby resulting in growth of $l_{\Delta}$. The
time-derivative of the uncertainty structure function
$\left\langle\mathcal{A}_{\delta}\right\rangle$ remains non-negligible
at all scales during the exponential of exponential growth regime, as
shown in figure \ref{fig:KHMHequation f2 2}.

In the power law growth regime, however, there appears a clear
tendency for $\partial
\left\langle\mathcal{A}_{\delta}\right\rangle/\partial t$ to approach
zero for all $r$ smaller than $l_{\Delta}$ as time increases (see
figures \ref{fig:KHMHequation f2 3} and \ref{fig:KHMHequation f2 4}),
thereby supporting the equilibrium $\partial
\left\langle\mathcal{A}_{\delta}\right\rangle/\partial t\approx0$ on
which the equilibrium model of uncertainty growth is built in
subsection \ref{sec:Production/dissipation scaling of uncertainty
	energy}. This tendency towards equilibrium at scales $r$ below
$l_{\Delta}$ and the fact that viscous diffusion
$\left\langle\mathcal{D}_{\delta}\right\rangle$ is negligible at
scales $r>l_{\lambda}^{(1)}$ (see figures \ref{fig:KHMHequation f2 3}
and \ref{fig:KHMHequation f2 4}) supports the prediction made in
subsection \ref{sec:Production/dissipation scaling of uncertainty
	energy} of a self-similar equilibrium cascade in the uncertainty
field, and equation (\ref{eq:simplified strcuture scale-by-scale
	uncertainty budget}) in particular, in the range
$l_{\lambda}^{(1)}<r<l_{\Delta}$ during the power law growth regime.


We close this subsection with some observations about the
eigenvalues/eigenvectors of $\boldsymbol{\Xi}^{(1)}$ and
$\boldsymbol{\Xi}_{\Delta}$.  Figure \ref{fig:eigenvalueXi}
illustrates the dependence on scale and time of the spatially averaged
eigenvalues of $\boldsymbol{\Xi}^{(1)}$ and
$\boldsymbol{\Xi}_{\Delta}$ normalized by the spatially averaged
modulus of the single-point strain rate tensor (only
  one single realisation of $\varepsilon_{0}=0.1$, but results hold
  for other realisations and the other F2 cases). In agreement with
our analysis in subsection \ref{sec:Scale-by-scale uncertainty energy
  balance in forced homogeneous turbulence}, both
$\boldsymbol{\Xi}^{(1)}$ and $\boldsymbol{\Xi}_{\Delta}$ have only two
non-zero eigenvalues and their spatial averages sum up to zero. Our
DNS data reveals that these spatial averages decrease with increasing
scale and are approximately statistically stationary in time.
Figures \ref{fig:averagealignment of deltaw in xi} and
\ref{fig:averagealignment of overlinew in xi} depict the spatial
alignments of $\delta\boldsymbol{w}$ and $\overline{\boldsymbol{w}}$ with the
eigenvectors of $\boldsymbol{\Xi}^{(1)}$ for various scales and times (also one single realisation of $\varepsilon_{0}=0.1$). We
observe that these spatially averaged alignments remain quasi-constant
across scales. A significant portion of the uncertainty energy is
consistently aligned with the compression direction across both small
and large scales, throughout all observed times. Additionally, figures
\ref{fig:alignment of deltaw in xi} and \ref{fig:alignment of
	overlinew in xi} demonstrate that whereas weak alignments of the
uncertainty field with the compression and stretching direction of
$\boldsymbol{\Xi}^{(1)}$ are comparably probable, strong alignments of the
uncertainty field with the compression direction of $\boldsymbol{\Xi}^{(1)}$
are significantly more probable that strong alignments of the
uncertainty field with the stretching direction of $\boldsymbol{\Xi}^{(1)}$.


\begin{figure}
	\centering
	\includegraphics[width=0.7\textwidth]{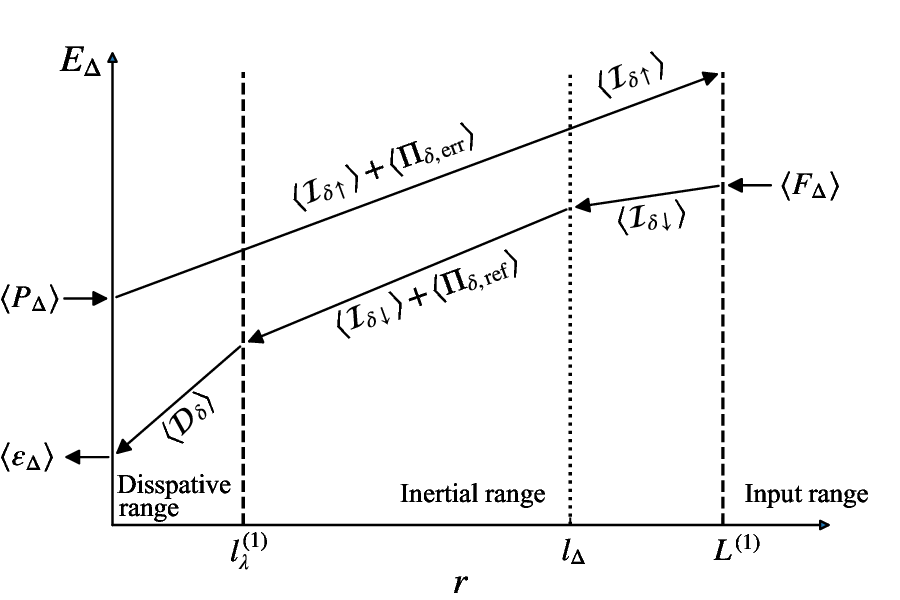}
	\caption{Qualitative schematic plot for the uncertainty transfer in scales} 
	\label{fig:schematicplot} 
\end{figure}

\subsection{Dual cascades hypothesis\label{sec:Validation of equilibrium dual cascades of uncertainty energy}}

We summarise the main results of the previous subsection. Local
one-point compressions of the reference field's strain rate tensor
enhance uncertainty by a production mechanism described by
\citet{ge2023production}. Then, compressive motions of the reference
field's relative deformation rate tensor generate an inverse cascade
based on their interactions with the uncertainty half sum velocity
field which transfers uncertainty energy from zero length scales $r$
to larger ones (average interscale transfer rate
$\left\langle\mathcal{I}^{a}_{\delta\uparrow}\right\rangle$). These
compressive motions also interact with the uncertainty half difference
velocity field to generate a forward cascade which transfers energy
from large to small scales (average interscale transfer rates
$\left\langle\mathcal{I}^{a}_{\delta\downarrow}\right\rangle$ and
$\left\langle\Pi^{a}_{\delta,\text{ref}}\right\rangle$). This forward
transfered uncertainty energy originates from the forcing input rate,
if there is one, and from the uncertainty production-driven inverse
cascade. There is also non-linear interscale transfer which is fully
within the uncertainty field and operates independently from the
reference field by uncertainty field stretching actions which generate
a very slow inverse cascade (very small negative average interscale
transfer rate
$\left\langle\Pi^{a}_{\delta,\text{err}}\right\rangle$). In total,
$\left\langle\mathcal{I}^{a}_{\delta\uparrow}\right\rangle$ dominates
and the interscale transfer is overall inverse with uncertainty energy
going from small to large scales, reflecting the dominance of
uncertainty production over uncertainty dissipation. Note that at
scales $r$ larger than $l_{\Delta}$ the two weakest interscale
transfer rates, $\left\langle\Pi_{\delta,\text{ref}}\right\rangle$ and
$\left\langle\Pi_{\delta,\text{err}}\right\rangle$, quickly attenuate
to near-zero values compared to
$\left\langle\mathcal{I}_{\delta\downarrow}\right\rangle$ and
$\left\langle\mathcal{I}_{\delta\uparrow}\right\rangle$. This summary
is schematically represented in figure \ref{fig:schematicplot}.

During the stochastic power-law growth regime, the interscale
transfers are in equilibrium at scales $r$ smaller than $l_{\Delta}$
and viscous diffusion is negligible at scales $r$ larger than
$l_{\lambda}^{(1)}$, giving rise to the self-similar equilibrium
cascade represented by the balance (\ref{eq:simplified strcuture
	scale-by-scale uncertainty budget}) in the inertial range
$l_{\lambda}^{(1)} < r < l_{\Delta}$. It is natural to ask whether
this self-similar equilibrium cascade consists of two
separate-on-average cascades in this scale-range, one forward, driven
by uncertainty dissipation and represented by the balance
\begin{equation}
	\label{eq:simplified DOWN scale-by-scale uncertainty budget}
	\left\langle \varepsilon_{\Delta}\right\rangle \approx
	\left\langle\Pi^{a}_{\delta,\text{ref}}\right\rangle+
	\left\langle\mathcal{I}^{a}_{\delta\downarrow}\right\rangle
\end{equation}
and one inverse, driven by uncertainty production and represented by
\begin{equation}
	\label{eq:simplified UP scale-by-scale uncertainty budget}
	-\left\langle P_{\Delta}\right\rangle \approx
	\left\langle\Pi^{a}_{\delta,\text{err}}\right\rangle +
	\left\langle\mathcal{I}^{a}_{\delta\uparrow}\right\rangle .
\end{equation}
This is the dual cascades hypothesis for the stochastic power-law
growth regime. A look at figures \ref{fig:KHMHequation f2 3} and
\ref{fig:KHMHequation f2 4} does not support this hypothesis since
both
$\left|\left\langle\Pi_{\delta,\text{ref}}\right\rangle+\left\langle\mathcal{I}_{\delta\downarrow}\right\rangle\right|$
and
$\left|\left\langle\mathcal{I}_{\delta\uparrow}\right\rangle+\left\langle\Pi_{\delta,\text{err}}\right\rangle\right|$
decrease with increasing scale in the range
$l_{\lambda}^{(1)}<r<l_{\Delta}$ (as actually also happens in the
reference field's energy cascade \citep{yasuda2018spatio}).
However, it may be that the Reynolds numbers of our DNS are not
sufficiently large.



\begin{figure}
	\centering \subfigure[]{
		\label{fig:dualcascadeepsilon1}
		\includegraphics[width=0.48\textwidth]{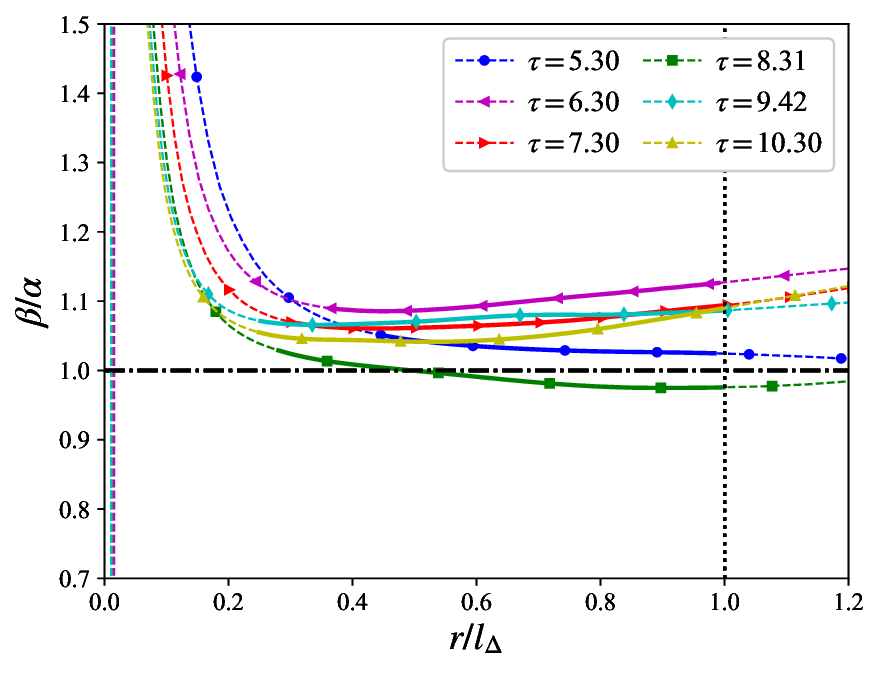}}
	\subfigure[]{
		\label{fig:dualcascadeepsilon2}
		\includegraphics[width=0.48\textwidth]{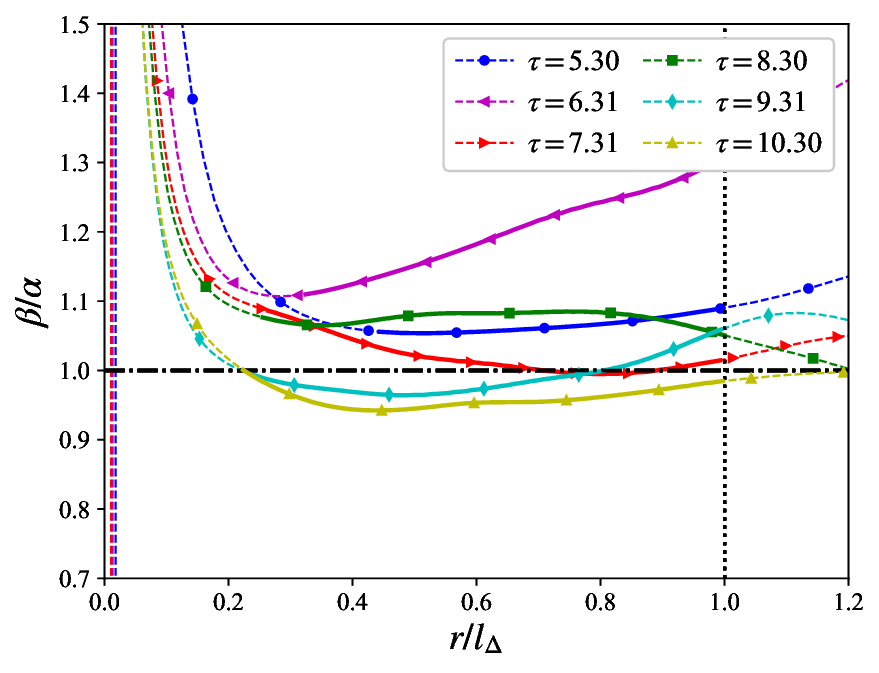}}
	\subfigure[]{
		\label{fig:dualcascadeepsilon3}
		\includegraphics[width=0.48\textwidth]{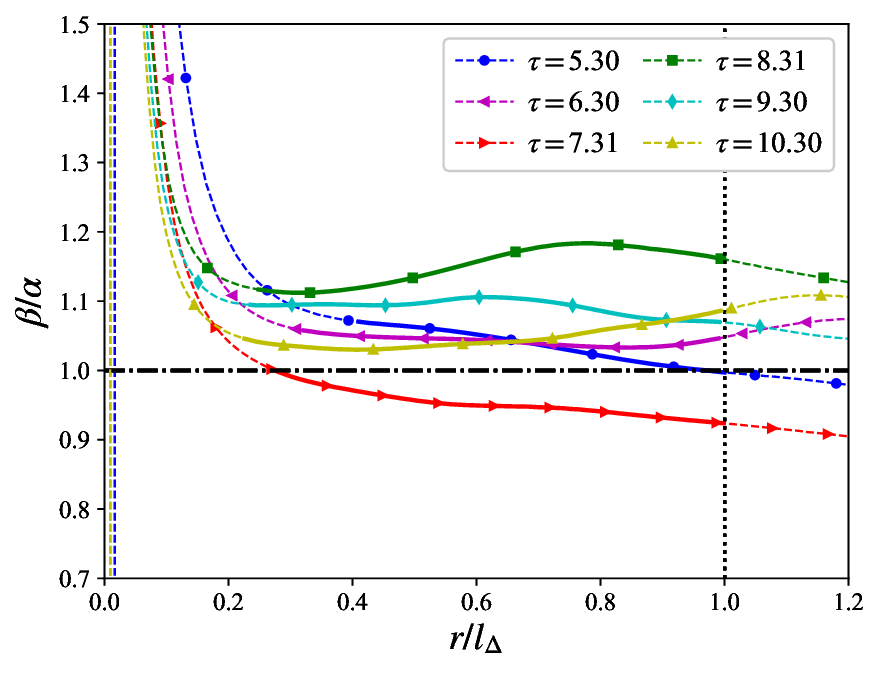}}
	\caption{Scale evolutions of $\beta/\alpha$ at different times
          during the power law growth regime for
            cases F2 with (a) $\varepsilon_{0}=0.1$; (b)
            $\varepsilon_{0}=0.15$; (c) $\varepsilon_{0}=0.2$. For
            $\varepsilon_{0}=0.1$, the data are ensemble averaged over
            six realisations, whereas for the values of
            $\varepsilon_0$ the data are taken from one single
            realisation.  The parts of the curves within the range of
          $l_{\lambda}^{(1)}<r<l_{\Delta}$ are plotted as solid lines
          whereas the parts before and after this range are dotted
          lines.}
	\label{fig:dualcascade} 
\end{figure}

We attempt an indirect examination of the dual cascades hypothesis on
the basis of a necessary condition for its validity which may be valid
even at Reynolds numbers where the dual cascades hypothesis is
not. This necessary but not sufficient condition for the validity of
equations (\ref{eq:simplified DOWN scale-by-scale uncertainty budget})
and (\ref{eq:simplified UP scale-by-scale uncertainty budget}) in the
scale- range $l_{\lambda}^{(1)}\ll r\ll l_{\Delta}$, is that the
following equality holds in that range:
\begin{equation}
	\label{eq:necessary condition approximation simplified strcuture scale-by-scale uncertainty budget}
	\frac{\left\langle
		P_{\Delta}\right\rangle}{\left\langle\varepsilon_{\Delta}\right\rangle}\approx-\frac{\left\langle\Pi^{a}_{\delta,\text{err}}\right\rangle+\left\langle\mathcal{I}^{a}_{\delta\uparrow}\right\rangle}{\left\langle\Pi^{a}_{\delta,\text{ref}}\right\rangle+\left\langle\mathcal{I}^{a}_{\delta\downarrow}\right\rangle}.
\end{equation}

\begin{figure}
	\centering \subfigure[]{
		\label{fig:scale average of b2a}
		\includegraphics[width=0.48\textwidth]{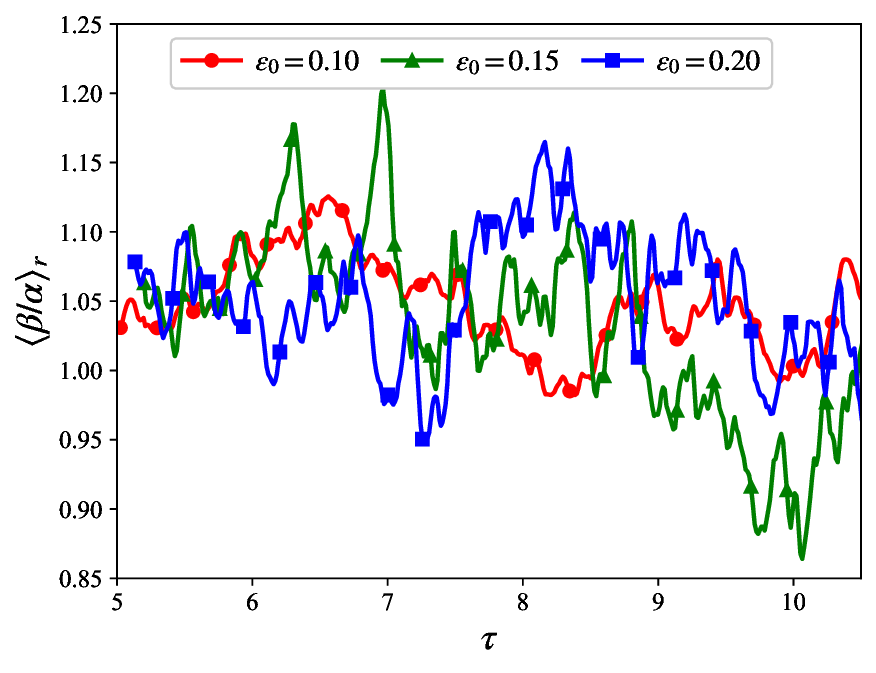}}
	\subfigure[]{
		\label{fig:standard deviation of b2a}
		\includegraphics[width=0.48\textwidth]{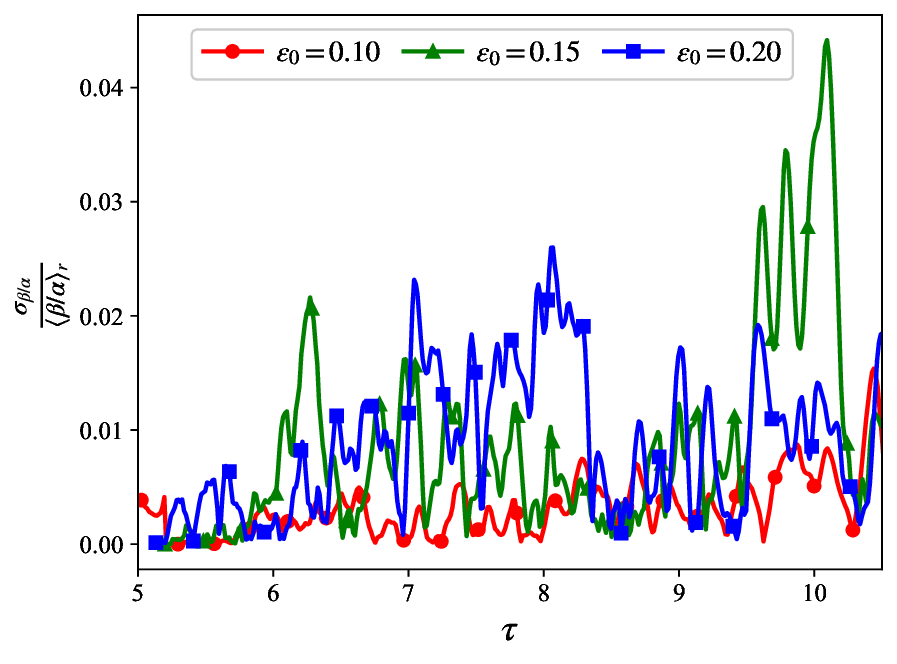}}
	\caption{(a) Time evolutions of
		$\left\langle\beta/\alpha\right\rangle_{r}$ and (b) its
		normalised standard deviation
		$\sigma_{\beta/\alpha}/\left\langle\beta/\alpha\right\rangle_{r}$
		for F2 cases. }
	\label{fig:b2a} 
\end{figure}

We define $\alpha(t)=\left\langle
P_{\Delta}\right\rangle/\left\langle\varepsilon_{\Delta}\right\rangle$
and
$\beta(r,t)=-\left(\left\langle\Pi^{a}_{\delta,\text{err}}\right\rangle+\left\langle\mathcal{I}^{a}_{\delta\uparrow}\right\rangle\right)/\left(\left\langle\Pi^{a}_{\delta,\text{ref}}\right\rangle+\left\langle\mathcal{I}^{a}_{\delta\downarrow}\right\rangle\right)$
and in figure \ref{fig:dualcascade} we plot the scale evolution of
$\beta/\alpha$ at different times for the three F2 cases. In most cases and times, the $\beta/\alpha$ curve
reaches an approximate constant plateau in the scale-range
$l_{\lambda}^{(1)}<r<l_{\Delta}$ with a value reasonably close to
1. To quantitatively evaluate this value, we define the average of
$\beta/\alpha$ over the scales $r$ in the range
$1.5l_{\lambda}^{(1)}<r<0.7l_{\Delta}$ and denote it
$\left\langle\beta/\alpha\right\rangle_{r}$.
We present its time evolution in figure \ref{fig:scale average of
  b2a}. In all three F2 DNS cases,
$\left\langle\beta/\alpha\right\rangle_{r}$ appears statistically
stationary around a time-average value
$\left\langle\left\langle\beta/\alpha\right\rangle_{r}\right\rangle_{\tau}=1.05\pm0.03$,
$1.03\pm0.07$ and $1.05\pm0.05$ for the cases $\varepsilon_{0}=0.10$,
$0.15$ and $0.20$, respectively
($\left\langle\cdot\right\rangle_{\tau}$ represents the time average
within $\tau\in[5.0,10.5]$). The standard deviation of
$\left\langle\beta/\alpha\right\rangle_{r}$, defined as
$\sigma_{\beta/\alpha}(\tau)=\sqrt{\frac{1}{0.7l_{\Delta}-1.5l_{\lambda}^{(1)}}\int_{r=1.5l_{\lambda}^{(1)}}^{0.7l_{\Delta}}\left(\beta/\alpha(r,\tau)-\left\langle\beta/\alpha\right\rangle_{r}(\tau)\right)^{2}\mathrm{d}r}$,
is plotted in figure \ref{fig:standard deviation of b2a}, where it can
be seen that it is smaller than $4\%$ of
$\left\langle\beta/\alpha\right\rangle_{r}$ for all three cases and
all times. (Note that the $\varepsilon_{0}=0.1$ case
  exhibits the most stable plateau value close to $1$, presumably
  because it has been ensemble-averaged.) Whereas we are not able
with our DNS to validate the dual cascade hypothesis, our DNS does
permit us to validate its necessary condition (\ref{eq:necessary
  condition approximation simplified strcuture scale-by-scale
  uncertainty budget}).

\subsection{Uncertainty production and dissipation scalings\label{sec:Non-equilibrium correction}}

An underpinning of the uncertainty production and dissipation scalings
in subsection \ref{sec:Production/dissipation scaling of uncertainty
	energy} is the self-similar equilibrium cascade represented by the
balance (\ref{eq:simplified strcuture scale-by-scale uncertainty
	budget}). Having found support for the self-similar equilibrium
cascade in the range $l_{\lambda}^{(1)}< r < l_{\Delta}$ with our DNS
data in subsection \ref{sec:Scale-by-scale uncertainty energy budget},
we now investigate the uncertainty production and dissipation scalings
with these data. In subsection \ref{sec:Time evolution of single-point
	uncertainty} we have concluded that our DNS coherently favour
$l_{\Delta} \sim t$ and $\left\langle E_{\Delta}\right\rangle \sim
t^{2/3}$ over $l_{\Delta} \sim t^{3/2}$ and $\left\langle
E_{\Delta}\right\rangle \sim t$ in the power law time range. For the
picture to be truly coherent, our DNS should also favour the
scalings (\ref{eq:production/disspation scaling law2}) over the
scalings (\ref{eq:production/disspation scaling law}) because the
former imply $l_{\Delta} \sim t$ and $\left\langle
E_{\Delta}\right\rangle \sim t^{2/3}$ whereas the latter imply
$l_{\Delta} \sim t^{3/2}$ and $\left\langle E_{\Delta}\right\rangle
\sim t$ in the power law regime.

\begin{figure}
	\centering 
	\includegraphics[width=0.9\textwidth]{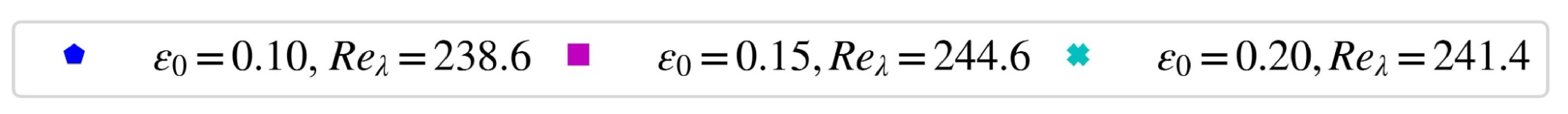}
	\subfigure[]{
		\label{fig:ProductionScalingLawClassical}
		\includegraphics[width=0.48\textwidth]{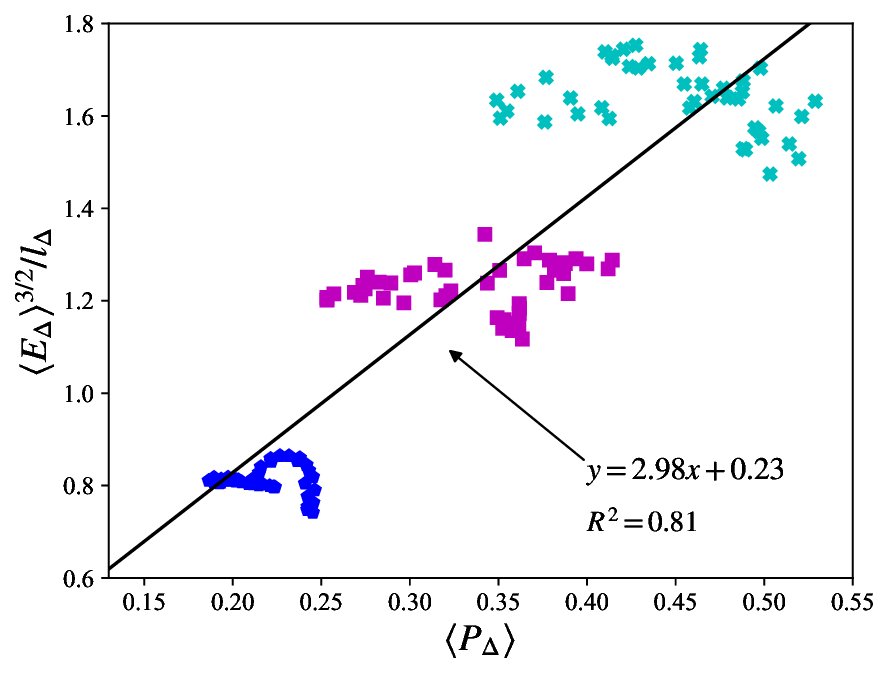}}
	\subfigure[]{
		\label{fig:DisspationScalingLawClassical}
		\includegraphics[width=0.48\textwidth]{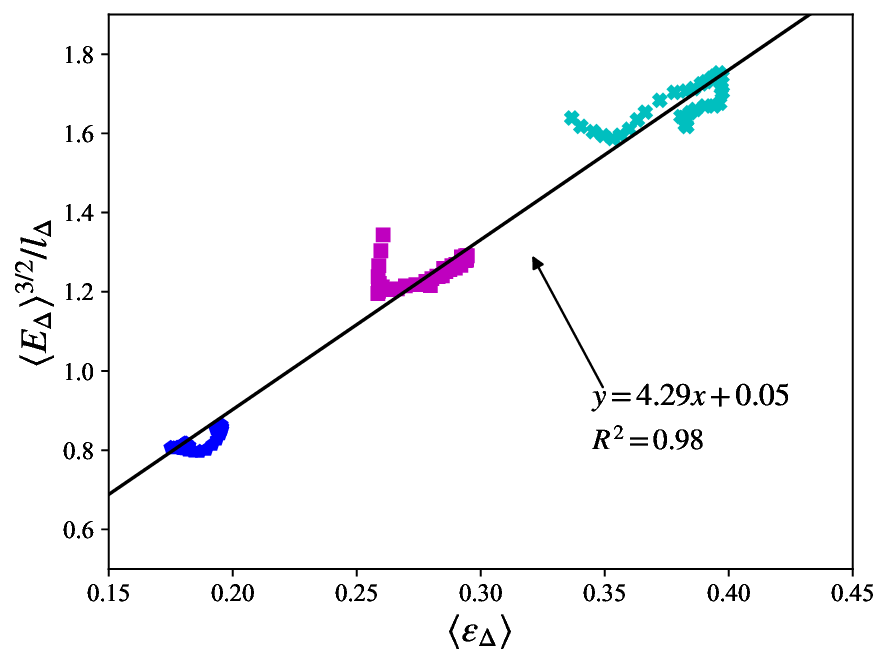}}
	\subfigure[]{
		\label{fig:ProductionScalingLaw}
		\includegraphics[width=0.48\textwidth]{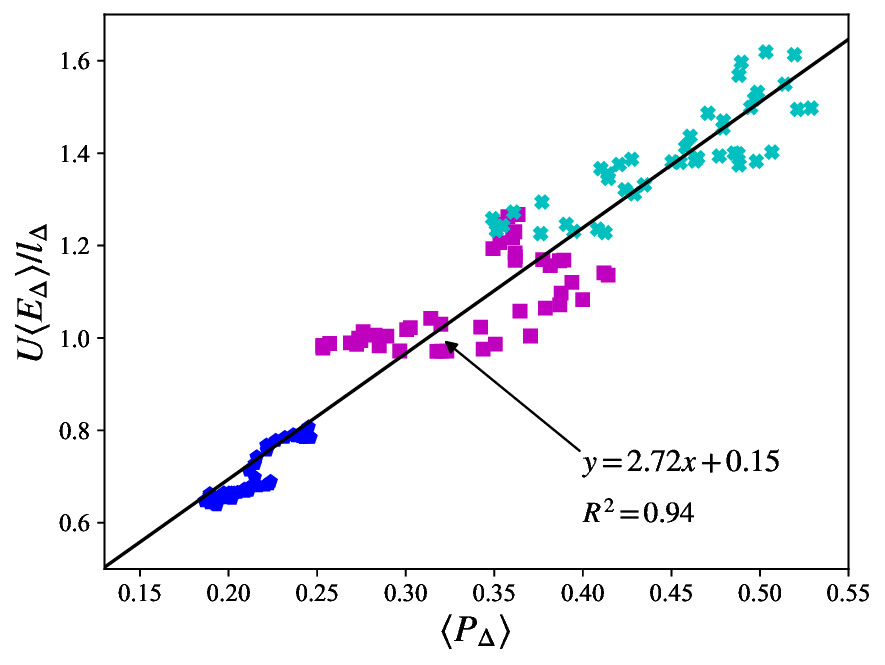}}
	\subfigure[]{
		\label{fig:DisspationScalingLaw}
		\includegraphics[width=0.48\textwidth]{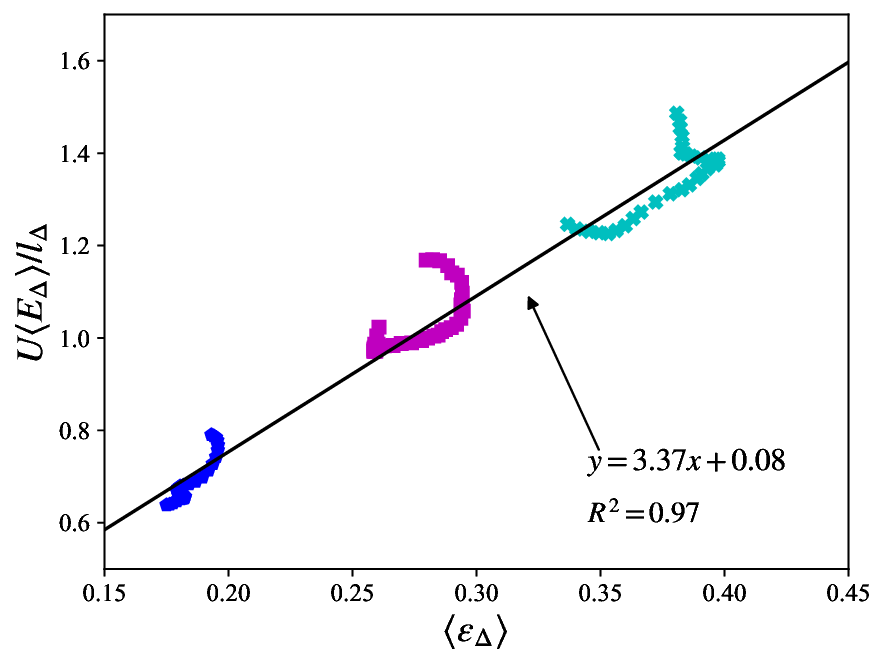}}	
	\caption{Scatter plots for (a) $\left\langle
          P_{\Delta}\right\rangle$ versus $\left\langle
          E_{\Delta}\right\rangle^{3/2}/l_{\Delta}$, (b) $\left\langle
          \varepsilon_{\Delta}\right\rangle$ versus $\left\langle
          E_{\Delta}\right\rangle^{3/2}/l_{\Delta}$ (c) $\left\langle
          P_{\Delta}\right\rangle$ versus $U^{(1)}\left\langle
          E_{\Delta}\right\rangle/l_{\Delta}$ and (d) $\left\langle
          \varepsilon_{\Delta}\right\rangle$ versus
          $U^{(1)}\left\langle E_{\Delta}\right\rangle/l_{\Delta}$
          during the time range $\tau\in[5.0,10]$ for F2 cases, where
          the linear regression and its $R^{2}$ value of the set of
          points of all six cases are given. For
            $\varepsilon_{0}=0.1$, the averaging includes ensemble
            averaging over six realisations, whereas for the other
            values of $\varepsilon_0$, the data are taken from one
            single realisation.}
	\label{fig:scalinglaw} 
\end{figure}

Figure \ref{fig:scalinglaw} presents four scatter plots generated with
data recording the evolutions in the stochastic time range $\tau \in
[5.0,10]$ of $\left\langle P_{\Delta}\right\rangle$, $\left\langle
\varepsilon_{\Delta}\right\rangle$, $\left\langle
E_{\Delta}\right\rangle^{3/2}/l_{\Delta}$ and $U^{(1)}\left\langle
E_{\Delta}\right\rangle/l_{\Delta}$ across all three F2 cases.
The scatter plot of $\left\langle P_{\Delta}\right\rangle$ and
$\left\langle E_{\Delta}\right\rangle^{3/2}/l_{\Delta}$ in figure
\ref{fig:ProductionScalingLawClassical} does not support the linear
relation between these two quantities advocated by the scalings
(\ref{eq:production/disspation scaling law}). On the other hand, the
scatter plot of $\left\langle P_{\Delta}\right\rangle$ and
$U^{(1)}\left\langle E_{\Delta}\right\rangle/l_{\Delta}$ in figure
\ref{fig:ProductionScalingLaw} does return a much better linear
relation and supports the scaling $\left\langle
P_{\Delta}\right\rangle \sim U^{(1)}\left\langle
E_{\Delta}\right\rangle/l_{\Delta}$ in (\ref{eq:production/disspation
	scaling law2}). Concerning the average uncertainty dissipation
scaling, the scatter plot of $\left\langle
\varepsilon_{\Delta}\right\rangle$ and $U^{(1)}\left\langle
E_{\Delta}\right\rangle/l_{\Delta}$ in figure
\ref{fig:DisspationScalingLaw} exhibits a same linear
relation as the scatter plot of $\left\langle
\varepsilon_{\Delta}\right\rangle$ and $\left\langle
E_{\Delta}\right\rangle^{3/2}/l_{\Delta}$ in figure
\ref{fig:DisspationScalingLawClassical}. All in all, our data return a
coherent picture where the uncertainty production and dissipation
rates follow the scalings (\ref{eq:production/disspation scaling
	law2}) and the uncertainty energy and integral length grow as
$l_{\Delta} \sim t$ and $\left\langle E_{\Delta}\right\rangle \sim
t^{2/3}$ in the stochastic time range.

\section{\label{sec:Conclusion}Conclusion}

In this paper we have studied the backward contamination of
uncertainty from smaller to larger scales in the inertial range of
three-dimensional, incompressible and statistically stationary
homogeneous/periodic turbulence where the turbulence cascade is
forward. Our focus has been on the growth of uncertainty by
stochasticity rather than chaoticity
\citep{lorenz1969predictability,thalabard2020butterfly}, and therefore
on the power law growth some time after the chaotic exponential growth
of average uncertainty energy $\left\langle
E_{\Delta}\right\rangle$.

At high enough Reynolds number
the external large-scale uncertainty input cannot influence the growth
of uncertainty in the inertial range ($F_{\Delta}=0$). The growth of
$\left\langle E_{\Delta}\right\rangle$ is fully controlled by internal
production ($\left\langle P_{\Delta}\right\rangle$) and dissipation
($\left\langle \varepsilon_{\Delta}\right\rangle$), according to
equation (\ref{eq:single point uncertainty equation}). To investigate
the interscale transport of uncertainty energy, we developed a
scale-by-scale uncertainty energy budget equation (\ref{eq:average
	difference KHMH difference/average}) based on spatially averaged
energy of two-point velocity half differences
$\left\langle\left|\delta\boldsymbol{w}\right|^{2}\right\rangle$,
and we applied it to DNS of statistically stationary periodic
turbulence. While the reference turbulence is statistically
stationary, the uncertainty field is not until it eventually fully
decorrelates from the reference field and reaches total (i.e. maximum)
uncertainty. Nevertheless, in the stochastic time range of power-law
growth of uncertainty, all uncertainty field length scales between the
Taylor length (above which viscous diffusion is negligible) and the
uncertainty field's integral length-scale can be considered to be
effectively decorrelated from the reference field and therefore
statistically stationary. With their periodicity/homogeneity this
implies that they are in a self-similar two-point equilibrium where
the uncertainty field interscale transfer rates sum up to be
independent of length-scale and equal to the difference between the
average uncertainty dissipation and production rates. This equilibrium
cascade of decorrelation is predominantly backward whereby uncertainty
produced by local strain rate compressions at zero-length scale, as
already demonstrated by \citet{ge2023production}, is
transfered to progressively larger scales by
$\left\langle\mathcal{I}_{\delta\uparrow}\right\rangle$. This transfer
to larger scales happens via compressions of the reference field's
relative deformation tensor (which we can refer to as two-point
compressions) and their alignments with the half sum uncertainty
field. These two-point compressions also align with the half
difference uncertainty field giving rise to positive
$\left\langle\mathcal{I}_{\delta\downarrow}\right\rangle$ and also
correlate with its local energy giving rise to positive
$\left\langle\Pi_{\delta,\text{ref}}\right\rangle$ so as to generate a
forward transfer of uncertainty from large to small scales fed by
large scale forcing input of uncertainty and/or the predominant
inverse uncertainty cascade. There is also a small, negligible for
most of the time, interscale transfer
($\left\langle\Pi_{\delta,\text{err}}\right\rangle$) of uncertainty by
and within the uncertainty field itself which is from small to large
scales. Unlike the other three interscale transfer mechanisms which
are linear and dominated by two-point compressions of the reference
field, this one is fully non-linear and dominated by stretching
motions of the relative deformation tensor (two-point stretchings) of
the uncertainty field itself when significantly non-zero.

To summarise the main points, one-point compressive motions
amplify/produce uncertainty and chaoticity locally in physical space
as previously demonstrated by \cite{ge2023production}; then, during
the time range of stochasticity, two-point compressions by the
reference field transfer uncertainty from zero to small to larger
scales predominantly via a linear inverse interscale transfer/cascade
mechanism involving the uncertainty field half sum (low pass
uncertainty field). A linear forward interscale transfer/cascade
occurs concurrently via reference field two-point compressions acting
on the uncertainty field half difference (high pass uncertainty field)
and feeding the non-zero dissipation rate of uncertainty. {\it
	Compressions are key to both chaoticity and stochasticity.}

In the stochastic time range of power-law growth of uncertainty our
DNS support the power laws $l_{\Delta} \sim t$ and $\left\langle
E_{\Delta}\right\rangle \sim t^{2/3}$ when there is
  no forcing-generated uncertainty. These power laws are based on the
uncertainty dissipation and production scalings
(\ref{eq:production/disspation scaling law2}). The difference between
these scalings (\ref{eq:production/disspation scaling law2}) and the
more conventional scalings (\ref{eq:production/disspation scaling
  law}) (which lead to different power laws for $l_{\Delta}$ and
$\left\langle E_{\Delta}\right\rangle$) is rooted in a presently naive
attempt to account for the effect that correlations between the
reference and uncertainty fields has on these scalings. This is an
important issue which requires substantial future research.
\backsection[Acknowledgements]{Jin Ge acknowledges financial support
  from the China Scholarship Council. We are grateful for the access
  to the computing resources supported by the Zeus supercomputers
  (Mésocentre de Calcul Scientifique Intensif de l'Université de
  Lille); Cluster Austral (Centre Régional
    Informatique et d'Applications Numériques de Normandie) and
    Lantuxinsuan (Chengdu) Technology Co., Ltd.. We also thank
    Dr. Yucang RUAN from Peking University and Mr. Xinli GE from BROAD
    Group for the help in obtaining computational resources.  }

\backsection[Funding]{This research received no specific grant from any funding agency, commercial or not-for-profit sectors. }

\backsection[Declaration of interests]{The authors report no conflict of interest.}

\bibliographystyle{jfm}
\bibliography{jfm}



\end{document}